\documentclass[a4paper,fleqn,usenatbib]{mnras}

\usepackage{newtxtext,newtxmath}

\usepackage[T1]{fontenc}
\usepackage{ae,aecompl}

\usepackage{graphicx}	
\usepackage{amsmath}	
\usepackage{siunitx}	
\usepackage{subfig}

\usepackage{ulem} 

\newcommand{\lya}{Ly\,$\upalpha$}       
\newcommand{\ha}{H\,$\upalpha$}         
\newcommand{\hb}{H\,$\upbeta$}          
\newcommand{\mg}{Mg\,\textsc{ii}}      
\newcommand{\oiii}{[O\,\textsc{iii}]}  
\newcommand{\kms}{km\,s$^{-1}$}        
\newcommand{\xmm}{\textit{XMM-Newton}} 
\defcitealias{Rakshit17}{R17}
\defcitealias{Rakshit20}{R20}
\defcitealias{Mejia16}{MR16}
\defcitealias{Jin1}{J12a}
\defcitealias{Jin2}{J12b}
\defcitealias{Jin3}{J12c}
\defcitealias{Paper1}{C15}
\defcitealias{Paper2}{C17}
\newcommand{\ts}{\textsuperscript}

\title[The SOUX AGN Sample -- Sample definition]{
The SOUX AGN Sample: SDSS--\textit{XMM-Newton} Optical, Ultraviolet and X-ray selected active galactic nuclei spanning a wide range of parameter space -- Sample definition}

\author[D. Kynoch et al.]{
Daniel Kynoch,$^{1,2,3}$\thanks{E-mail: d.kynoch@soton.ac.uk}
Jake A. J.\ Mitchell,$^{3}$
Martin J.\ Ward,$^{3}$
Chris Done,$^{3}$
\newauthor
Elisabeta Lusso$^{4,5}$ and
Hermine Landt$^{3}$
\\
$^{1}$School of Physics and Astronomy, University of Southampton, University Road, Southampton, SO17 1BJ, UK \\
$^{2}$Astronomical Institute, Czech Academy of Sciences, Bo\v{c}n\'{i} II 1401, 141 00 Prague, Czech Republic \\
$^{3}$Centre for Extragalactic Astronomy, Department of Physics, Durham University, South Road, Durham, DH1 3LE, UK \\
$^{4}$Dipartimento di Fisica e Astronomia, Universita di Firenze, via G. Sansone 1, I-50019 Sesto Fiorentino, Firenze, Italy\\
$^{5}$INAF – Osservatorio Astrofisico di Arcetri, L.go Enrico Fermi 5, I-50125 Firenze, Italy\\
}

\date{Accepted XXX. Received YYY; in original form ZZZ}

\pubyear{2023}

\begin{document}
\label{firstpage}
\pagerange{\pageref{firstpage}--\pageref{lastpage}}
\maketitle

\begin{abstract}
We assemble a sample of \textcolor{black}{696} type 1 AGN up to a redshift of $z=2.5$, all of which have an SDSS spectrum containing at least one broad emission line (\ha, \hb\ or \mg) and an \textit{XMM-Newton} X-ray spectrum containing at least 250 counts in addition to simultaneous optical/ultraviolet photometry from the \textit{XMM} Optical Monitor.
Our sample includes quasars and narrow-line Seyfert 1s: thus our AGN span a wide range in luminosity, black hole mass and accretion rate.
We determine single-epoch black hole mass relations for the three emission lines and find that \textcolor{black}{they} provide broadly consistent mass estimates whether the continuum or emission line luminosity is used as the proxy for the broad emission line region radius.
We explore variations of the UV/X-ray energy index $\alpha_\mathrm{ox}$ with the UV continuum luminosity and with black hole mass and accretion rate, and make comparisons to \textcolor{black}{the physical quasar spectral energy distribution (SED)} model \textcolor{black}{\textsc{qsosed}}.
The majority of the AGN in our sample lie in a region of parameter space with $0.02<L/L_\mathrm{Edd}<2$ as defined by \textcolor{black}{this} model, with narrow-line type 1 AGN offset to lower masses and higher accretion rates than typical broad-line quasars.
We find differences in \textcolor{black}{the dependence of} $\alpha_\mathrm{ox}$ \textcolor{black}{on UV luminosity} between both narrow/broad-line and radio-loud/quiet subsets of AGN\textcolor{black}{: $\alpha_\mathrm{ox}$ has a slightly weaker dependence on UV luminosity for broad-line AGN and radio-loud AGN have systematically harder $\alpha_\mathrm{ox}$.}
\end{abstract}




\section{Introduction}
Active galactic nuclei (AGN) are powered by mass accretion onto the supermassive black holes at the centres of galaxies.
Most of the gravitational potential energy of the infalling material is liberated as radiation across the full electromagnetic spectrum.
For typical, luminous AGN, the majority of their power emerges in the ultraviolet and optical bands as thermal radiation from a viscous accretion disc (\citealt{SS73}; \citealt{NT73}).
The non-thermal, hard X-ray emission emerges via the Compton upscattering of soft seed photons from the disc in a hot ($T_\mathrm{e}\sim100$~keV) plasma above the disc (the `corona' e.g.\ \citealt{Haardt91}).
A strong relationship between the ultraviolet and X-ray emission of AGN was found in early studies (e.g.\ \citealt{Tananbaum79}) and its nature has been explored further in more recent investigations (e.g.\ \citealt{Lusso17}, \citealt{Salvestrini19}).
The non-linear and remarkably tight relation between the ultraviolet and X-ray luminosities enables luminous AGN to be used as `standardisable candles' to test cosmological models (e.g.\ \citealt{Sacchi22}).
Clearly the disc and the corona are intimately related, but a detailed description of how has yet to be formulated.

For the majority of AGN, the peak of their energy output occurs in the far-UV/soft X-ray band which is unobservable because of Galactic extinction. 
This complicates the determination of the energetics and the disc-corona relationship.
One means of circumventing this problem is to make studies of high-$z$ AGN for which the spectral peak is redshifted to observable optical-UV wavelengths, as was done by \cite{Paper1}.
Making detailed studies of the optical and X-ray spectra of 11 sources, \cite{Paper1} were able to accurately determine the bolometric luminosities and accretion rates of the AGN, demonstrating estimates of the bolometric luminosity made by scaling from a single wavelength or narrow-band luminosity (e.g.\ at 2500\,\si\angstrom\ or 2--10~keV) are generally inaccurate (see also \citealt{Netzer19}).
However, obtaining high-quality multiwavelength spectra for high-$z$ AGN necessarily restricts the sample to a small number of bright quasars, thereby limiting the parameter space which can be explored.
Another approach is to sample the far-UV and soft X-ray emission on either side of the peak and use a physical model (e.g.\ \citealt{Done12}) to `bridge the gap' and recover the intrinsic spectral energy distribution (SED).
This method was adopted by \cite{Jin1}, who explored the optical and X-ray spectroscopic properties of 51 AGN in a series of works (\citealt{Jin1,Jin2,Jin3}).
They showed that the broadband SEDs of AGN could generally be fit with three components: a standard accretion disc, a hot corona, and an intermediate `warm corona' responsible for the observed excess of soft X-ray emission (e.g.\ \citealt{Magdziarz98}; \citealt{GD04}).
The findings of \cite{Jin1,Jin2,Jin3} have enabled the development and refinement of new, physical models of AGN SEDs (\citealt{Kubota18,Kubota19}).
These models have been constructed with reference to a small number of representative AGN SEDs.
In the decade since the work of \cite{Jin1,Jin2,Jin3} the continuing SDSS (\citealt{Blanton17}; \citealt{SDSS-DR14Q}) and \xmm\ (\citealt{4XMM-DR9}) surveys have greatly increased the number of AGN with quality optical/UV and X-ray spectra.
We now have the opportunity to test the predictions made by current physical models against the observed properties of a large and diverse sample of AGN. 

Achieving a more rigorous understanding of the accretion flow and disc-corona relationship will also enable us to better address another outstanding problem in astrophysics: the origin of relativistic jets. 
How AGN launch and power relativistic jets of outflowing matter is an active area of research.
Progress towards a fuller understanding of the nature of jets and their relation to the accretion flow may be made via comparisons to accreting stellar-mass black holes (X-ray binaries: XRBs, e.g.\ \citealt{Fender04}) in which the disc-corona system likely regulates the launching of jets.
Many studies have explicitly investigated the analogies between AGN and XRBs in terms of the accretion flow properties 
(e.g. \citealt{Noda18}; \citealt{Ruan19}; \citealt{Arcodia20}) and the disc-jet coupling
(e.g.\  \citealt{Merloni03}; \citealt{Kording06}; \citealt{Svoboda17}; \citealt{FO21}; \citealt{Moravec22}).
Generally these studies are supportive of a unified scheme of mass accretion and ejection across the mass scale, although many open questions have yet to be resolved.
Studies of high accretion rate AGN in the local Universe (narrow-line Seyfert 1s: NLS1s) have found a relative deficit of radio-loud sources: only $\approx5$~per cent of NLS1s are radio-loud (e.g.\ \citealt{Komossa06}; \citealt{Rakshit17}) compared with $\approx15$--20~per cent of quasars (e.g.\ \citealt{Kellerman89}), again suggesting a relationship between the accretion flow (the disc-corona configuration) and the presence of a radio jet. 

Previous studies have mostly either investigated very large and diverse AGN samples without a detailed exploration of their spectroscopic properties (e.g.\ \citealt{Svoboda17}) or have presented a thorough analysis of much smaller, focussed samples of AGN with high-quality spectroscopic data (e.g.\ \citealt{Jin1,Jin2,Jin3} and \citealt{Paper1,Paper2}).
In this first paper in a series, we expand on the work of \cite{Jin1,Jin2,Jin3} and \cite{Paper1,Paper2} and make use of recent optical, UV and X-ray catalogs to compile a large sample of AGN (696 unique sources) with both optical and X-ray spectra in addition to broad multi-wavelength coverage.
This new sample has the advantage of being large enough to investigate population statistics whilst still having good quality multi-wavelength data.
It has been selected to probe a diverse range of AGN properties; as we will demonstrate, the sample spans several orders of magnitude in black hole mass and luminosity (accretion rate).
The sample is a well-suited selection with which to perform detailed investigations of the evolution of the disc-corona system with black hole mass and accretion rate (Mitchell et al., submitted).
In this paper we make key measurements of AGN properties and take a preliminary look at the evolution of the spectral energy distribution with both mass and luminosity. 
In Section~\ref{sec:selection} we describe the assembly of the sample; in Section~\ref{sec:mwl} we obtain multi-wavelength measurements for the selected AGN; in Section~\ref{sec:mass} we determine black hole masses from optical/UV data; we investigate the UV/X-ray energy index $\alpha_\mathrm{ox}$ and make comparisons with a physical model in Section~\ref{sec:aox}; finally in Sections~\ref{sec:disc} and \ref{sec:conc} we discuss our results and summarise our conclusions.
To convert fluxes to luminosities we have assumed a flat $\Lambda$CDM cosmology with $H_0=70$~km\,s$^{-1}$\,Mpc$^{-1}$, $\Omega_\mathrm{m}=0.3$ and $\Omega_\Lambda=0.7$.

\section{Sample selection}
\label{sec:selection}
\subsection{The parent catalogues}
\subsubsection{Optical spectra: SDSS-DR14Q}
The Quasar Catalog of the Fourteenth SDSS Data Release (SDSS-DR14Q, \citealt{SDSS-DR14Q}) contains 526356 quasars.
Each quasar has been the subject of \textcolor{black}{at} least one optical spectroscopic observation recorded on or before 2016 December 5.
The catalogue also contains SDSS photometric data from imaging observations.
Multiwavelength data from \textit{ROSAT}, \xmm, \textit{GALEX}, 2MASS, \textit{WISE}, UKIDSS and FIRST are included, where available.
The \xmm\ data are taken from the 3XMM-DR7 catalogue (an earlier version of the serendipitous survey catalogue: see below).
14736 of the SDSS-DR14Q quasars ($\approx3$~per cent) have an X-ray source detection within a 5~arcsec matching radius.
Although the catalogue contains some X-ray data obtained from an \xmm\ catalogue (3XMM-DR7), it does not contain any optical/UV photometry recorded with the \xmm\ Optical Monitor (OM). 

\subsubsection{X-ray data: 4XMM-DR9}
4XMM-DR9 is the first iteration of the fourth source catalogue compiled from the \textit{XMM-Newton} Serendipitous Survey;
this catalogue, released in 2019 December, is the ninth data release from the survey overall (\citealt{4XMM-DR9}). 
The catalogue contains 810795 X-ray detections pertaining to 550124 unique sources from observations made between 2000 February 3 and 2019 February 26.

\subsubsection{Optical and ultraviolet photometry: XMM-SUSS4.1}
We take \textit{XMM-Newton} optical/UV photometric data from the fourth \textit{XMM-Newton} Serendipitous Ultraviolet Source Survey (SUSS: \citealt{XMM-OM}) catalogue, XMM-SUSS4.1.
This catalogue, released in 2018 January, compiles data from observations made up to 2017 July.
It contains 8176156 entries from 5503765 sources.

\subsection{Initial source selection criteria}
\subsubsection{Quasars}
\label{sec:qso-selection}
\begin{figure*}
    \begin{tabular}{cc}
        \includegraphics[width=\columnwidth]{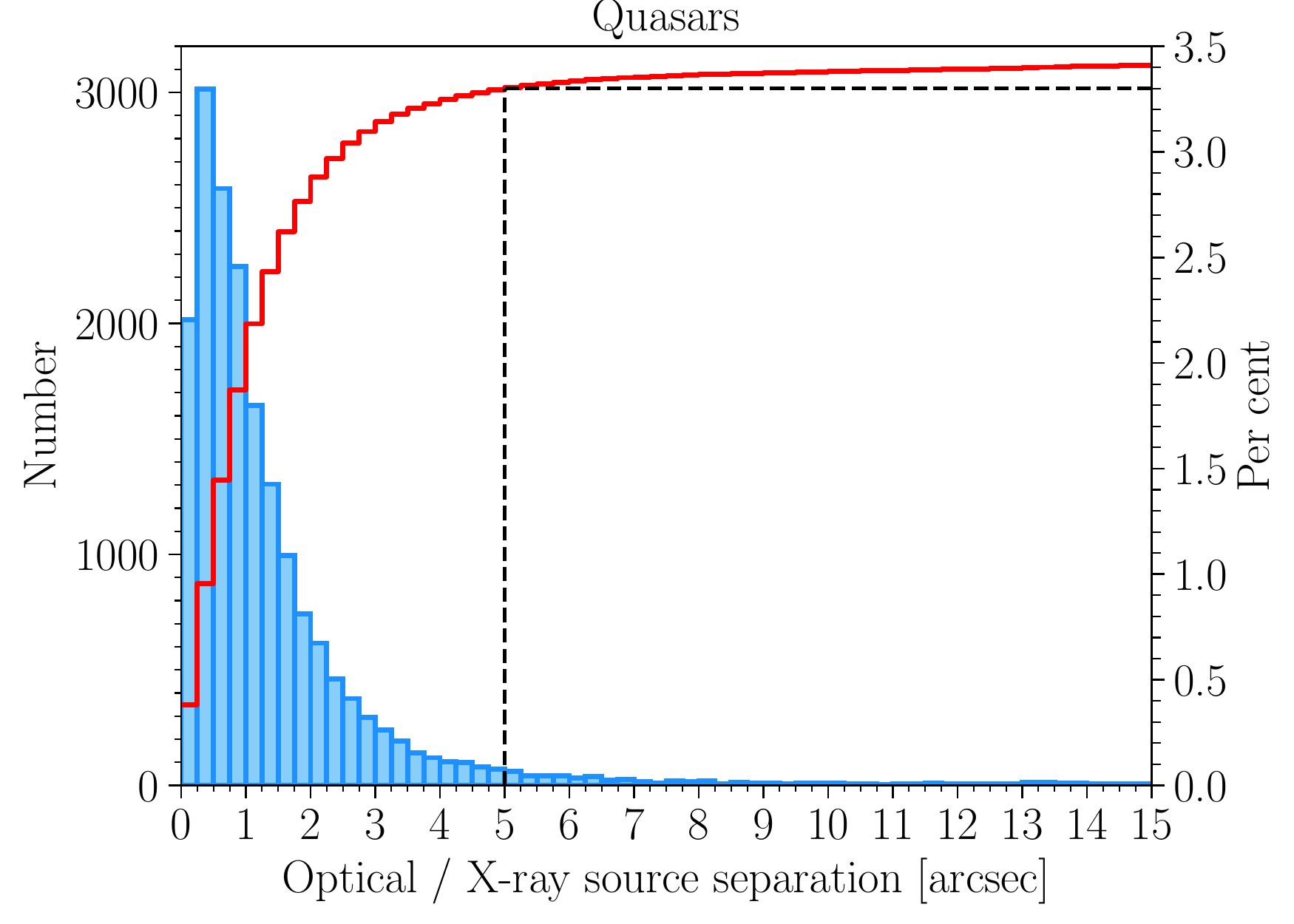} &
        \includegraphics[width=\columnwidth]{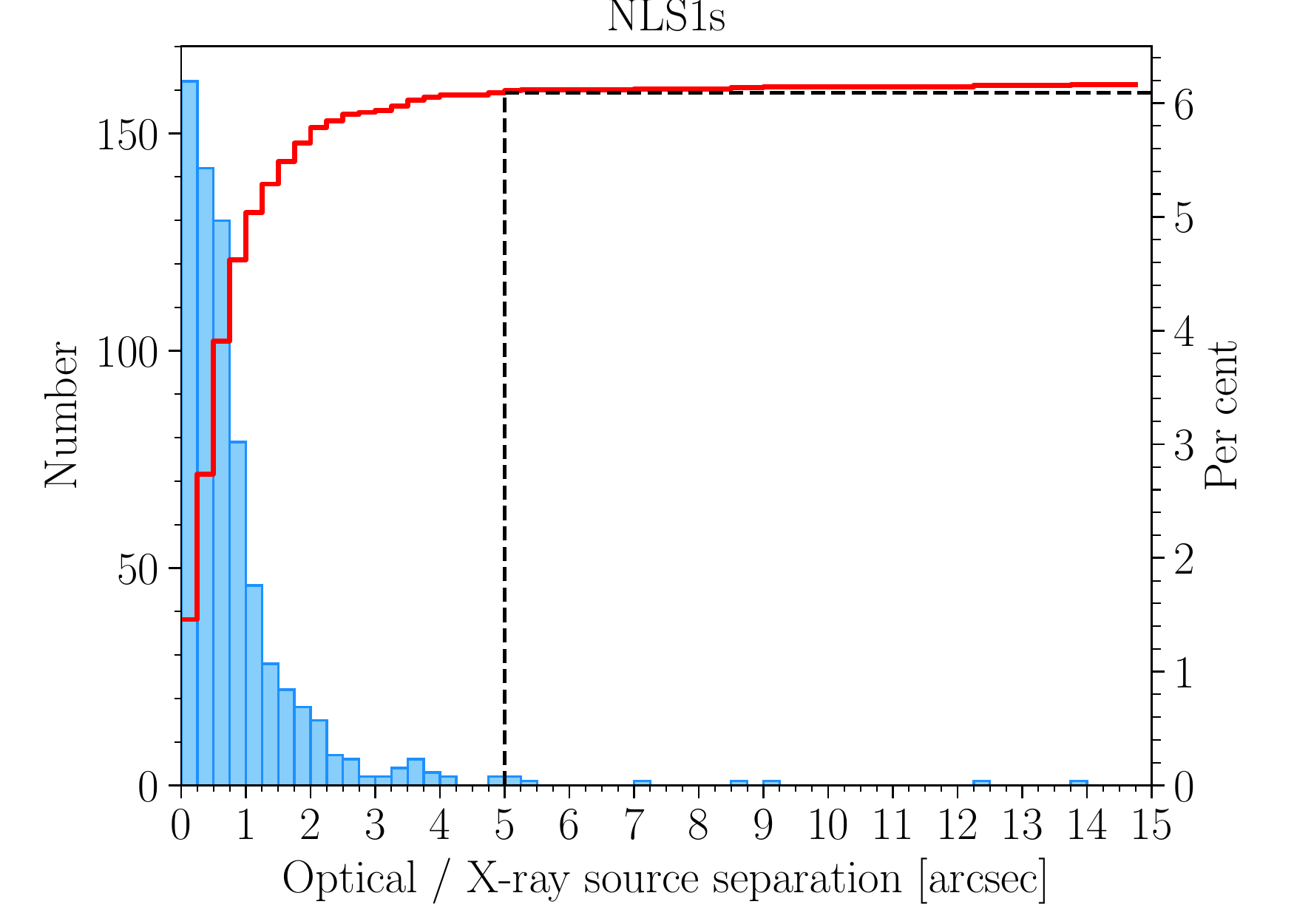} \\
    \end{tabular}
    \caption{The left-hand ordinate axis indicates the number of matched quasars. 
    The right-hand ordinate axis gives the percentage of all 526356 quasars, with the cumulative distribution shown in red. 
    To select our sample, we chose a matching radius of 5~arcsec which returns 17336 quasars, approximately 3.3 per cent of all SDSS-DR14Q quasars.
    \textit{Right}: The same, for \citealt{Rakshit17} NLS1s.  
    Of 11101 SDSS NLS1s, 676 (6.1 per cent) have a clean X-ray detection within 5~arcsec.}
    \label{fig:sepradius}
\end{figure*}

We first establish which optical quasars have an X-ray source detection in the latest \xmm\ data release.
To do this, we cross-match the optical catalogue with the slimline version of the X-ray catalogue.
The slimline 4XMM-DR9 catalogue contains just one row per unique X-ray source (rather than one row per detection as in the full catalogue).
There are 550050 X-ray sources in the catalogue, excluding 74 which are flagged as `CONFUSED' (i.e.\ with a nonzero probability of being associated with more than one distinct source).  
Following \citealt{SDSS-DR14Q}, we choose a match radius of 5~arcsec and retain all of the best (nearest) matching X-ray sources.  
Fig.~\ref{fig:sepradius} shows the distribtion of angular separations between the optical quasars and X-ray sources.
Our cross-match returns 17336 X-ray detected optical quasars (approximately 3.3 per cent of those in SDSS-DR14Q).
We record the unique X-ray source ID (`SRCID') of each quasar.
We do not remove broad absorption line (BAL) quasars from the optical catalogue initially, but will weed these out of our sample by visually inspecting the optical spectra.

Before performing a cross-match to the full X-ray catalogue, we clean it by excluding the sources which:
\begin{itemize}
\item have a poor source summary flag (SUM\_FLAG~$\geqslant3$);
\item are flagged as `CONFUSED';
\item were observed with a high background.
\end{itemize}
The clean catalogue then contains 88 per cent of the total number of records, with 713866 X-ray detections from 484726 unique sources.

\begin{figure}
	\includegraphics[width=\columnwidth]{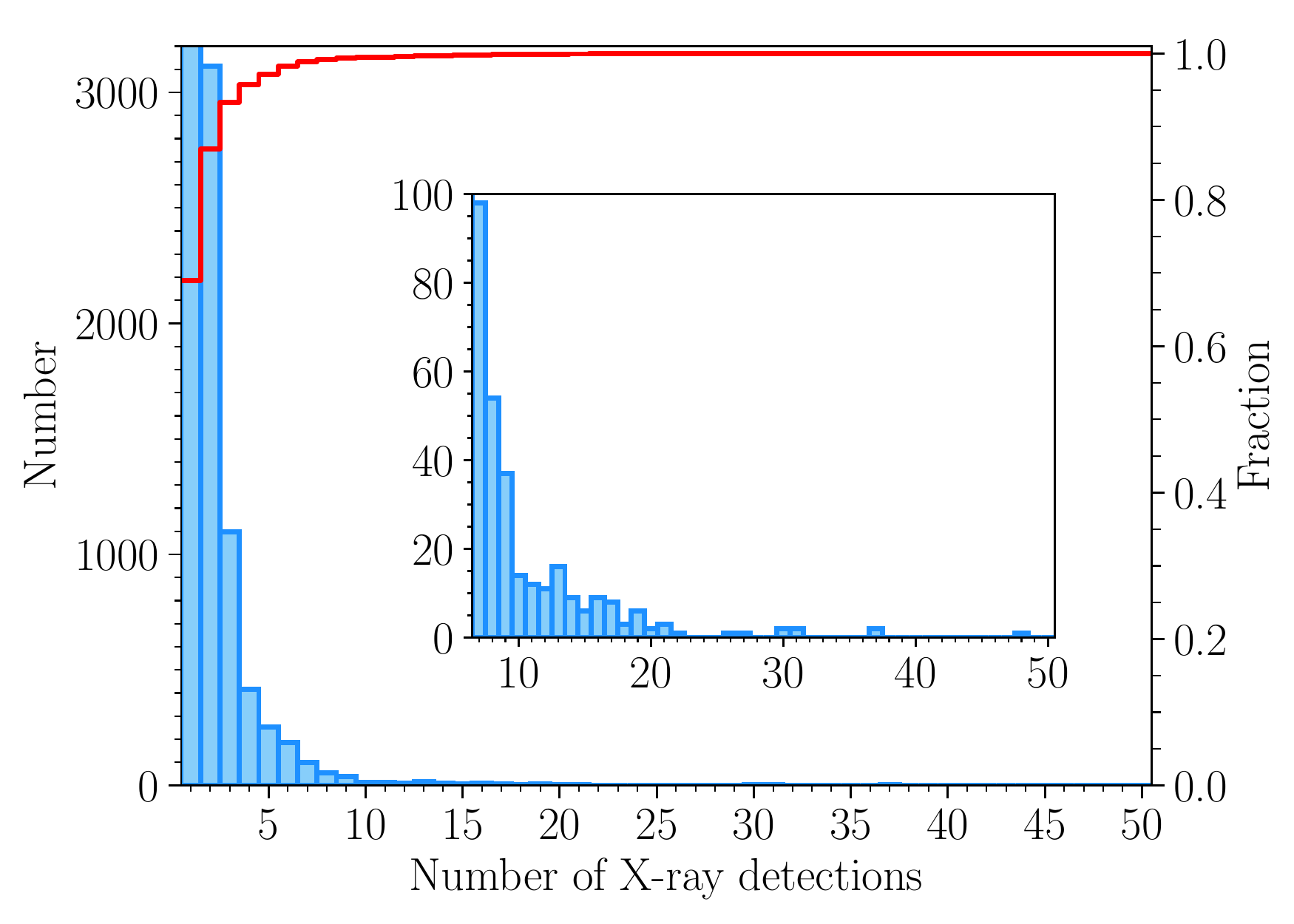}
    \caption{The number of clean X-ray detections for 17299 X-ray detected optical quasars.
    11935 quasars (69 per cent) have only 1 clean X-ray detection.
    Only 109 have 10 or more detections.
    The red line shows the cumulative fraction, with the fraction indicated on the right-hand ordinate axis.}
    \label{fig:nx}
\end{figure}

Using the X-ray source IDs, we find all clean observations of the optical quasars. 
We match all records in the cleaned, full 4XMM-DR9 catalogue to our pre-selected optical quasars.
We find that 17299 of the pre-selected 17336 quasars have a clean X-ray detection. 
Of these, 5364 (31 per cent) were observed by \xmm\ more than once; Fig.~\ref{fig:nx} shows the number of clean X-ray observations per optical quasar.

We then match with sources in the OM catalog. 
This matching requires that the OM source is within 1~arcsecond of the optical quasar coordinates and that the observation ID
of the OM record matches that of the X-ray observation (i.e.\ our OM and EPIC data were recorded simultaneously).
4499 quasars meet these criteria of which 868 have more than one simultaneous OM-EPIC observation.

At $z\gtrsim0.4$, the OM photometry may be compromised by the strong \lya~$\lambda1216$ UV emission line and \lya\ forest absorption blueward of $912$~\AA\ (rest frame). 
To determine which OM filters would be free of \lya\ emission and absorption, we calculated the quasar redshifts for which the red wing of \lya\ would fall just outside of the effective bandpass of each filter.
Then, for each redshift bin we require an OM detection in one of the following:
\begin{itemize}
\item $z<0.40$: any filter;
\item $0.40\leqslant z<0.55$: V, B, U, UVW1 or UVM2 filters;
\item $0.55\leqslant z<0.90$: V, B, U or UVW1 filters;
\item $0.90\leqslant z<1.35$: V, B, or U filters;
\item $1.35\leqslant z<1.95$: V or B filters;
\item $1.95\leqslant z<2.50$: V filter only.
\end{itemize}
1768 quasars have at least one useful OM photometry point. 
 
\begin{figure}
	\includegraphics[width=\columnwidth]{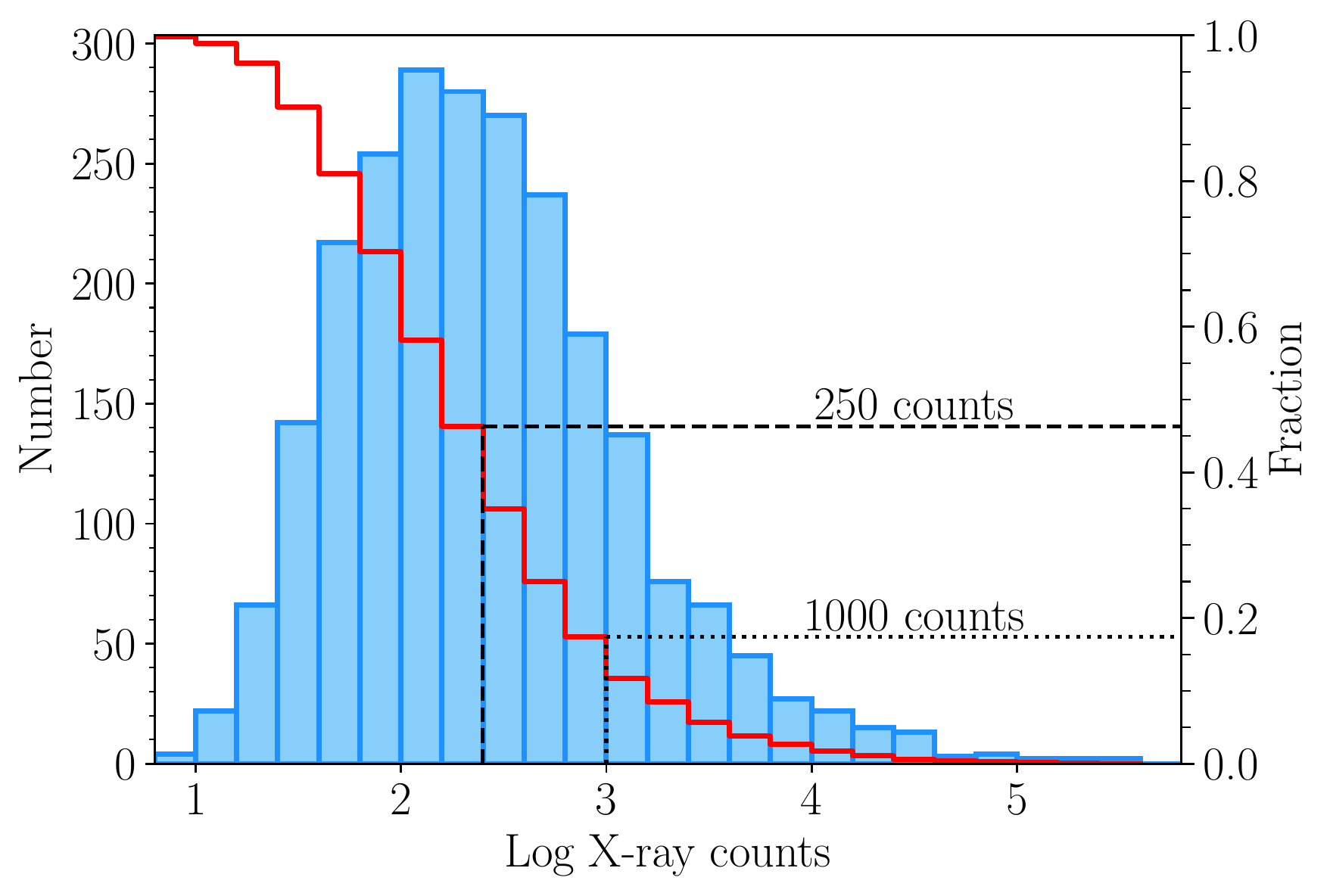}
    \caption{
    The X-ray counts distribution for 2374 `clean' X-ray observations with simultaneous \textit{XMM}-OM photometry redward of \lya. 
    For each observation, we take X-ray counts to be the greatest number of counts recorded by any of the three EPIC detectors (pn, MOS1 or MOS2).
    The left-hand ordinate axis gives the total number of observations and the right gives the fraction with a minimum number of counts (shown by the red line).}
    \label{fig:xcts}
\end{figure}

We wish to obtain reasonable constraints on the X-ray spectral parameters of these quasars, for which we require relatively high-quality spectra.
We therefore make a quality cut on the number of X-ray counts recorded by an EPIC detector in the 0.2--12~keV band.  
The number of sources as a function of X-ray counts is shown in Fig.~\ref{fig:xcts}.
It can be seen in Fig.~\ref{fig:xcts} that the counts distribution peaks $\gtrsim200$~counts and that only 17 per cent of detections are made with 1000 counts or more.
We chose to make a cut at 250 counts, which includes 46 per cent of the clean detections. 
For quasars with more than one X-ray observation, we retain the one with the greatest number of counts in any detector (so that we now have only one simultaneous EPIC and OM observation of each quasar).
After applying this cut, we are left with 782 quasars.

To make an estimate of the black hole mass, we require spectral coverage of at least one of the emission lines \ha, \hb\ or \mg.
\ha\ or \hb\ will be visible in the spectra of low-redshift sources and the \mg\ emission line will be visible in the wavelength range of the BOSS spectrograph up to $z \lesssim 2.5$ \textcolor{black}{(see Fig.~\ref{fig:z})}.
We therefore consider only the SDSS quasars with $z\leqslant2.5$.
Following this redshift cut, our sample contains 768 quasars.

\subsubsection{Narrow-line Seyfert 1s}
\label{sec:nls1-selection}
Narrow-line Seyfert 1s (NLS1s) are a subset of unobscured AGN with relatively narrow broad lines.
They are generally defined as optical type 1 AGN with FWHM(\hb)~$\lesssim2000$~\kms, weak {\oiii}$\lambda{5007}$ relative to \hb\ and strong Fe\,\textsc{ii} emission (e.g.\ \citealt{OP85}, \citealt{Goodrich89}).
NLS1s are typically found in spiral galaxies in the local Universe.
Because Seyfert AGN are less luminous than quasars, many NLS1s are absent from the SDSS quasar catalogues.
For example, of the 12 NLS1s in the sample of \cite{Jin1}, only 2 are present in our selection of quasars.
Nevertheless, NLS1s represent a unique and important region of parameter space to explore, and enable contrasts to be made with typical broad-line AGN and quasars (e.g. \citealt{Jin1,Jin2,Jin3}). 

\cite{Rakshit17} (hereafter \citetalias{Rakshit17}) assembled a sample of 11101 NLS1s from the SDSS-DR12 spectroscopic database, by analysing the spectra of `QSO'-type sources and retaining those with $\mathrm{FWHM}(\mathrm{H}\,\upbeta)\leqslant2200$~\kms\ and flux ratio $[\mathrm{O}\,\textsc{iii}]\lambda5007/\mathrm{H}\,\upbeta<3$.
We identify NLS1s with \xmm\ OM and EPIC data by matching the \citetalias{Rakshit17} catalog to the X-ray and OM serendipitous catalogs, following the method described in Section~\ref{sec:qso-selection}.
Our criteria return 147 NLS1s: 63 new sources and 84 that were already included from the quasar catalogue.

\subsection{Optical spectral fitting}
\label{sec:pyqsofit}
\cite{Rakshit20} (hereafter \citetalias{Rakshit20}) performed fits to the optical spectra of all SDSS DR14Q quasars using the the publicly-available \textsc{python} code \textsc{PyQSOFit} (\citealt{Guo18}, \citealt{Guo19}, \citealt{Shen19}).
For the quasars in our sample, we can take many emission line and continuum measurements from \citetalias{Rakshit20}.
However, the 63 NLS1s not present in the quasar catalog have only optical spectral measurements from \citetalias{Rakshit17} who used a private, custom spectral fitting routine to obtain their measurements of NLS1s.
These two routines differ in several respects. 
The broad emission lines were often modelled with a single Lorentzian profile in \citetalias{Rakshit17}; Lorentzian profiles are not currently an option in \textsc{PyQSOFit} and instead one or more Gaussians may be used to fit a line.  
\citetalias{Rakshit20} chose to use 3 Gaussians to model broad \hb\ and \ha, for example.
The host galaxy subtraction, which is often required for the low-luminosity NLS1s, is also approached in different ways.
\citetalias{Rakshit17} removed the host galaxy contribution by performing simple stellar population modelling whereas principal component analysis (PCA: e.g.\ \citealt{Yip04a,Yip04b}) is employed in \textsc{PyQSOFit}.
Naturally these different methods will yield somewhat different results for the continuum and emission line parameters.
Derived quantities such as the virial black hole mass and Eddington ratio will then also differ.

For consistency of the optical spectral measurements of our sample, we adapted \textsc{PyQSOFit} to replicate the spectral fitting procedure of \citetalias{Rakshit20} and refit the spectra of the sources drawn from \citetalias{Rakshit17}.
We briefly summarise the adopted \textsc{PyQSOFit} spectral fitting routine here.
In this setup, the \cite{SFD98} dust map is queried to determine the line-of-sight extinction and the spectrum is dereddened using the extinction curve of \cite{Fitzpatrick99}. 
The spectrum is then transformed to the rest-frame.
The PCA method is used to decompose stellar and AGN emission with 5 and 20 eigenspectra used to construct the host galaxy and quasar contributions, respectively.
The host galaxy subtracted continuum is then modelled with a power-law plus Fe\,\textsc{ii} templates and a functional Balmer continuum.
Emission line complexes are then modelled separately in several broad windows: 
\begin{itemize}
    \item 6400--6800~\AA: \ha, the [N\,\textsc{ii}] doublet ($\lambda=6549,6585$~\AA) and the [S\,\textsc{ii}] doublet ($\lambda=6718, 6732$~\AA);
    \item 4640--5100~\AA: \hb, the \oiii\ doublet ($\lambda=4959, 5007$~\AA) and He\,\textsc{ii};
    \item 2700--2900~\AA: \mg.
\end{itemize}
Following the fitting procedure of \citetalias{Rakshit20}, the broad components of \ha\ and \hb\ and He\,\textsc{ii} are each modelled with 3 Gaussian components;
broad \mg\ is modelled with 2 Gaussians.
The broad Gaussians have a minimum FWHM of 900~\kms\ and maximum velocity offset $\pm3000$~\kms\ from their centre.
Narrow Gaussians are used to model forbidden lines and the narrow components of the permitted lines; these are allowed a width in the range $100\leqslant\mathrm{FWHM}\leqslant900$~\kms. 
\oiii$\lambda\lambda4959,5007$ additionally has Gaussian wings that may be broader than the narrow cores (in the range $400\leqslant\mathrm{FWHM}\leqslant2000$~\kms).
The velocity offsets of the narrow lines are not allowed to exceed $\pm1000$~\kms.
The flux ratios of the \oiii\ and [N\,\textsc{ii}] doublets are fixed to their theoretical value 1:3.
In each emission line complex the widths and offsets of the narrow lines are tied together.
Uncertainties on the emission line and continuum paramaters are estimated using the Monte Carlo method, resampling each spectrum 50 times.
Following the procedures of \citetalias{Rakshit20} and \cite{Calderone17}, we added a quality flag for each measured emission line and continuum point; the flags are listed in Appendix~\ref{sec:qual}.

\textsc{PyQSOFit} uses optical galaxy eignenspectra to perform the host galaxy subtraction; these eigenspectra do not extend into the UV and therefore the fitted spectrum is truncated to match the wavelength range of the eigenspectra ($\approx3500$--9000~\si\angstrom).
Consequently, there are no measurements of the UV continuum and emission lines for sources in which host galaxy subtraction was performed in the optical.
To obtain the UV spectral measurements for these AGN we checked for spectra that contained a \mg\ line that was not initially measured and then fit the short-wavelength part of these spectra ($\lambda_\mathrm{rest}=1000$--4000~\AA) without performing host-galaxy subtraction.
The measurements for \mg\ and the UV continuum were added to our catalogue, and we have added a flag (`MGII\_SEP') to indicate the AGN for which the \mg\ region was fitted separately from the optical region.

\subsubsection{Identification of broad- and narrow-line type 1 AGN}
\label{sec:bl_nl}
\begin{figure}
	\includegraphics[width=\columnwidth]{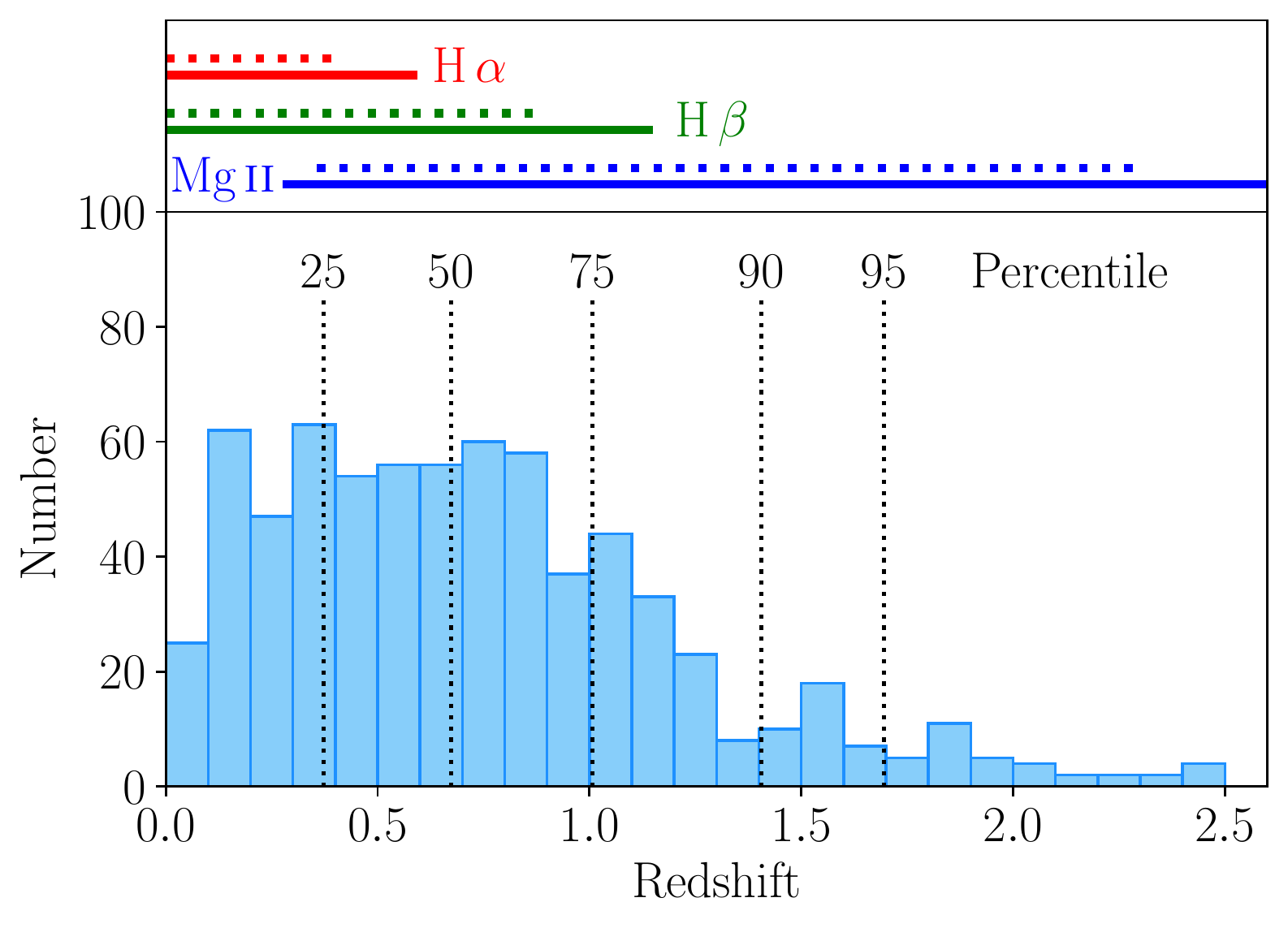}
    \caption{
    The redshift distribution of the 696 AGN in our sample.
    Black dotted lines indicate the 25\ts{th}, 50\ts{th}, 75\ts{th}, 90\ts{th} and 95\ts{th} percentiles of the distribution.
    The top panel shows the redshift range for which the centroids of prominent broad emission lines (\ha, \hb\ and \mg) are just visible in the wavelength coverage of the SDSS (dotted lines) and BOSS (solid lines) spectrographs.}
    \label{fig:z}
\end{figure}

To ease the comparison between AGN with relatively narrow and broad permitted emission lines, we assigned each AGN a linewidth flag (`B' for broad-line type 1 AGN; `N' for narrow-line type 1 AGN).
The determination is based on the fitted width of \hb\ where it is available, \mg\ for the higher-redshift sources without \hb, and \ha\ only if neither \hb\ or \mg\ were measured.  
\textcolor{black}{(Spectroscopic coverage of these emission lines as a function of redshift is shown in Fig~\ref{fig:z}.)}
We set the division between the subsets at the usual 2000~\kms.
The differences in fitting routine and the division between narrow- and broad-line AGN mean that our classifications differ from those of \citetalias{Rakshit17} (those authors measured lines by fitting a Lorentzian profile and chose a limit of FWHM[\hb]$\leqslant2200$~\kms).
Consequently, not all AGN drawn from \citetalias{Rakshit17} are flagged `N' in our sample.
Typically, the broad line FWHMs measured by \textsc{PyQSOFit} are slightly broader than those measured by \citetalias{Rakshit17}.
However, the differences in linewidth measured by the two routines are relatively minor:
the median difference in FWHM(\hb) is $\approx+100$~\kms\ and 
for approximately two-thirds of the AGN from the NLS1 catalogue in our sample, FWHM(\hb) as measured by \textsc{PyQSOFit} is within $\pm350$~\kms\ of the width measured by \citetalias{Rakshit17} (350~\kms\ is the median uncertainty on FWHM[\hb] from \textsc{PyQSOFit}).
We note that the narrow-line type 1 AGN in our sample are not necessarily NLS1s by the standard definition described in Section~\ref{sec:nls1-selection}: we have not always used broad \hb\ as the characteristic line width and we have made no assessment of the strengths of \oiii~$\lambda{5007}$ or the optical [Fe\,\textsc{ii}] emission.

\subsection{Quality checks and selection of the final SOUX sample}
\label{sec:curated}
We performed a visual inspection of all 831 optical spectra and removed any sources which had low signal to noise or any absorption features particularly affecting the line regions from which black hole mass measurements were derived. In addition any objects with a diffuse emission flag in their X-ray spectra were removed from the sample.  Following this quality control, our final sample consists of 696 AGN\footnote{In this work, we consider only one optical spectrum per source: that listed as the `science primary' spectrum at the time of the catalogue compilation.  However, 116 AGN have multiple spectra recorded on different nights; we describe these in Appendix~\ref{sec:multispec}}.  Of these, 636 are drawn from SDSS-DR14Q, 60 unique sources are added from \citetalias{Rakshit17} and 72 are common to both catalogues.

\section{Multiwavelength measurements}
\label{sec:mwl}
We will present a detailed study of the broadband spectral energy distributions (SEDs) of our AGN in a future paper (Mitchell et al., submitted).
However, the SED shape can be characterised by two common parameters: the radio loudness and UV-X-ray energy index. 
The radio-loudness parameter indicates the power of AGN outflows relative to the accretion flow, with radio-loud AGN exhibiting powerful jets (Section~\ref{sec:radio}).
The optical luminosity at 4400~\AA\ and radio luminosity at 5~GHz are required to calculate the radio-loudness.
The UV-X-ray energy index ($\alpha_\mathrm{ox}$: Section~\ref{sec:aox}) indicates the relative power of the UV-emitting accretion disc to the X-ray emitting corona.
To calculate $\alpha_\mathrm{ox}$ we require a measurement of the UV luminosity at 2500~\si\angstrom\ and the X-ray luminosity at 2~keV.
In this Section, we describe how the measurements of these quantities were obtained.

\subsection{UV and optical luminosities}
\label{sec:uv_lum}
For sources at $z\gtrsim0.5$ the rest-frame monochromatic luminosity at 2500~\AA\ is available directly from the SDSS optical spectrum.
For nearer sources, this wavelength falls outside the SDSS spectrograph range, but photometry covering it may be available from either \textit{XMM} OM or SDSS.
For AGN at lacking spectroscopic coverage of 2500~\AA\ we first attempt obtain an estimate from the \textit{XMM} OM photometry.
\textcolor{black}{We determine which OM bands sample 2500~\AA\ in the rest frame and from these we calculate the luminosity from the flux in the band with an effective wavelength closest to rest frame 2500~\AA.}
If there is no \textit{XMM} OM coverage of 2500~\AA\ we take the PSF flux from the appropriate Sloan filter (either the \textit{u} or \textit{g} band). 
We deredden the photometry, taking the line-of-sight colour excess from the \cite{SFD98} dust maps as was done for the spectral analysis and convert the fluxes to rest-frame luminosities. 
Finally, $L_{2500}$ is obtained from the optical spectrum of 421 AGN, from \textit{XMM} OM photometry of \textcolor{black}{237} AGN and from SDSS photometry of \textcolor{black}{27} AGN; overall, we have a measurement of $L_{2500}$ for \textcolor{black}{685} AGN (\textcolor{black}{98}~per cent of the full sample). 
The AGN lacking a $L_{2500}$ measurement are all at the low-redshift end of our sample ($z<0.2$\textcolor{black}{1}).

The optical continuum at rest-frame 4400~\si\angstrom\ was not measured by \citetalias{Rakshit20} or \citetalias{Rakshit17} but, similarly to $L_{2500}$, $L_{4400}$ can be measured for the majority of our AGN either from the SDSS spectrum, or from the Sloan or \textit{XMM} OM photometry.
When making a measurement from the optical spectrum, we avoid contamination of the continuum flux measurement by the blended H\,$\upgamma$, \oiii\ and Fe\,\textsc{ii} emission lines by performing a linear interpolation between line-free continuum windows at 4200 and 5100~\si\angstrom, where possible.
In spectra where these two windows are not present (or are too near the end of the spectrum to be reliable), we measure the 4400~\si\angstrom\ flux directly.
For AGN lacking optical spectral coverage of 4400~\si\angstrom\ we took the flux from the photometric band containing rest-frame 4400~\si\angstrom, corrected for Galactic reddening and converted the flux to a rest-frame luminosity, as was done for the $L_{2500}$ measurements.
In our final sample of 696 AGN, 482 have $L_{4400}$ measured by interpolating the spectral continuum; 111 from a direct measurement of the 4400~\si\angstrom\ spectral flux; 38 estimates from SDSS photometry; and 65 have no 4400~\si\angstrom\ coverage in either the spectrum or photometry.
None of the 65 sources lacking $L_{4400}$ from the SDSS spectrum or photometry had coverage in the \textit{XMM} OM data because they were higher-\textit{z} AGN and the OM filters do not cover long enough wavelengths to contain the rest-frame 4400~\AA.
The AGN lacking a $L_{4400}$ measurement are all at the high-redshift end of our sample ($z>1$).

\textcolor{black}{The mixture of spectroscopically- and photometrically-derived UV and optical luminosities is not ideal.
Whilst the \textit{XMM} OM photometry is contemporaneous with the X-ray data, broad-band photometry is susceptible to contamination from non-continuum emission, particularly at lower redshifts.
The SDSS and \textit{XMM} data are not contemporaneous, so AGN variability will introduce some scatter between spectroscopic and photometric fluxes.
Additionally, aperture effects may also introduce systematic differences.
In Appendix~\ref{sec:mix_lum} we assess the effects of mixing different measures of these luminosities.
We find that on average the differences between spectroscopic and photometric luminosities are small, with $\langle\Delta\log(L_{4400})\rangle=-0.03\pm0.13$ and $\langle\Delta\log(L_{2500})\rangle=-0.02\pm0.16$.
As Fig.~\ref{fig:mix_lum} shows, the spectroscopic and photometric luminosity measures are approximately equal across several orders of magnitude in luminosity, although there is substantial scatter. 
We discuss the potential impact on our results in more detail in Appendix~\ref{sec:mix_lum} and Section~\ref{sec:disc-aox}.}

\subsection{X-ray luminosities}
\label{sec:xray_lum}
A detailed analysis of the X-ray spectra of our sources is beyond the scope of this paper but will be presented in future work.
However, it is possible to estimate the X-ray luminosities from data provided in the DR9 catalog.
Our sources span a wide range of redshifts and therefore the EPIC fluxes reported for various (observed) energy bands are not directly comparable since they sample different parts of the emitted spectrum.

To calculate the monochromatic X-ray flux at rest-frame 2~keV, we follow a similar approach to that of \cite{Lusso16} (see section 2.1 of that paper).
We determine the EPIC flux in soft (0.5--2~keV) and hard (2--12~keV) energy ranges.
An approximate correction for Galactic photoelectric absorption is applied to the X-ray fluxes\footnote{The line-of-sight Galactic neutral hydrogen column density is obtained from the maps of the Leiden/Argentine/Bonn (LAB) Galactic H\,\textsc{I} Survey (\citealt{Kalberla05}) using the Python tool \texttt{gdpyc} ({\url{https://pypi.org/project/gdpyc/}}).}.
From these band fluxes we estimate the monochromatic fluxes at 1 and 3.5~keV, assuming a photon index $\Gamma=1.7$ in each energy band.
We then translate these fluxes at observed energies to luminosities at rest-frame frequencies in $\log(\nu)$-$\log({\nu}L_\nu)$ form.
Using these two points, the rest-frame monochromatic 2~keV luminosity is calculated by linear interpolation (for sources with $z<1$) or extrapolation (for sources with $z>1$).  An estimate of the X-ray photon index $\Gamma$ is also made \textcolor{black}{by computing the gradient between the absorption-corrected monochromatic fluxes at 1 and 3.5~keV\footnote{\textcolor{black}{We note that our results do not depend very strongly on the assumption of a standard $\Gamma=1.7$ in the soft and hard bands to estimate the 1 and 3.5~keV fluxes.
If we compare the dervied distributions of $\Gamma$ and $\log(L_\mathrm{2keV})$ that result from instead assuming $\Gamma=1.5$ and 1.9, we find that the mean $\Gamma$ differ by $\approx0.1$ and the mean $\log(L_\mathrm{2keV})$ differ by only $\approx0.01$~dex.}}}.

\begin{figure}
	\includegraphics[width=\columnwidth]{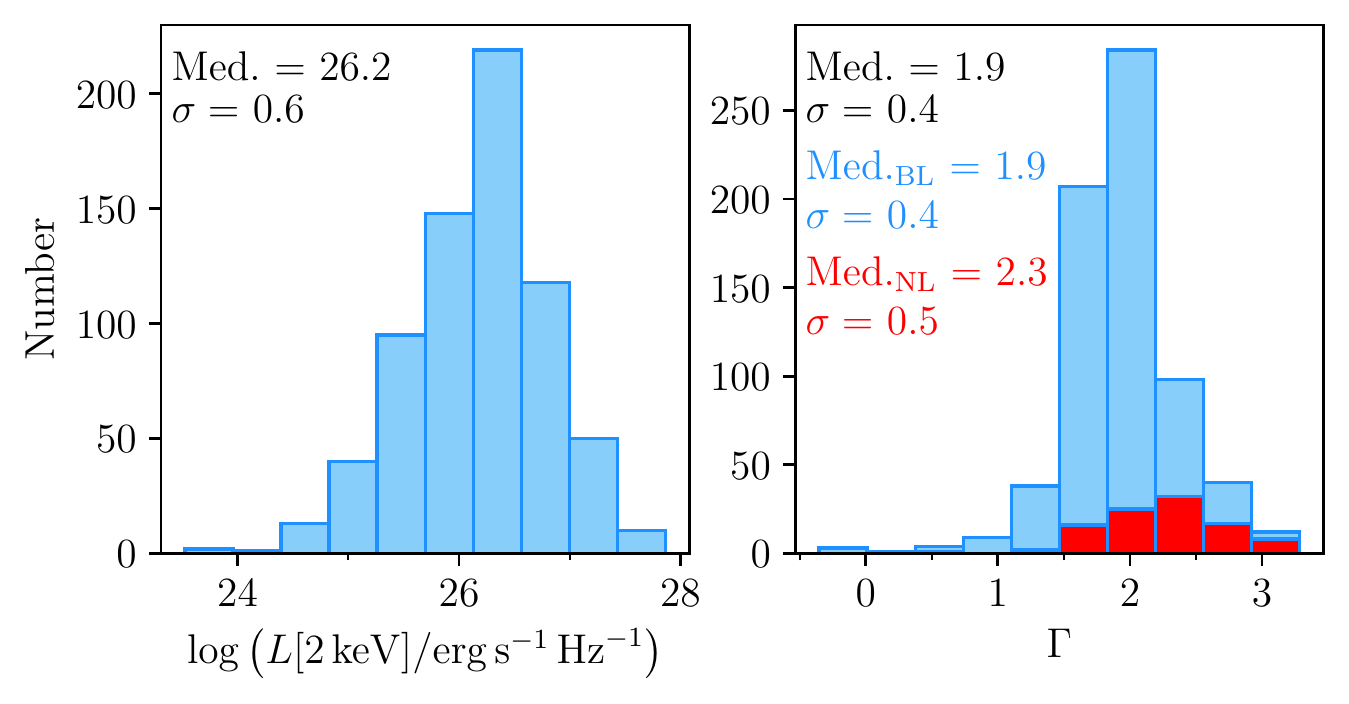}
    \caption{
    Distributions of the monochromatic luminosity at 2~keV in the rest frame (left) and the estimated X-ray photon index $\Gamma$ (right).
    In the right-hand panel the subsample of 101 AGN with relatively narrow permitted lines (`NL') are shown in red, and the 595 broad-line sources (`BL') in blue. 
    }
    \label{fig:l2kev}
\end{figure}

In Fig.~\ref{fig:l2kev} we show the distributions of $\log(L_\mathrm{2\,keV})$ and $\Gamma$.
Our sample spans approximately 4 dex in X-ray luminosity, with median $\log(L_\mathrm{2\,keV})=26.2$ and $\sigma=0.6$.
The median estimated photon index is $\Gamma=1.9$ with $\sigma=0.4$.
673 of the AGN (97~per~cent) have $1\leqslant\Gamma\leqslant3$.  
Less than 2~per~cent of sources have $\Gamma<1$, perhaps indicating absorption of soft X-rays in the AGN or host galaxy.
The $\Gamma$ distribution of the 101 narrow-line AGN is different from that of the 595 broad-line AGN.  
The narrow-line AGN have a median $\Gamma=2.3$ and $\sigma=0.5$, indicating that their X-ray spectra are generally softer than typical quasars.
Kolmogorov-Smirnov and Anderson-Darling tests both indicate that the photon indices of the narrow-line and broad-line AGN are drawn from different distributions.  \textcolor{black}{For a null hypothesis that the two samples are drawn from the same distribution we determine a KS statistic 0.41 and $p=8\times10^{-14}$ and AD statistic 44 and $p<0.001$.}

\subsection{Radio properies}
\label{sec:radio}
\begin{figure}
	\includegraphics[width=1\columnwidth]{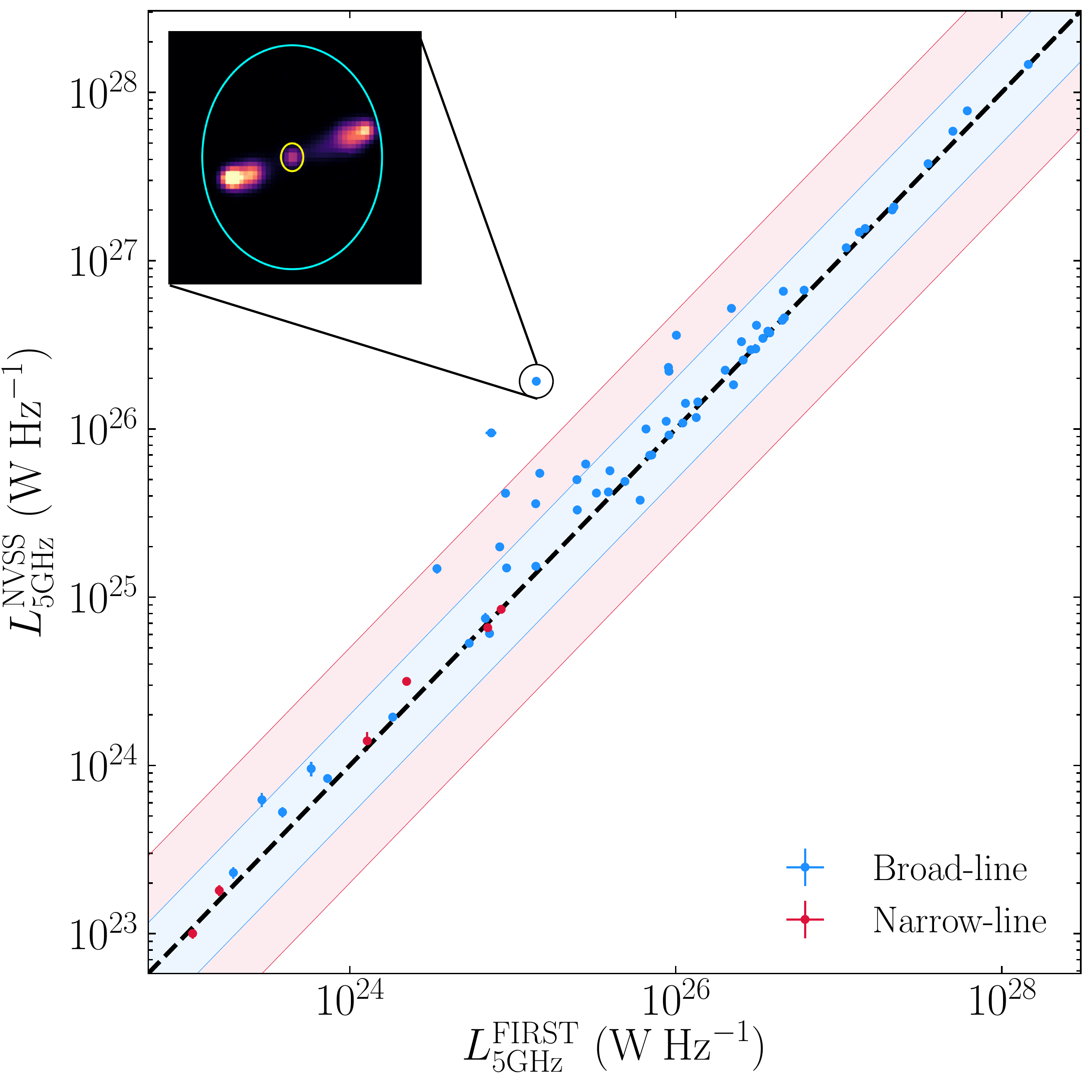}
    \caption{A comparison of the rest-frame luminosities at 5~GHz derived from FIRST and NVSS fluxes at 1.4~GHz, as described in the text. The dashed line marks the one-to-one relation, the shaded blue region being a factor of 2 deviation, and the pink shaded region a factor of 5.
    The top-left inset figure shows the radio map for the FR-II source SDSS\,J151443.07$+$365050.4, which has a much higher luminosity in NVSS than FIRST, as a result of the larger beam size of the former survey.
    The 45" beamwidth of NVSS is in blue, and the 5.6" beamwidth of FIRST is in yellow.
    }
    \label{fig:lum_nvss_first}
\end{figure}

Our sample of 696 AGN was cross-matched with both FIRST \citep{FIRST} and NVSS \citep{NVSS}, using a matching radius of 10$^{\prime\prime}$ that has been shown to have a false association rate of only 0.2~per~cent with FIRST \citep{lu07}. Both FIRST and NVSS sample the sky at 1.4~GHz, with beam-widths of 5.6$^{\prime\prime}$ and 45$^{\prime\prime}$ respectively.  679 of our AGN were within the FIRST footprint, of which we found a matching radio source for 107.  73 AGN also have an NVSS radio detection (all of which were FIRST-detected).      
We converted the 1.4~GHz fluxes to rest-frame 5~GHz luminosities using the method of \cite{Alexander03}, assuming 
a common radio spectral index ($\alpha_\mathrm{R}$) of 0.6. 
A comparison of the NVSS and FIRST luminosities for the 73 of objects matched with both catalogues can be seen in Fig.~\ref{fig:lum_nvss_first}. It is clear that for the majority of sources the FIRST and NVSS luminosities match to within a factor of 2. The population is skewed towards higher luminosities in NVSS due to its larger beam width; this is most pronounced in objects which possess extended morphologies as can be seen by the radio map of the FR-II source SDSS\,J151443.07$+$365050.4 (Fig.~\ref{fig:lum_nvss_first}, inset).

\begin{figure}
	\includegraphics[width=1\columnwidth]{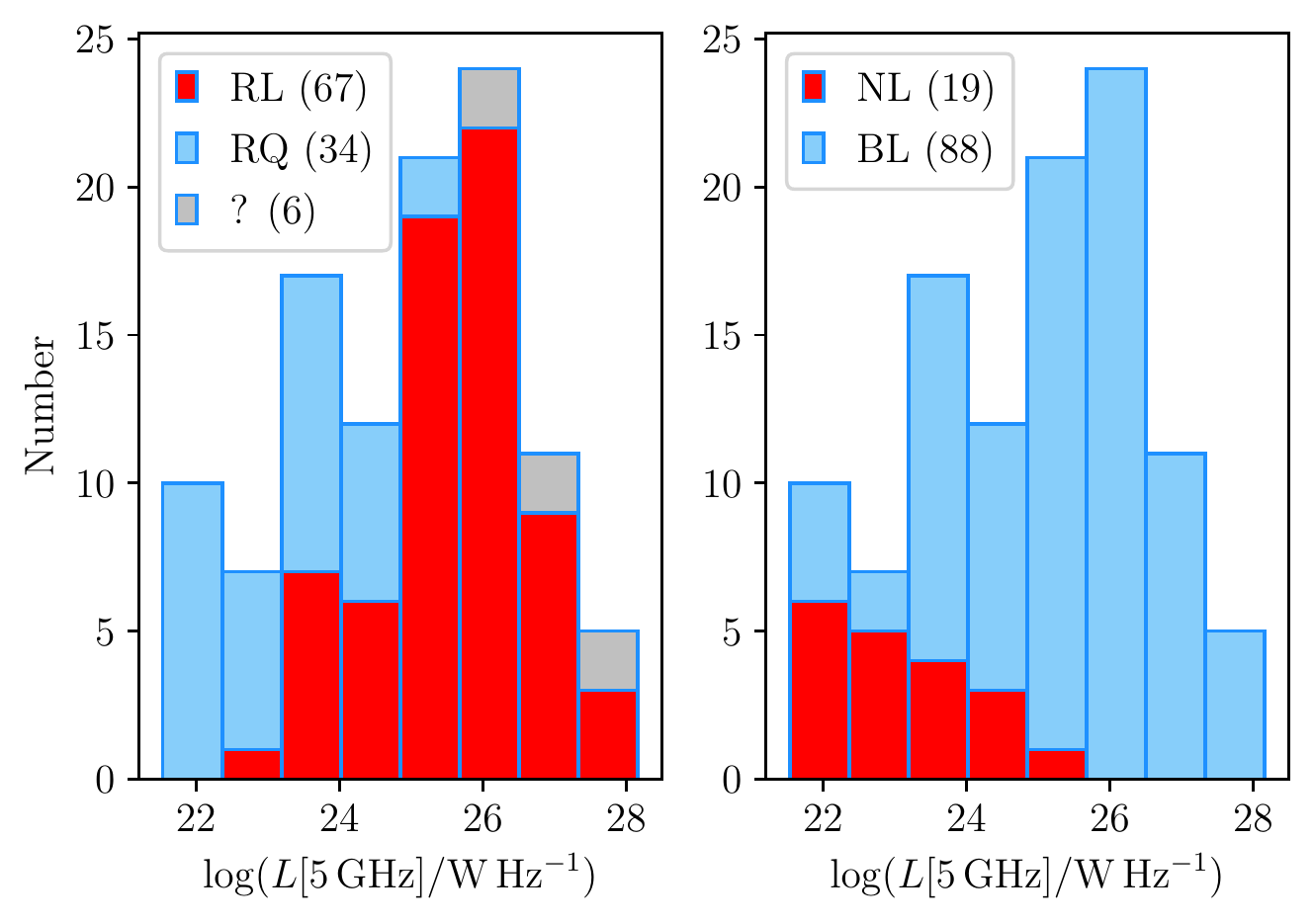}
    \caption{For the 107 FIRST radio detected AGN in our sample we show \textcolor{black}{the} distribution of 5~GHz rest-frame radio luminosities derived from the observed 1.4~GHz fluxes, as described in the text. 
    In the left panel, the distribution is divided into radio-loud and quiet (`RL' and `RQ') subsamples in addition to 6 sources lacking an optical $L_\mathrm{4400}$ measurement with unknown radio loudness (`?').
    In the right panel the distribution is divided into narrow-line and broad-line (`NL' and `BL') subsamples on the basis of the relative widths of the permitted optical emission lines. 
    }
    \label{fig:first_radio_lum}
\end{figure}

\begin{figure}
    \centering
    \includegraphics[width=\columnwidth]{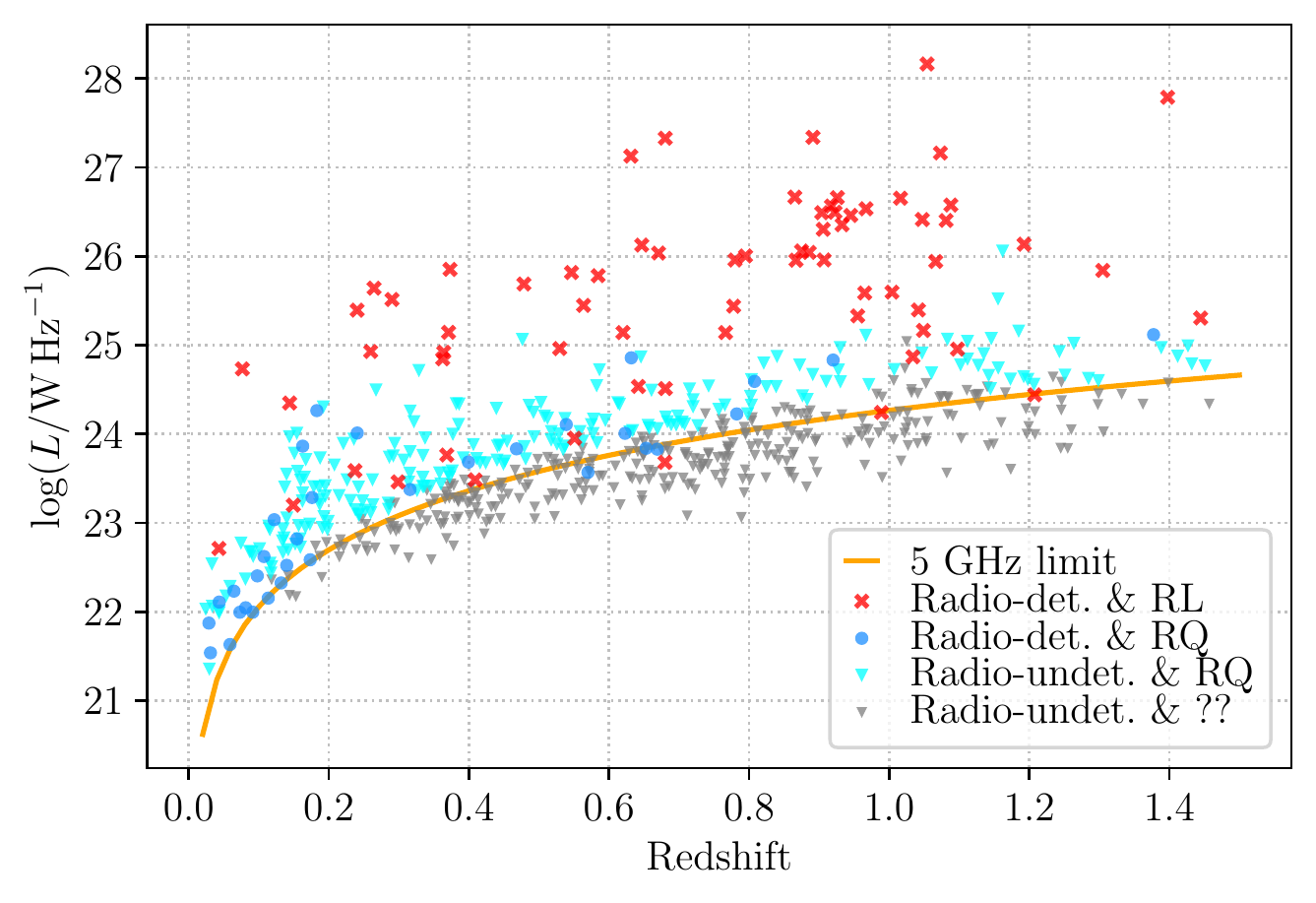}
    \caption{The orange curve shows the limiting rest-frame 5~GHz luminosity for a source to be detected by FIRST at 1.4~GHz, assuming a survey flux density threshold of 1~mJy and radio spectral index $\alpha=0.6$.
    Radio-detected and radio-loud (radio-quiet) sources are shown with red crosses (blue circles).
    For radio-undetected sources, we plot the luminosity corresponding to $R=L_{5\mathrm{GHz}}/L_{4400}=10$.
    Undetected sources more than 3$\sigma$ above the curve (cyan triangles) are considered radio-quiet.
    Undetected sources within 3$\sigma$ of the curve and below (grey triangles) could be either radio-quiet or radio-loud. 
    }
    \label{fig:first_lim}
\end{figure}

\begin{figure}
	\includegraphics[width=\columnwidth]{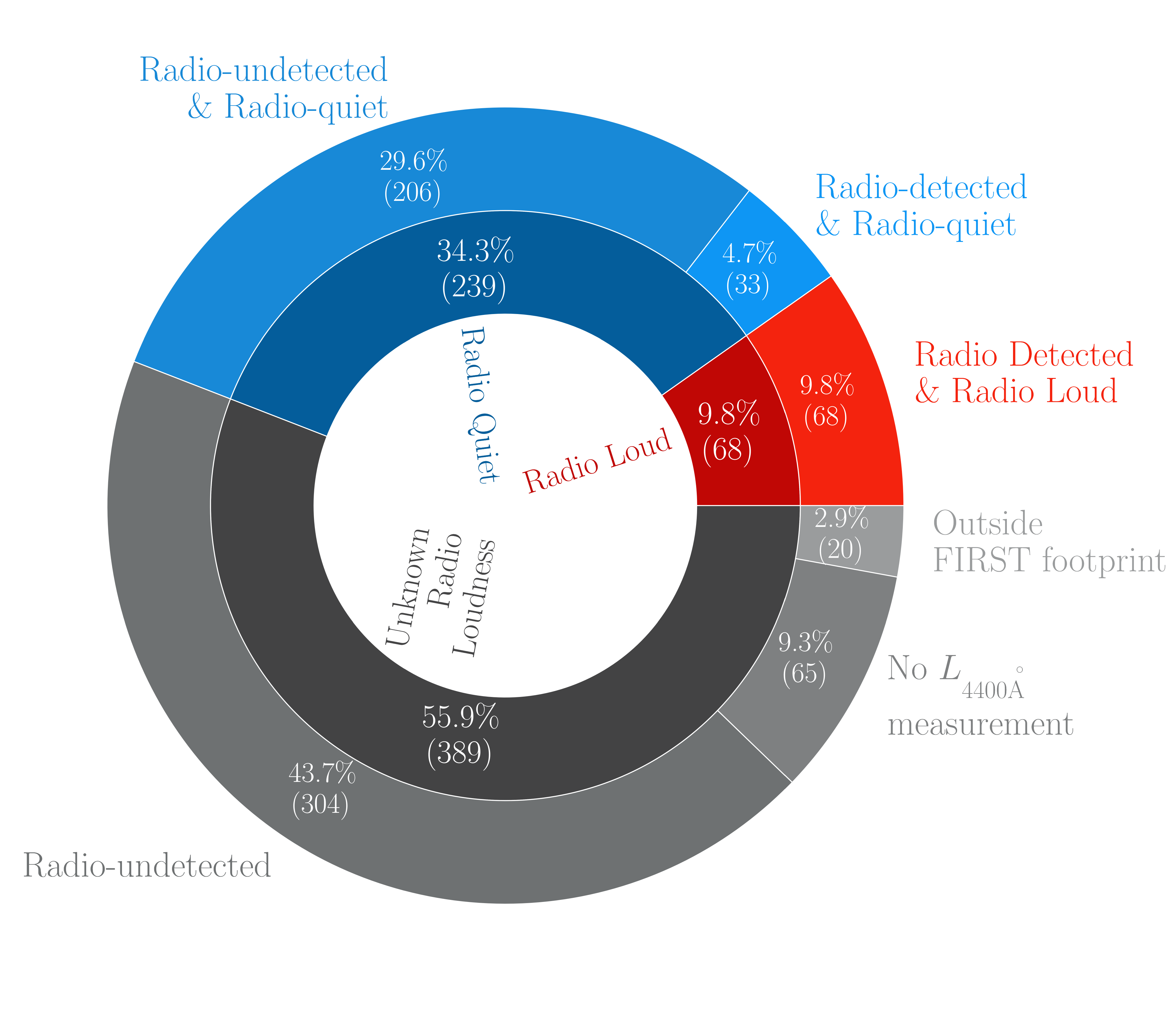}
    \caption{Radio detection and radio loudness fractions for the AGN in our sample. 
    }
    \label{fig:first_radio_loud}
\end{figure}

For the 101 radio-detected AGN with a $L_{4400}$ measurement we calculated the radio loudness according to the common definition of $L_{5\,\mathrm{GHz}} / L_{4400}$ \citep{Kellerman89}.
Fig.~\ref{fig:first_radio_lum} displays the distribution of the 5~GHz luminosities of the radio-detected AGN. 
The distribution has been separated into distinct populations: the left-hand panel shows the radio-loud and radio-quiet subsets and the right-hand panel shows the narrow-line and broad-line subsets (see Section~\ref{sec:bl_nl}).
For many undetected radio sources, we can still estimate whether they are likely to be radio-quiet by considering the FIRST detection threshold as an upper limit to their radio flux.  
Assuming a FIRST survey flux density threshold of 1~mJy \citep{FIRST}, we calculate a curve corresponding to the lowest rest-frame 5~GHz luminosity we would expect for a source to be detected; we show this curve in Fig.~\ref{fig:first_lim}.
For radio undetected sources we plot the radio luminosity corresponding to $R=L_{5\,\mathrm{GHz}}/L_{4400}=10$ (the luminosity above which they would be considered radio-loud).
Radio-undetected AGN with luminosities more than 3$\sigma$ above the curve (2\textcolor{black}{06} AGN) are considered to be radio-quiet, since we would expect them to have been radio-detected if they were radio-loud.
However, as Fig.~\ref{fig:first_lim} illustrates, there remain 30\textcolor{black}{4} AGN below the curve which are too faint for us to determine whether they are either radio-loud or radio-quiet.
Following this exercise, we determine that our sample contains 6\textcolor{black}{8} radio-loud AGN, 2\textcolor{black}{39} radio-quiet AGN and a total of 38\textcolor{black}{9} AGN for which we are unable to determine the radio-loudness.
The radio-loudness classifications are summarised in Table~\ref{tab:radio} \textcolor{black}{and illustrated in Fig.~\ref{fig:first_radio_loud}}.

\begin{table*}
    \caption{Radio-loudness of the sample}
    \label{tab:radio}
    \centering
    \begin{tabular}{l|cll}
    \hline
        Subset                                                & $N$    & \multicolumn{2}{l}{Per cent} \\
    \hline
        No radio coverage (Outside FIRST footprint)           & 20                      & 2.8 \%  & of the full sample \\
        No 4400~\si\angstrom\ coverage                        & 65                      & 9.3 \%  & of the full sample \\
        Radio-detected and radio-loud                         & 6\textcolor{black}{8}                      & 11.\textcolor{black}{1} \% & of AGN with radio and 4400~\si\angstrom\ coverage \\
        Radio-detected and radio-quiet                        & 3\textcolor{black}{3}                      &  5.\textcolor{black}{4} \% & of AGN with radio and 4400~\si\angstrom\ coverage\\
        Radio-undetected and radio-quiet                      & 2\textcolor{black}{06}    & 3\textcolor{black}{3.7} \% & of AGN with radio and 4400~\si\angstrom\ coverage\\
        Radio-undetected, undetermined radio-loudness         & 30\textcolor{black}{4}    & 49.\textcolor{black}{8} \% & of AGN with radio and 4400~\si\angstrom\ coverage\\
    \hline
    \end{tabular}
\end{table*}

Overall, $\sim10$~per cent of our sample with both FIRST and 4400~\si\angstrom\ coverage is found to be radio-loud, similar to previous estimates for bright quasars \citep{Kellerman89,miller90}.
However, we caution that this radio-loud fraction is a lower limit because of the large number of sources in our sample without any radio-loudness determination.
We tentatively find differences in the radio-loud fractions of broad-line and narrow-line type 1 AGN.
Whilst narrow-line AGN make up $\approx15$~per cent of the sample overall, they constitute only $\approx9$~per cent of the radio-loud sources.
This is consistent with previous findings that the radio-loud fraction of NLS1s is lower than that of typical broad-line AGN and quasars \citep{Komossa06,Zhou06,Rakshit17}.
Again, selection biases and incomplete data mean it is not possible to draw firm conclusions from our sample.

\section{Black hole masses}
\label{sec:mass}
The widths and luminosities of the broad permitted emission lines, as well as the continuum luminosities can be used to derive the masses of the black holes in our AGN.
In this Section we assess the broad emission line properties and calculate the black hole masses using new scaling relations.
\subsection{Broad emission line properties}
\label{sec:lines}
\begin{figure}
	\includegraphics[width=\columnwidth]{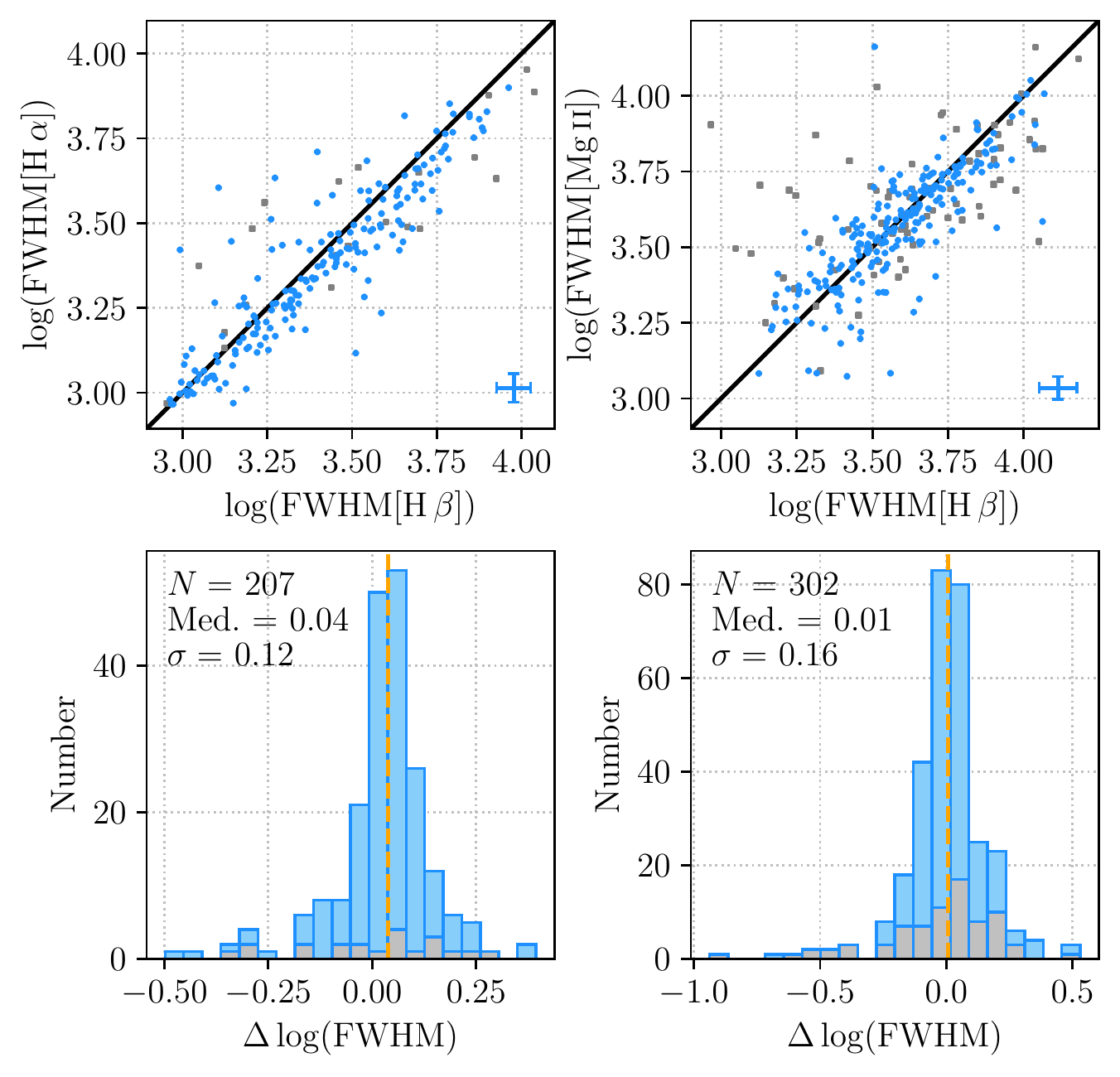}
    \caption{
    A comparison of the full widths at half maxima of \ha, \hb\ and \mg.
    Upper panels show the values for sources with at least two lines in their spectra.
    Representative errorbars are shown in the lower right corners of the plots.
    The lower panels show the logarithm of the line width ratios.}
    \label{fig:fwhms}
\end{figure}

We compare the broad line widths for sources with spectra containing multiple lines.
The line width comparisons are shown in Fig.\ref{fig:fwhms}. 
We find that \hb\ is slightly broader than both \ha\ and \mg.
207 spectra contain both \ha\ and \hb\ and from the median of the logarithmic distribution we find that \hb\ is on average $\approx10$~per~cent broader than \ha.  
There is greater scatter between the widths of \hb\ and \mg, and we do not find strong evidence of a systematic difference in the line widths (\hb\ is on average only $\approx2$~per~cent broader than \mg) as measured in 302 spectra.
Qualitatively, these results somewhat are similar to those of previous studies, e.g.\ \cite{OS82} found \hb\ to be $\approx16$~per~cent broader than \ha\ in Seyfert galaxies; \cite{Mejia16} (hereafter \citetalias{Mejia16}) also found \hb\ to be broader than both \ha\ and \mg.

\begin{figure}
	\includegraphics[width=\columnwidth]{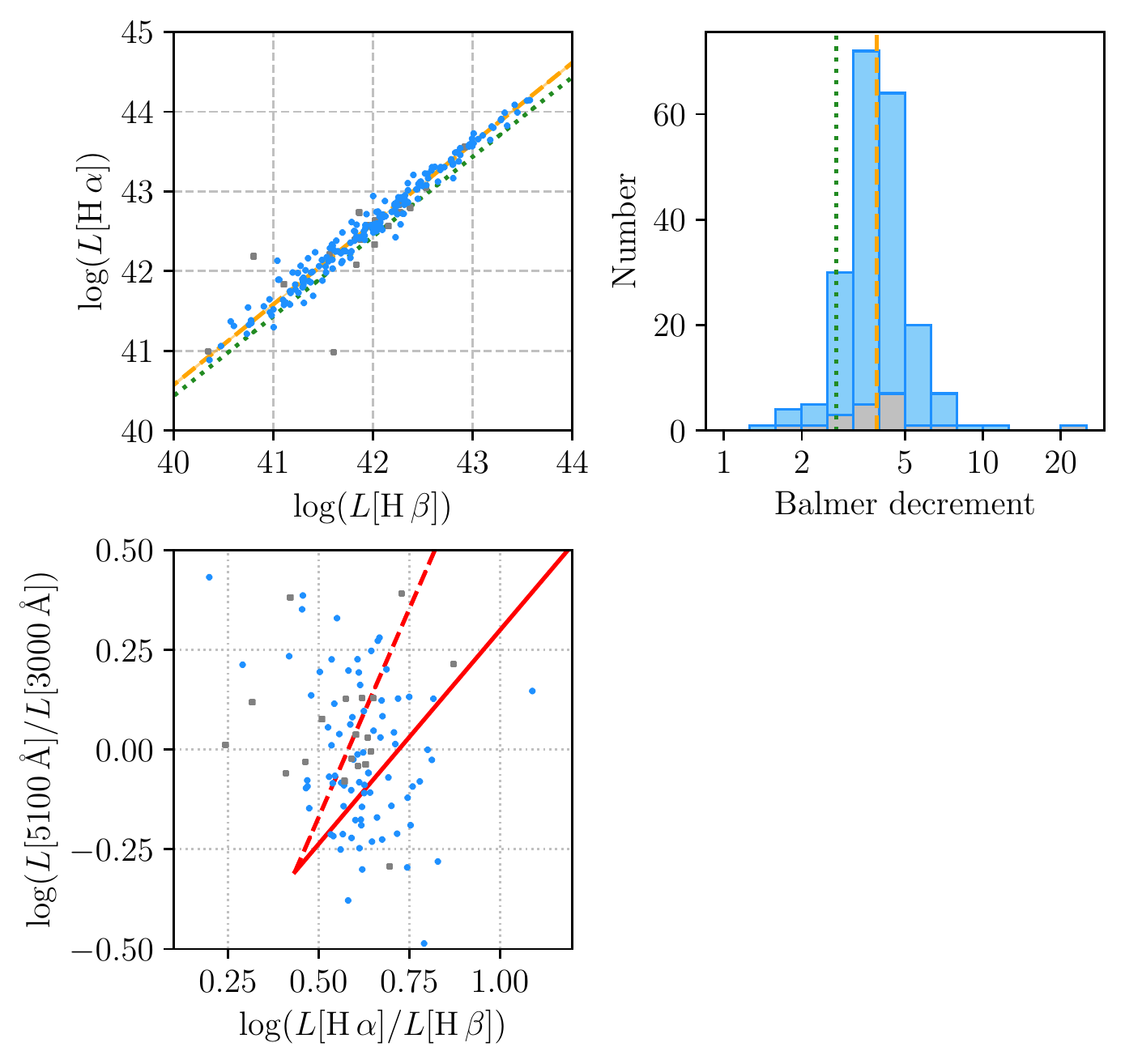}
    \caption{
    \textit{Top left:} A comparison of the \ha\ and \hb\ broad line luminosities for low-$z$ sources with spectra containing both lines.
    \textit{Top right:} The distribution of the Balmer decrements of these sources.
    The orange dashed lines indicate the best-fitting linear regression, corresponding to $L_\mathrm{H\,\upalpha}/L_\mathrm{H\,\upbeta}=3.9$.
    The green dotted lines correspond to $L_\mathrm{H\,\upalpha}/L_\mathrm{H\,\upbeta}=2.72$ \protect\citep{Gaskell17}.
    \textit{Bottom:}  Balmer decrements versus ratios of the monochromatic continuum lumionsities at 5100 and 3000~\AA.
    The \textcolor{black}{solid} red line shows the expected trend from the \protect\cite{Gaskell04} \textcolor{black}{AGN} reddening curve\textcolor{black}{; the predicted trend for the Milky Way reddening curve of \protect\cite{CCM89} is shown by the dashed line, for comparison}.
    Data with good quality measurements of all quantities are shown as blue circles; the grey squares represent data in which at least one measurement has a quality flag raised.
    \textcolor{black}{All luminosities have been dervied from the SDSS optical spectrum and are therefore contemporaneous.}}
    \label{fig:balmer_decs}
\end{figure}

In Fig.\ref{fig:balmer_decs} we show the ratio of the \ha\ to \hb\ luminosities (the Balmer decrement).
\textcolor{black}{For the logarithmic mean value, w}e find $\langle \mathrm{H}\,\upalpha / \mathrm{H}\,\upbeta \rangle=3.9$ which is higher than the Case B value 2.74 (\citealt{OF06}) and the intrinsic value $2.72\pm0.04$ suggested by \cite{Gaskell17}, based on observations of blue AGN.
This Balmer decrement is also greater than the average found in previous studies of type 1 AGN:  $\langle \mathrm{H}\,\upalpha / \mathrm{H}\,\upbeta \rangle=3.16$ (\citealt{Lu19}); $\langle \mathrm{H}\,\upalpha / \mathrm{H}\,\upbeta \rangle=3.06$ (\citealt{Dong08}); $\langle \mathrm{H}\,\upalpha / \mathrm{H}\,\upbeta \rangle=3.45$ (\citealt{LaMura07}).
From the \cite{Shen11} Catalog of SDSS DR7 Quasar Properties, we determine $\langle \mathrm{H}\,\upalpha / \mathrm{H}\,\upbeta \rangle=3.55$ with $\sigma=0.14$~dex.
Taken as a reddening indicator, this large average Balmer decrement implies a high degree of \textcolor{black}{intrinsic} dust extinction in our sample of AGN: on average the colour excess $E(B-V)\approx0.4$~mag. 
However, if a large Balmer decrement is due to reddening, we would expect the optical\textcolor{black}{-UV} continuum to be similarly affected such that AGN with a higher Balmer decrement will have a correspondingly greater value of $L_{5100}/L_{3000}$.
\textcolor{black}{Assuming an intrinsic (Case B) broad line Balmer decrement of 2.72 \citep{Gaskell17} and an AGN continuum slope $\alpha=1/3$, we have used the AGN reddening curve of \cite{Gaskell04} to calculate Balmer decrements and the corresponding continuum luminosity ratios for increasing dust extinction: this trend line is shown in the lower panel of Fig.~\ref{fig:balmer_decs}.
The \cite{Gaskell04} AGN curve is much flatter in the UV than Milky Way curves; for comparison, Fig.~\ref{fig:balmer_decs} also shows the expected luminosity ratio trend derived from the \cite{CCM89} Galactic reddening curve.}
\textcolor{black}{Each data point in the figure has been obtained from a single spectrum; it} can be seen, \textcolor{black}{that} these data do not follow \textcolor{black}{either} trend.

\begin{figure*}
    \begin{tabular}{c}
    \includegraphics[width=0.75\textwidth]{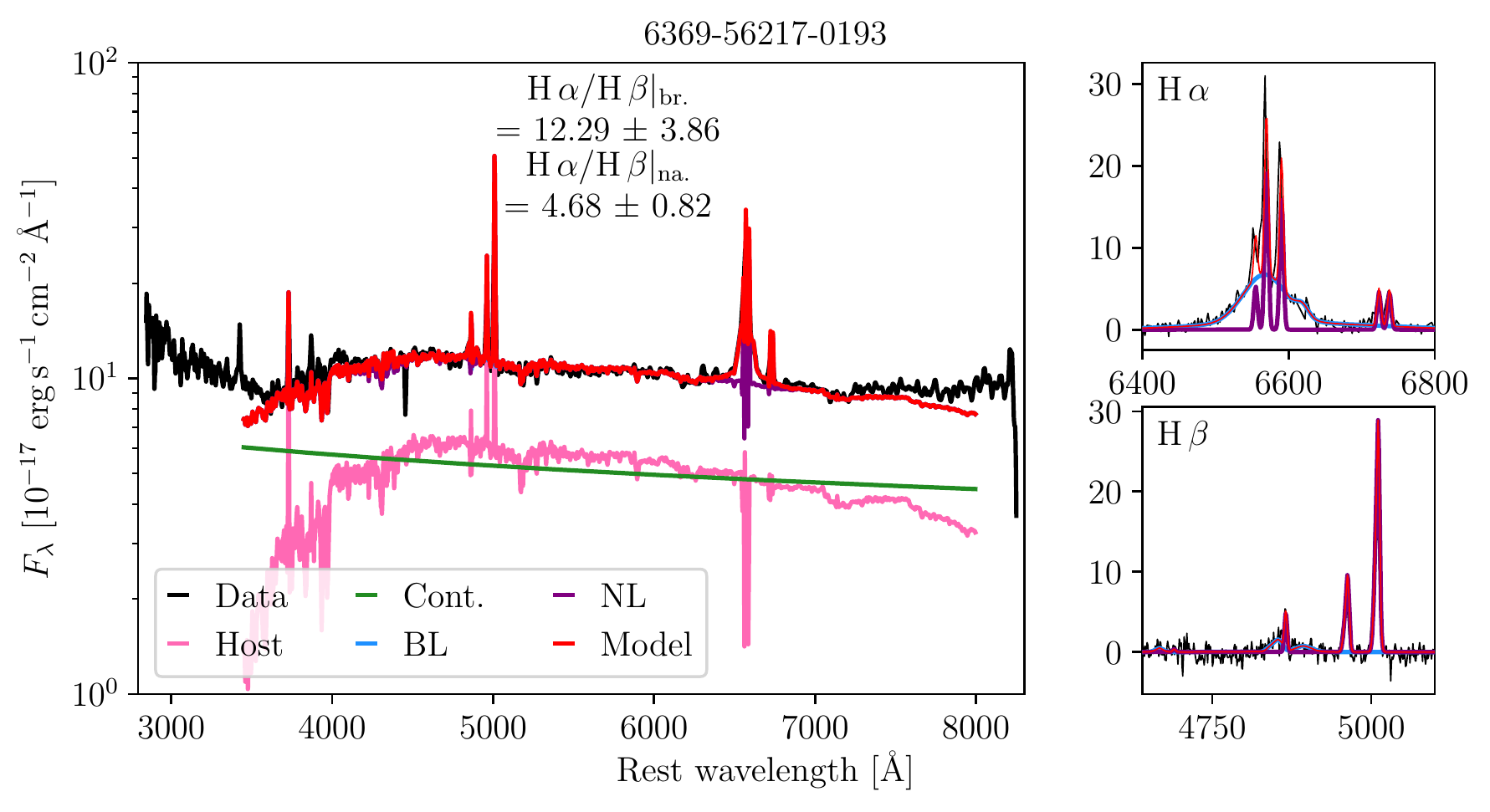} \\
    \includegraphics[width=0.75\textwidth]{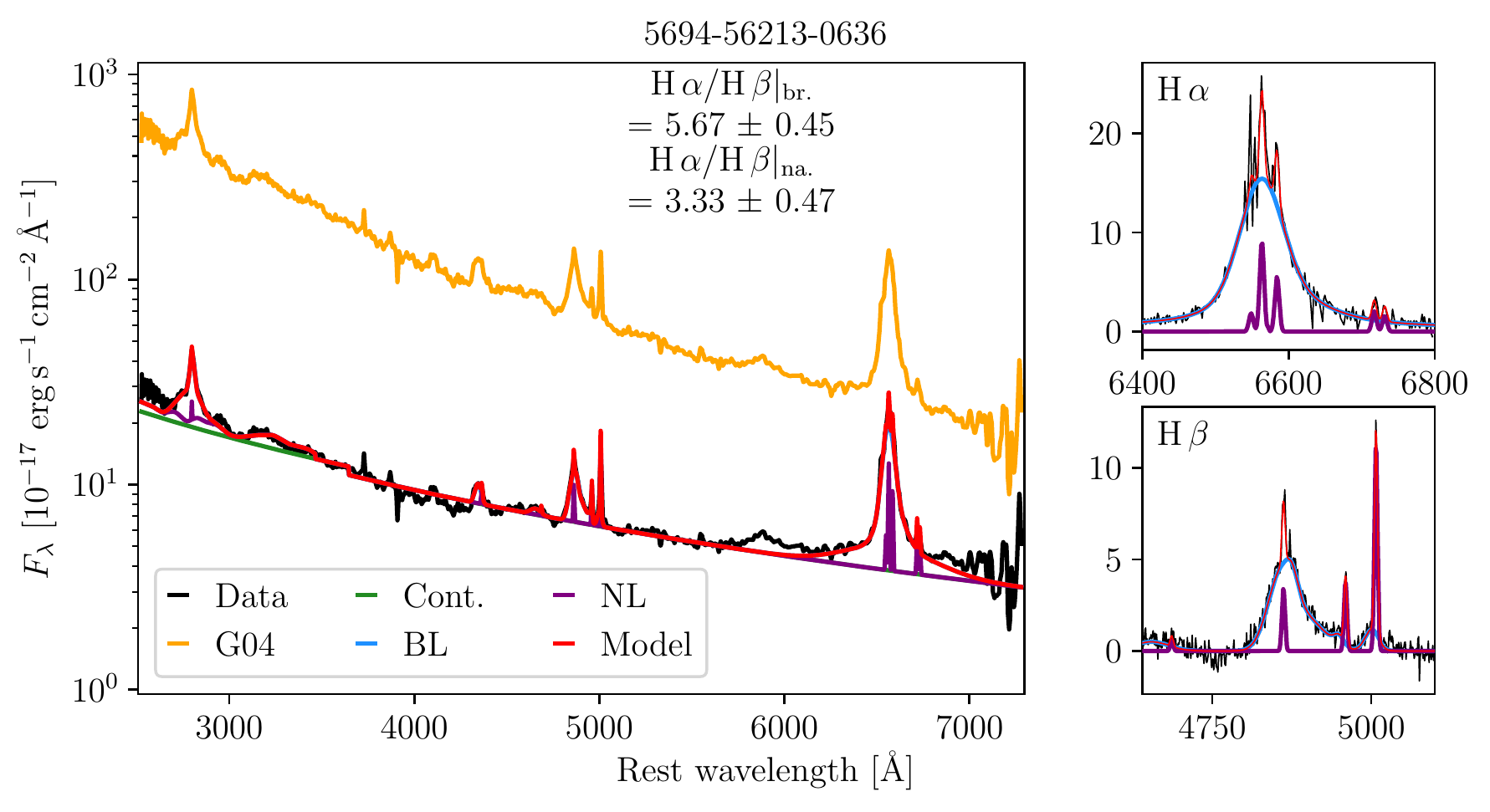} \\
    \end{tabular}
    \centering
    \caption{Example spectra with high broad Balmer line decrements.
    The H\,$\upalpha$/H\,$\upbeta$ flux ratios of both broad and narrow lines are given in the inset text.
    Each spectrum (shown in black) has been corrected for Galactic reddening and transformed to the rest frame.
    The inset legend indicates the components of the total \textsc{PyQSOFit} model (red): the host galaxy emission (pink); the AGN continuum (green) and selected, modelled broad and narrow emission lines (`BL' and `NL', coloured blue and purple, respectively).
    In the lower panel, we also show in orange the spectrum 5694-56213-0636 additionally dereddened with the \protect\cite{Gaskell04} curve assuming an $A_\mathrm{v}=5.3$ derived from the broad Balmer line decrement.}
    \label{fig:hbd}
\end{figure*}

\textcolor{black}{We further investigated whether spectra with large Balmer decrements may be reddened by visually inspecting those with ($\mathrm{H}\,\upalpha / \mathrm{H}\,\upbeta>5$): two examples are shown in Fig.~\ref{fig:hbd}.
Whilst reddening is a plausible explanation for the large Balmer decrement in the spectrum 6369-56217-0193, visual inspection of the spectrum 5694-56213-0636 does not indicate any obvious reddening.
Since the UV continuum is not highly extinguished, if the AGN is reddened its reddening curve must flatten towards shorter wavelengths.
\cite{Gaskell04} presented such a curve for AGN, and we have used this curve to modify this spectrum using an $A_\mathrm{v}=5.3$ (estimated from the broad line Balmer decrement assuming the intrinsic line ratio is 2.72).
The modified spectrum is clearly much brighter than the one used in our analysis, but is still quasar-like so we cannot definitively rule out intrinsic reddening.
We discuss this point further in Section~\ref{sec:disc-blr}.}

\begin{figure}
    \centering
    \includegraphics[width=\columnwidth]{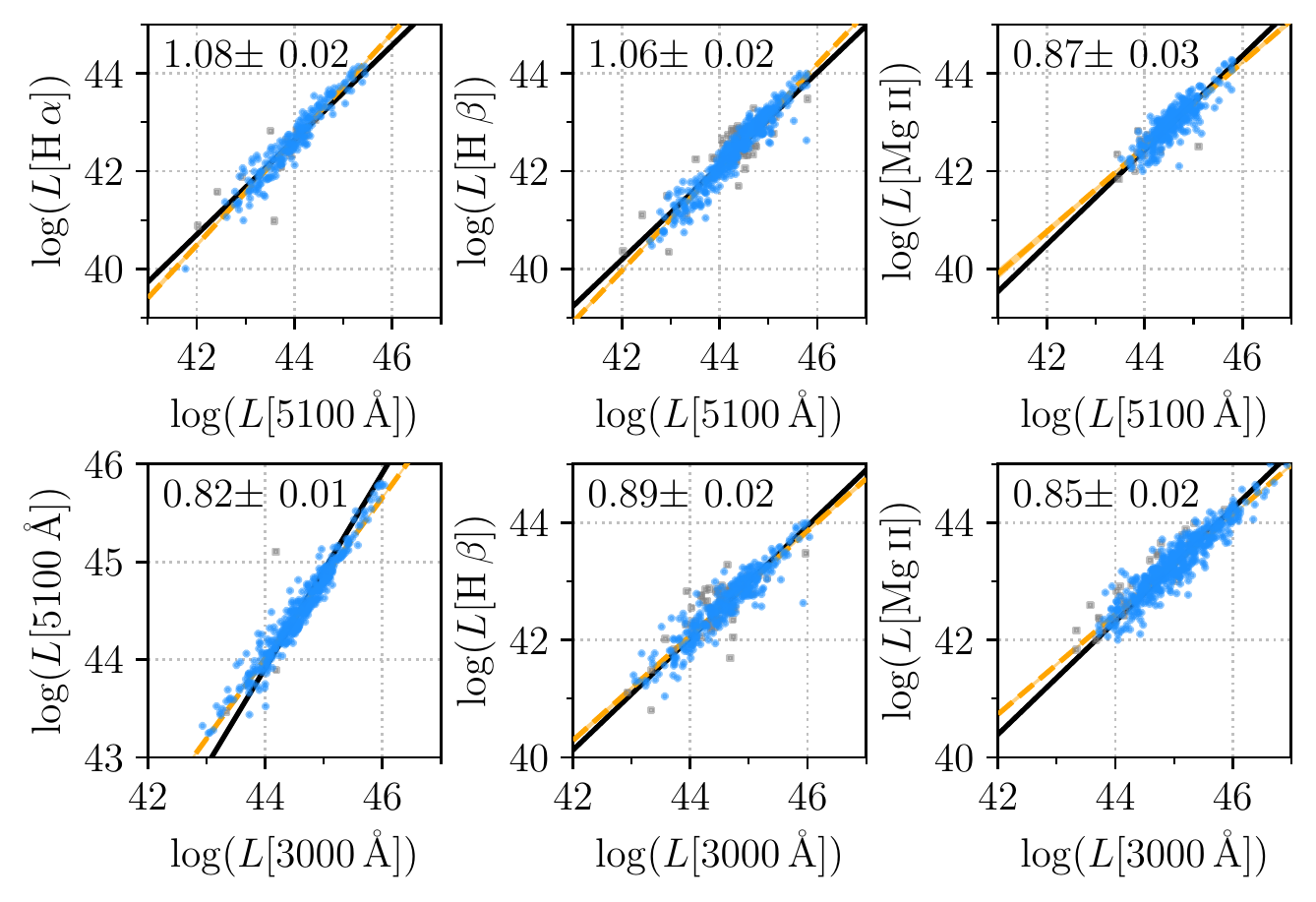}
    \caption{Correlations between emission line and continuum luminosities.
    Dashed orange lines show the best-fitting linear regression to the data, and the value of the slope ($m$) is given in the inset text.
    Relations with a gradient $m=1$ are shown for comparison (solid black lines).
    Data with good quality measurements of all quantities are shown as blue circles; the grey squares represent data in which at least one measurement has a quality flag raised.
    \textcolor{black}{All luminosities have been dervied from the SDSS optical spectrum and are therefore contemporaneous.}
    }
    \label{fig:line_cont_lum}
\end{figure}

We have computed the best-fitting linear regressions between \textcolor{black}{various luminosities} using the package \textsc{linmix} \citep{Kelly07}.
The results are shown in Fig.~\ref{fig:line_cont_lum} and the values are quoted in Table~\ref{tab:line_cont_lum}.
The Balmer lines show an almost 1:1 correspondence with the 5100\,\si\angstrom\ continuum luminosity, whereas the relation with \mg\ is slightly shallower with a gradient $m=0.87\pm0.03$.  
The relationships of both \hb\ and \mg\ luminosities with the 3000\,\si\angstrom\ continuum luminosity are shallower compared with those calculated for the 5100\,\si\angstrom\ luminosity.
The strongest correlation we find is between the two continuum luminosities: the best-fitting linear regression has a slope $m=0.82\pm0.01$ and a dispersion of only 0.11~dex (around half that of the emission lines).

\begin{table}
    \centering
    \caption{Correlations between emission line and continuum luminosities.
    We give the independent and dependent variables ($x$ and $y$) and the slope and intercept ($m$ and $c$) of the best-fitting linear regression.
    $\sigma$ is the dispersion of the data about the linear regression.}
    \begin{tabular}{llccc}
    \hline
    $x$              & $y$                                & $m$           & $c$          & $\sigma$  \\
    \hline
    $\log{L_{5100}}$ & $\log{L_\mathrm{H\,{\upalpha}}}$   & $1.08\pm0.02$ & $-5.0\pm1.0$ & 0.20 \\
    $\log{L_{5100}}$ & $\log{L_\mathrm{H\,\upbeta}}$      & $1.06\pm0.02$ & $-4.4\pm0.7$ & 0.18 \\
    $\log{L_{5100}}$ & $\log{L_\mathrm{Mg\,\textsc{ii}}}$ & $0.87\pm0.03$ & $4.4\pm1.3$  & 0.21 \\
    $\log{L_{3000}}$ & $\log{L_{5100}}$                   & $0.82\pm0.01$ & $7.8\pm0.5$  & 0.11 \\
    $\log{L_{3000}}$ & $\log{L_\mathrm{H\,\upbeta}}$      & $0.89\pm0.02$ & $2.8\pm0.9$  & 0.20 \\
    $\log{L_{3000}}$ & $\log{L_\mathrm{Mg\,\textsc{ii}}}$ & $0.85\pm0.02$ & $5.1\pm0.7$  & 0.19 \\
    \hline
    \end{tabular}
    \label{tab:line_cont_lum}
\end{table}

\subsection{Black hole mass estimates from broad emission lines}
\label{sec:mass_measurements}
Assuming that the gas in the BLR is virialised, an estimate of the mass of the black hole ($M_\mathrm{BH}$) can be made by applying the virial theorem:
\begin{equation}
    M_\mathrm{BH} = f\frac{R_\mathrm{BLR}V^2}{G}
    \label{eqn:virial}
\end{equation}
where $R_\mathrm{BLR}$ is the radius of the BLR, $V$ is the velocity of the gas and $G$ is the gravitational constant.
$R_\mathrm{BLR}$ is often determined from radius-luminosity relations derived for reverberation-mapped AGN (e.g.\ \citealt{Peterson04}; \citealt{Bentz06}; \citealt{Landt11}).
The virial factor $f$ accounts for the geometry and inclination to the line of sight of the BLR and cannot be easily determined for individual sources.
Where the broad line FWHM is used as a proxy for $V$ then $f\sim1$.

We first make estimates of the black hole masses of the AGN in our sample using recently-calibrated relations for virial black hole masses based on the broad \ha, \hb\ and \mg\ emission lines.
The relations used are listed in Table~\ref{tab:bh_mass_relations}.
It is common to use the AGN continuum luminosity as a proxy for the BLR radius, however \citetalias{Mejia16}, \cite{Greene10} and \cite{Woo18} provide relations using instead the broad emission line luminosity (of \ha, \hb, and \mg, respectively). 
Both \citetalias{Mejia16} and \cite{Woo18} provide relations to estimate the black hole mass directly from a luminosity and broad emission line FWHM, assuming $f=1$ and $f=1.12$, respectively. 
For the \citetalias{Mejia16} relations we use the values calibrated from the global fit to the continuum (the middle column of Table~7 in that paper); these are the most appropriate relations for our sample since \textsc{PyQSOFit} performs a global fit to the broadband AGN continuum, rather than a local fit under each separate emission line.
The \cite{Woo18} relation we adopt for \mg\ was derived with free $\beta$ and $\gamma$, which has least scatter with respect to fiducial black hole masses estimated using FWHM(\hb) and $L_{5100}$.
\cite{Greene10} provide a formula to calculate $R_\mathrm{BLR}$ from the \hb\ emission line luminosity; with an estimate of $R_\mathrm{BLR}$we then calculate the black hole mass using Eqn.~(\ref{eqn:virial}) assuming $f=1$.  

\begin{table}
    \centering
    \caption{The virial black hole mass relations from literature used.}
    \begin{tabular}{lllcc}
         \hline
         FWHM & $L$                & Relation ref.\ & $N_\mathrm{spectra}$ & $N_\mathrm{spectra}^\mathrm{good}$ \\ 
         \hline
         \ha\ & 5100\,\si\angstrom & MR16  & 219 & 189 (86\%) \\ 
         \ha\ & \ha\               & MR16  & 220 & 194 (88\%) \\ 
         \hb\ & 5100\,\si\angstrom & MR16  & 481 & 401 (83\%) \\ 
         \hb\ & \hb\               & G10   & 482 & 406 (84\%) \\ 
         \mg\ & 3000\,\si\angstrom & MR16  & 517 & 479 (93\%) \\  
         \mg\ & \mg\               & W18   & 521 & 483 (93\%) \\ 
         \hline 
    \end{tabular}
    \parbox[]{7cm}{
    We list the emission-line FWHM and luminosity $L$ used to calculate the black hole masses.
    $N_\mathrm{spectra}$ is the number of spectra in which the relevant quantities were measured.  
    $N_\mathrm{spectra}^\mathrm{good}$ is the subset of $N_\mathrm{spectra}$ (and the percentage) for which no quality flags were raised on the relevant emission line or luminosity measurements.
    \textit{References}: MR16 = \cite{Mejia16}; G10 = \cite{Greene10}; W18 = \cite{Woo18}.
    }
    \label{tab:bh_mass_relations}
\end{table}

\subsection{Comparisons of black hole mass estimates}
\begin{figure*}
    \centering
    \includegraphics[width=1.5\columnwidth]{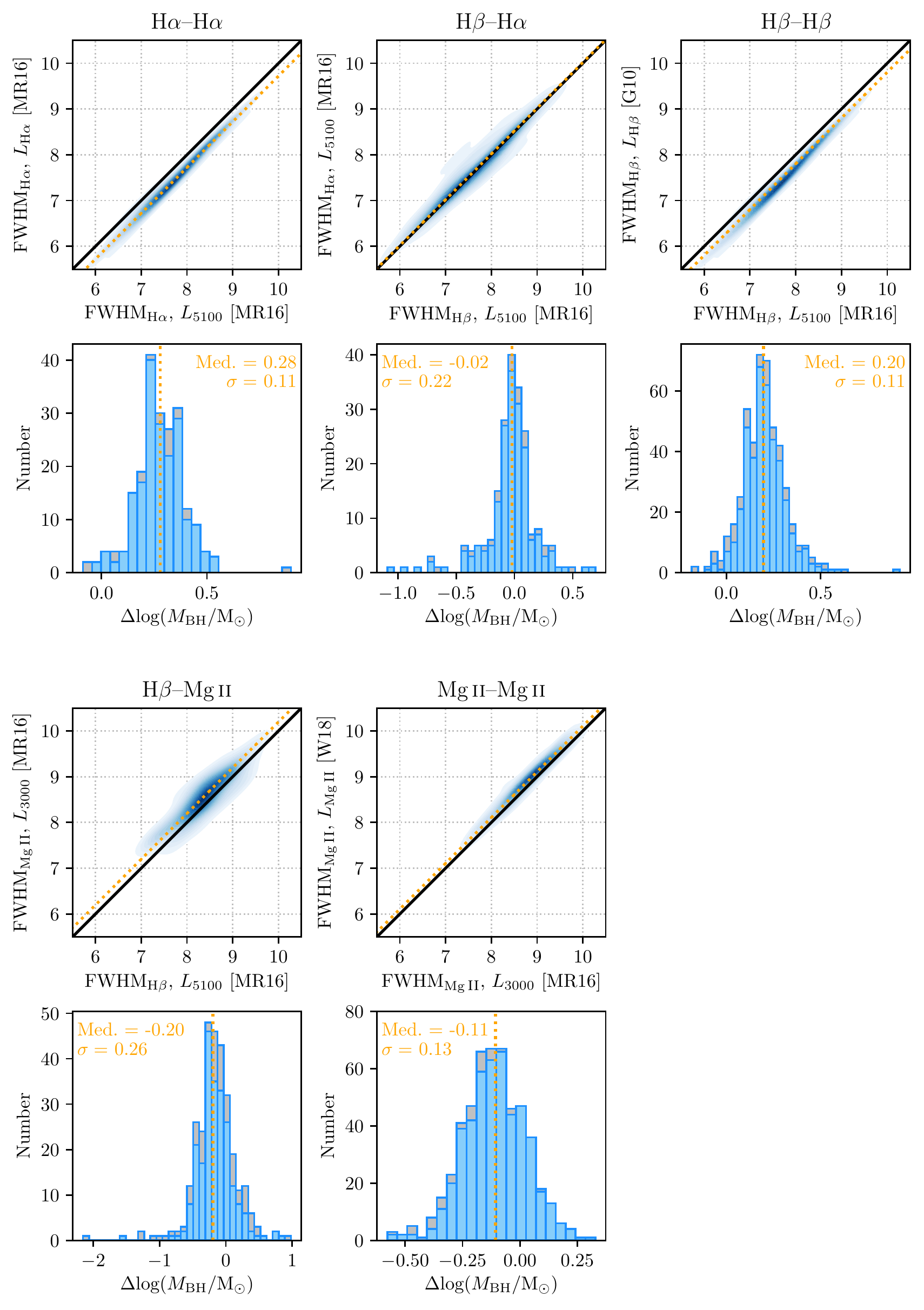}
    \caption{Comparisons of black hole mass estimates made using single-epoch relations taken from literature.
    The contour plots in the upper panels show $\log(M_\mathrm{BH}/\mathrm{M}_\odot)$, calculated using the broad emission line FWHM and the luminosity indicated on the axis label. The literature reference for the relation used is also given in square parentheses (see Table~\ref{tab:bh_mass_relations} in the text).
    The lower panels show the difference in $\log(M_\mathrm{BH}/\mathrm{M}_\odot)$.
    Reliable mass differences (with no quality flags raised on any of the quantities used to calculate the two masses) are shown in blue and the quality-flagged values are shown in grey.
    The median and standard deviations of the distributions for the reliable differences are given in the inset text.
    \textcolor{black}{The solid black lines in the upper panels show the 1:1 relation and} the orange dotted lines mark the median in both lower and upper panels.
    }
    \label{fig:mass_comparisons}
\end{figure*}

In Fig.~\ref{fig:mass_comparisons} we compare the mass estimates obtained using the relationships given in Table~\ref{tab:bh_mass_relations}.
Generally, we find good agreement between the different estimates.
For estimates using the continuum luminosity, we see excellent agreement between \ha\ and \hb\ masses;
on average the difference between the mass estimates are negligible ($\Delta{M}_\mathrm{BH}\approx5$~per~cent) and the scatter is $\sigma=0.22$~dex.
There is slightly greater scatter between the \hb\ and \mg\ estimates ($\sigma=0.26$), as has been noted in previous studies (e.g. \citetalias{Mejia16}).
The average agreement is less good, also, with the \mg\ masses being systematically higher than the \hb\ masses by $\approx60$~per~cent.
The greatest discrepancy we see is between the two \ha\ relations: masses calculated using the broad \ha\ line luminosity are systematically lower than those made with the continuum luminosity by almost a factor of two.
Mass estimates using the broad \hb\ line luminosity are also systematically lower than those involving the continuum luminosity, but the discrepancy is less than between the \ha\ estimates ($\approx60$~per cent).
In the case of \mg, there is a very minor systematic in the opposite sense with estimates using the continuum luminosity being $\approx20$~per~cent smaller than those using the line luminosity.

\begin{figure}
    \centering
    \includegraphics[width=\columnwidth]{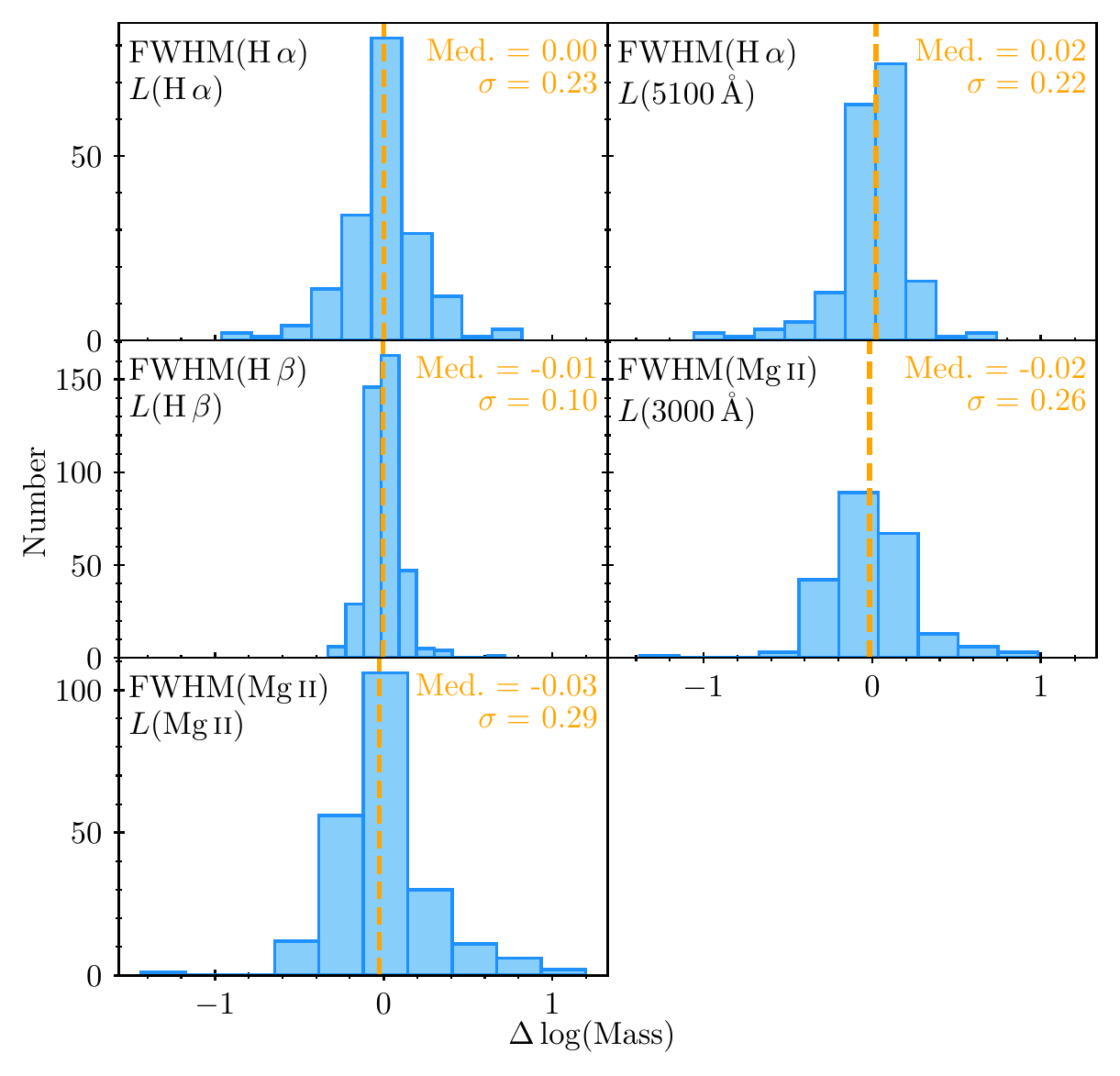}
    \caption{Comparisons of BH mass estimates from broad emission lines.
    $\Delta\log(\mathrm{Mass})=\log\left(\mathrm{Mass|}^{\mathrm{FWHM}(\mathrm{H}\,\upbeta)}_{L(5100\,\si\angstrom)}\right)-\log\left(\mathrm{Mass|}^{\mathrm{FWHM}(x)}_{L(y)}\right)$.}
    \label{fig:new_delta_mass}
\end{figure}

To obtain better consistency between the various mass estimates, we have calculated new virial mass relations.
We have assumed that the relation using the FWHM of broad \hb\ and the 5100~\AA\ luminosity provides the best and most reliable estimate of the mass.
Then, assuming a virial mass relation of the form
\begin{equation}
   \log\left(\frac{M_\mathrm{BH}}{\mathrm{M}_\odot}\right) = \log(K) + \alpha\log\left(\frac{L}{10^{44}~\mathrm{erg}\,\mathrm{s}^{-1}}\right)  + 2\log\left(\frac{\mathrm{FWHM}}{1000~\mathrm{km}\,\mathrm{s}^{-1}}\right),
\end{equation}
we determine the quantities $K$ and $\alpha$ which reduce the systematic offset and scatter in the relations with respect to the \hb-5100~\AA\ mass estimates.
These values are given in Table~\ref{tab:new_mass_rel} along with the median $\Delta\log(M_\mathrm{BH})$ with respect to the \hb-5100~\AA\ mass estimates, and the dispersion $\sigma$.
\textcolor{black}{Here, and in Fig.~\ref{fig:new_delta_mass}, it} can be seen that the systematic offsets between different relations are greatly reduced.

For additional quality assurance, we visually inspected all of the optical spectra of the AGN in our sample \textcolor{black}{to identify problematic line profiles (e.g.\ very noisy or low-contrast profiles; profiles affected by absorption features or data gaps, etc.)} and select the best broad line from which to estimate the black hole mass, \textcolor{black}{considering} the line profile shape and S/N in the vicinity of the line.
For each AGN in our sample we list the `preferred' mass estimate indicating our choice of broad emission line and luminosity scaling relation for that source.
We have adopted, wherever possible, the black hole mass from estimates using FWHM(\hb) and $L_{5100}$\textcolor{black}{,
since \hb\ is the broad emission line most commonly used to establish the BLR radius-luminosity relationship in optical reverberation mapping campaigns and this relationship forms the basis of single-epoch mass estimates such as those used here.
As noted in the literature, other line-luminosity relations are often associated with a greater overall uncertainty on the black hole mass and typically relations involving other lines are calibrated to that of \hb\ (e.g.\ \citealt{Netzer07}; \citealt{Shen11}; \citealt{Mejia16}; \citealt{Woo18}).
In cases where the \hb\ profile was unsuitable, we have chosen another broad line and nearby continuum luminosity (usually \ha-$L_{5100}$ for low-redshift sources and \mg-$L_{3000\,\si\angstrom}$ for high-redshift sources).
Spectra lacking any broad emission line from which a black hole mass could be reliably estimated were previously removed in our initial quality inspection (Section~\ref{sec:curated})}.

\begin{table}
    \centering
    \caption{Recalibrated virial black hole mass relations}
    \resizebox{\columnwidth}{!}{\begin{tabular}{cc|ccccc}
    \hline
    FWHM  & $L$      & $\log(K)$          & $\alpha$        & Median  & $\sigma$ & $N_\mathrm{spec}$ \\
          &          &                    &                 & $\Delta\log(M_\mathrm{BH})$ & & \\
    \hline
    \ha\  & \ha\     & $7.56\pm0.04$      & $0.47\pm0.02$   & $0.00$  & $0.21$ & 182 \\
    \ha\  & 5100~\AA & $6.91\pm0.02$      & $0.55\pm0.03$   & $0.02$  & $0.22$ & 182 \\
    \hb\  & \hb\     & $7.83\pm0.01$      & $0.494\pm0.007$ & $-0.01$ & $0.10$ & 401 \\
    \mg\  & \mg\     & $7.70\pm0.05$      & $0.46\pm0.04$   & $-0.03$ & $0.29$ & 224 \\
    \mg\  & 3000~\AA & $6.86\pm0.03$      & $0.51\pm0.03$   & $-0.02$ & $0.26$ & 224 \\
    \hline
    \end{tabular}}
    \parbox[]{\columnwidth}{The relations are of the form $\log(M_\mathrm{BH}) = \log(K) + \alpha\log(L) + 2\log(\mathrm{FWHM})$.
    Values of $K$ and $\alpha$ have been determined which minimise the systematic offset and scatter with respect to masses estimated from the relation involving the FWHM of broad \hb\ and the 5100~\AA\ luminosity.  
    The relations are calculated with the luminosity $L$ in units $10^{44}$~erg\,s$^{-1}$ and FWHM in units 1000~\kms.
    For each relation we have calculated we give the resulting systematic offset and dispersion with respect to the \hb-5100~\AA\ mass estimates.
    $N_\mathrm{spec}$ is the number of spectra used to establish the relation.}
    \label{tab:new_mass_rel}
\end{table}

\section{$\alpha_\mathrm{\lowercase{ox}}$: The UV/X-ray relationship}
\label{sec:aox}
We calculate the UV/optical-X-ray energy index
\begin{equation}
\alpha_\mathrm{ox} = -0.384 \log\left(\frac{L_\mathrm{2\,keV}}{L_{2500}}\right) 
\end{equation}
for the AGN in our sample.
An $\alpha_\mathrm{ox}$ determination is possible for \textcolor{black}{685} AGN in our sample (9\textcolor{black}{8} per cent), with the remaining AGN having no spectroscopic or photometric coverage of 2500~\si\angstrom.

\begin{figure*}
    \begin{tabular}{c}
	\includegraphics[width=2\columnwidth]{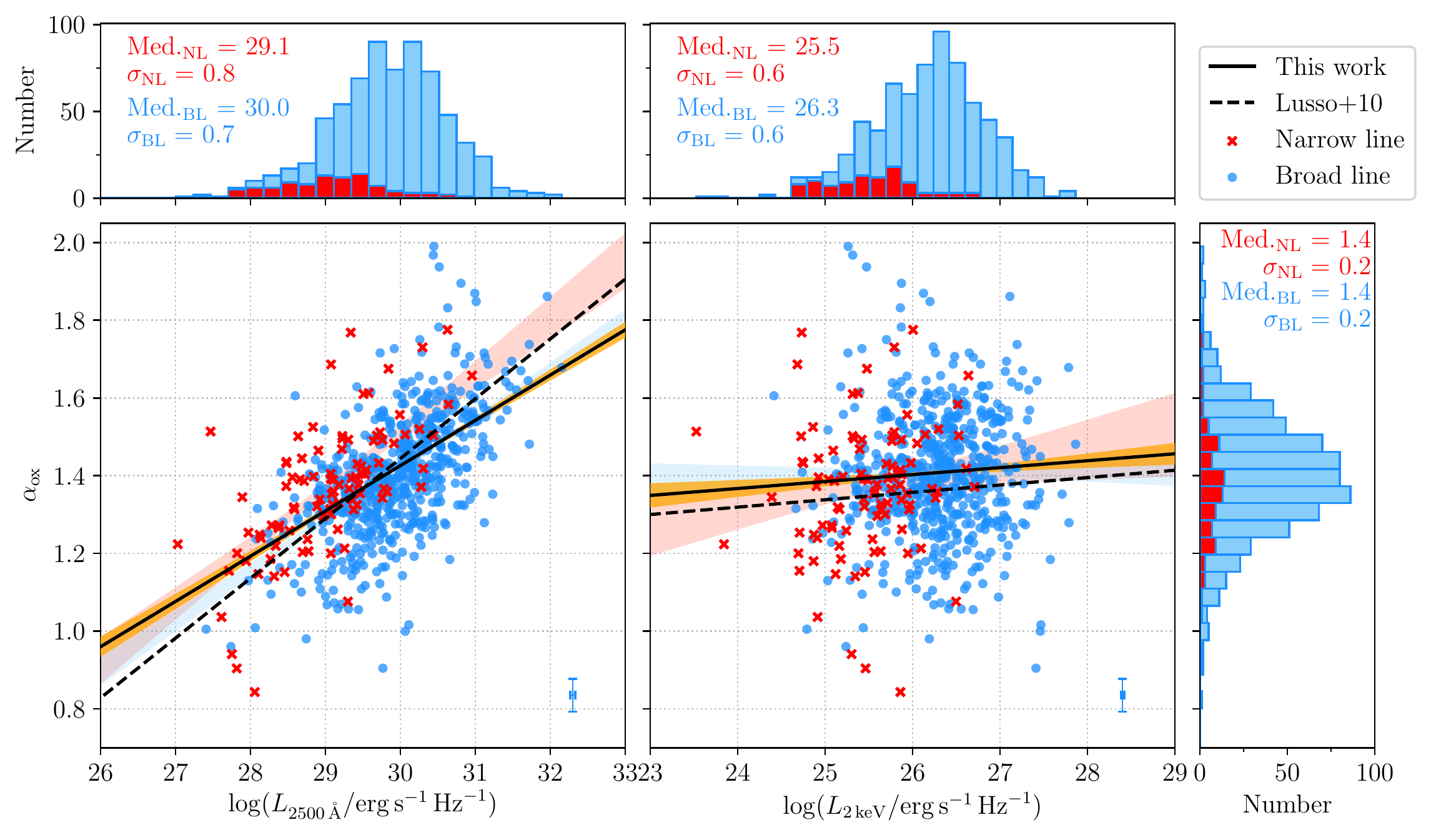} \\
	\includegraphics[width=2\columnwidth]{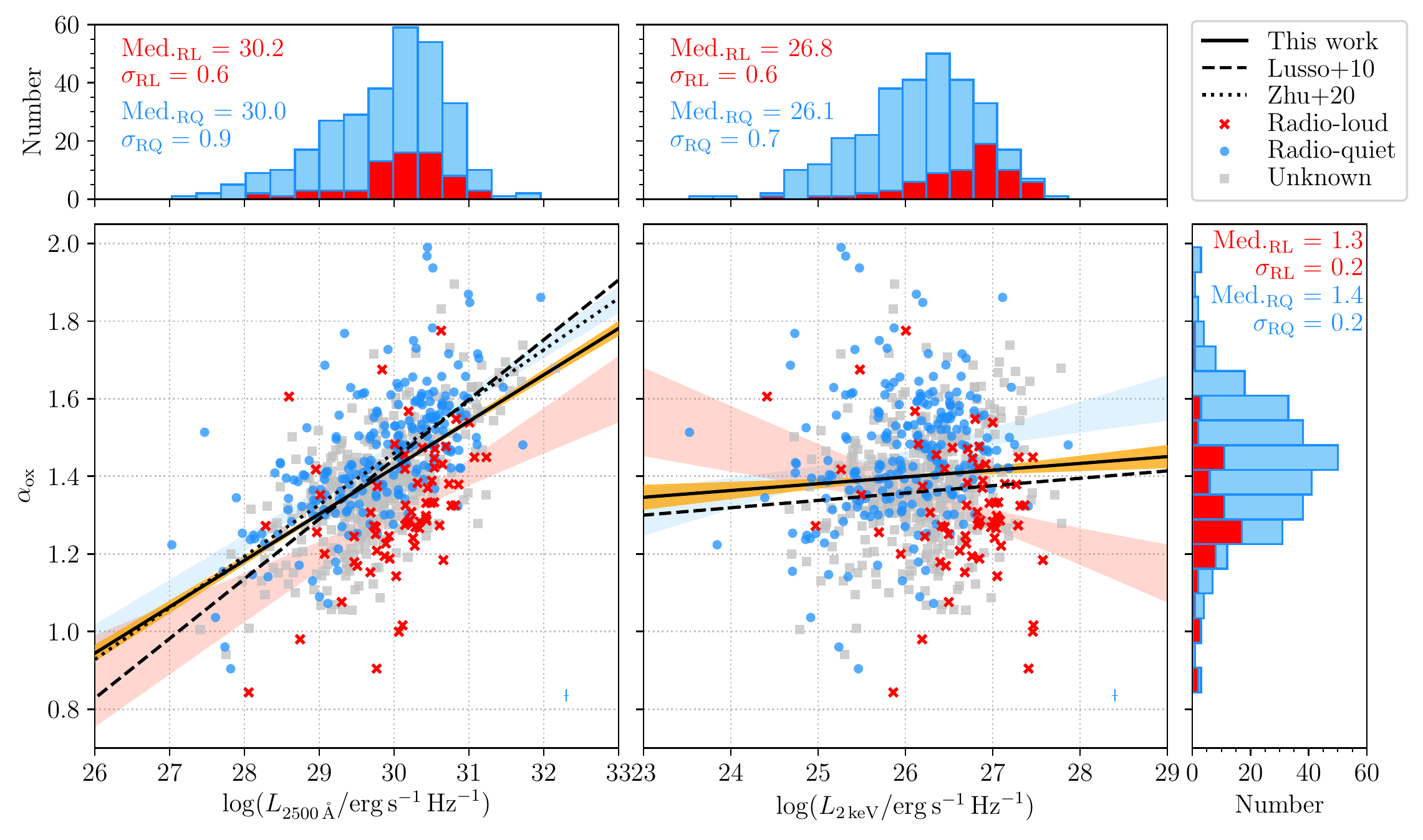} \\
	\end{tabular}
    \caption{
    The relationship between $\alpha_\mathrm{ox}$ and $L_\mathrm{2500}$ (left) and between $\alpha_\mathrm{ox}$ and $L_\mathrm{2~keV}$ (right).
    We show the best fit relation found in this work for all 673 AGN with an $\alpha_\mathrm{ox}$ determination (solid black line) and its 1$\sigma$ confidence region (in orange).
    For comparison we show the best fit relation\textcolor{black}{s} found by \protect\cite{Lusso10} (dashed black line) \textcolor{black}{and \protect\cite{Zhu20} (dotted black line)}.
    Stacked histograms show the distributions of the three quantities.
    In the top figure, narrow-line ($\mathrm{FWHM}<2000$~km\,s$^{-1}$) and broad-line sources are coloured red and blue, respectively; 
    the pale red and blue areas in the scatter plots are the linear regression 1$\sigma$ confidence regions for the respective subsets.
    In the bottom figure data pertaining to the radio-loud and radio-quiet sources are similarly coloured red and blue, respectively.
    \textcolor{black}{The median 1$\sigma$ uncertainty on each quantity is shown by the error bars in the lower right-hand corner of each scatter plot.}
    }
    \label{fig:aox}
\end{figure*}

We find a mean $\alpha_\mathrm{ox}=1.$\textcolor{black}{40} with standard deviation $\sigma=0.16$.
These values are in good agreement with those found by \cite{Lusso10} for a sample of quasars in the XMM-COSMOS field (mean 1.37 and dispersion 0.18).
\cite{Lusso16} also show that there is a correlation of
$\alpha_\mathrm{ox}$ with $L_{2500}$. 
Fig.~\ref{fig:aox} (upper, left-hand panel) shows this for the \textcolor{black}{685} AGN in our sample. 
Plainly there is a good correlation of these quantities in our data.
We use \textsc{linmix}, the Python version of the package \textsc{linmix\_err} \citep{Kelly07}, to perform a linear regression analysis.
We find \textcolor{black}{a statistically significant correlation with}
\begin{equation}
\alpha_\mathrm{ox} = -(2.\textcolor{black}{0}7\pm0.19) + (0.1\textcolor{black}{17}\pm0.00\textcolor{black}{7})\times \log L_{2500}.   
\end{equation}
The dispersion about this line is 0.12~dex, showing that including the luminosity dependence does indeed reduce the scatter, as required if this is to be used as a cosmological probe (\citealt{Lusso16}).
By contrast, there is much larger scatter of $\alpha_\mathrm{ox}$ 
with the monochromatic X-ray luminosity, $L_{2~\mathrm{keV}}$. 
For these quantities we find
\begin{equation}
\alpha_\mathrm{ox} =  (0.\textcolor{black}{94}\pm0.2\textcolor{black}{5}) + (0.0\textcolor{black}{18}\pm0.010)\times \log L_{2~\mathrm{keV}}\textcolor{black}{;}   
\end{equation}
\textcolor{black}{unlike the $\alpha_\mathrm{ox}$-$L_{2500}$ relation, this relation is not statistically significant at the $2\sigma$ level.}

\subsection{$\alpha_\mathrm{ox}$ related to emission line width and radio-loudness}
\begin{figure}
	\includegraphics[width=\columnwidth]{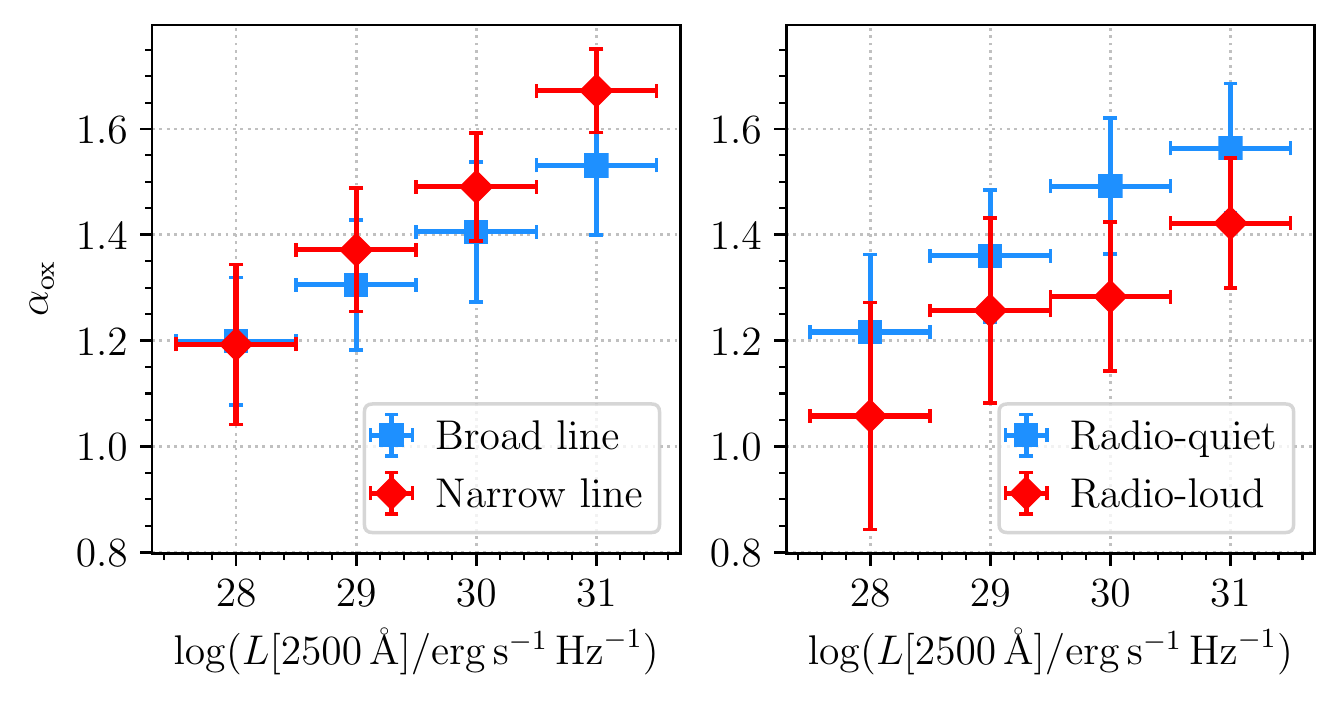} \\
    \caption{
    $\alpha_\mathrm{ox}$ versus UV luminosity for (left) narrow-line / broad-line and (right) radio-loud / radio-quiet subsamples.
    In each luminosity bin, narrow-line sources have higher $\alpha_\mathrm{ox}$ values (softer SEDs) than broad-line sources whereas
    radio-loud sources have lower $\alpha_\mathrm{ox}$ values (harder SEDs) than radio-quiet sources.
    }
    \label{fig:aox_subsamp_lumbin}
\end{figure}

In addition to values for the full sample, we have also calculated median $\alpha_\mathrm{ox}$ and UV/X-ray luminosities for the narrow- and broad-line and radio-loud and radio-quiet subsets.  
We have also determined the luminosity-$\alpha_\mathrm{ox}$ relations for these subpopulations.
The resultant values are listed in Table~\ref{tab:aox}.
It is clear that the narrow-line AGN are typically lower in luminosity (both UV and X-ray) than the broad-line AGN.
A \textcolor{black}{slightly} greater difference is seen in the UV luminosities (median $\Delta L_{2500}\approx\textcolor{black}{0.9}$~dex) than the X-ray luminosities (median $\Delta L_{2\,\mathrm{keV}}\approx0.\textcolor{black}{8}$~dex) between the narrow- and broad-line AGN.
We therefore see a \textcolor{black}{small} difference in the $\alpha_\mathrm{ox}$ distributions of the subsets, with the narrow-line AGN having, on average, slightly lower $\alpha_\mathrm{ox}$.
Fig.~\ref{fig:aox_subsamp_lumbin} shows that this is a consequence of the narrow-line sources having lower UV luminosities with $\alpha_\mathrm{ox}\propto L_{2500}$.
If the AGN are binned by $L_{2500}$ then it can be seen that the $\alpha_\mathrm{ox}$ values of the narrow-line AGN are systematically \textit{higher} than those of the broad-line AGN across the luminosity bins, and progressively so with increasing $L_{2500}$. 

The differences in luminosities between the RL and RQ populations are less marked than those seen in the narrow- and broad-line subsets.
The RL and RQ AGN have similar UV luminosities, although the RL subset is on average more luminous in X-rays (median $\Delta L_{2\,\mathrm{keV}}\approx0.\textcolor{black}{7}$~dex).
The radio-loud AGN have, on average, lower values of $\alpha_\mathrm{ox}$ than the radio-quiet AGN.
\textcolor{black}{In the lower-left panel of Fig.~\ref{fig:aox} we show the $L_{2500}$-$\alpha_\mathrm{ox}$ relation derived by \cite{Zhu20} from a sample of radio-quiet quasars.  We see that our relation for the RQ subset agrees very well with theirs, and that our RL sources have systematically lower $\alpha_\mathrm{ox}$.} 

\begin{table*}
    \centering
    \caption{Median, standard deviation and linear regression values from $\alpha_\mathrm{ox}$ analysis.
    For the whole sample of AGN with an $\alpha_\mathrm{ox}$ determination (673 AGN) we list the median (M) and standard deviation ($\sigma$) of $\alpha_\mathrm{ox}$, the logarithms of the 2500~\si\angstrom\ UV luminosity and 2~keV X-ray luminosity as well as the slope ($m$) and intercept ($c$) of the best-fitting linear regression of the luminosities with respect to the dependent variable $\alpha_\mathrm{ox}$ and the dispersion ($\sigma$) of the data about the best fit.
    These quantities are also given for the narrow-line (`NL'), broad-line (`BL'), radio-loud (`RL') and radio-quiet (`RQ') subsets of the sample.
    $N$ is the number of AGN in each set.
    Luminosities are in units erg\,s$^{-1}$\,Hz$^{-1}$.}
    \resizebox{\textwidth}{!}{
    \begin{tabular}{l|ccccccccccccc}
    \hline
         Set & $N$ & M$_{\alpha_\mathrm{ox}}$ & $\sigma_{\alpha_\mathrm{ox}}$ & M$_{\log L_{2500}}$ & $\sigma_{\log L_{2500}}$ & M$_{\log L_{2\,\mathrm{keV}}}$ & $\sigma_{\log L_{2\,\mathrm{keV}}}$ & $m(\log L_{2500})$ & $c(\log L_{2500})$ & $\sigma(\log L_{2500})$ & $m(\log L_{2\,\mathrm{keV}})$ & $c(\log L_{2\,\mathrm{keV}})$ & $\sigma(\log L_{2\,\mathrm{keV}})$ \\
    \hline
         All & 6\textcolor{black}{85} & 1.\textcolor{black}{40} & 0.16 & 29.8\textcolor{black}{5} & 0.7\textcolor{black}{7} & 26.2\textcolor{black}{3} & 0.6\textcolor{black}{3} & $0.1\textcolor{black}{17}\pm0.00\textcolor{black}{7}$    & $-2.\textcolor{black}{0}7\pm0.19$ & 0.12 & ~~$0.0\textcolor{black}{18}\pm0.010$    &  $0.\textcolor{black}{94}\pm0.2\textcolor{black}{5}$ & 0.15 \\
         NL  & 9\textcolor{black}{6}  & 1.3\textcolor{black}{7} & 0.17 & 29.0\textcolor{black}{9} & 0.7\textcolor{black}{6} & 25.5\textcolor{black}{1} & 0.5\textcolor{black}{7} &  $0.1\textcolor{black}{47}\pm0.017$   & $-\textcolor{black}{2.89}\pm0.\textcolor{black}{50}$ & 0.12 & ~~$0.0\textcolor{black}{38}\pm0.03\textcolor{black}{1}$    & $\textcolor{black}{0.40}\pm0.\textcolor{black}{79}$ & 0.1\textcolor{black}{7} \\
         BL  & 58\textcolor{black}{9} & 1.\textcolor{black}{40} & 0.16 & \textcolor{black}{30.00} & 0.7\textcolor{black}{0} & 26.3\textcolor{black}{1} & 0.5\textcolor{black}{6} &  $0.1\textcolor{black}{29}\pm0.008$   & $-2.\textcolor{black}{45}\pm0.2\textcolor{black}{3}$ & 0.12 & ~~$0.0\textcolor{black}{04}\pm0.012$    &  $\textcolor{black}{1.31}\pm0.3\textcolor{black}{0}$ & 0.15 \\
         RL  & 6\textcolor{black}{8}  & 1.2\textcolor{black}{9} & 0.16 & 30.1\textcolor{black}{9} & 0.6\textcolor{black}{5} & 26.8\textcolor{black}{1} & 0.60 &  $0.10\textcolor{black}{1}\pm0.0\textcolor{black}{30}$   & $-1.\textcolor{black}{72}\pm0.\textcolor{black}{92}$ & 0.15 & $-0.0\textcolor{black}{74}\pm0.033$   &  $\textcolor{black}{3.28}\pm0.8\textcolor{black}{7}$ & 0.16 \\
         RQ  & 2\textcolor{black}{29} & 1.45 & 0.1\textcolor{black}{6} & \textcolor{black}{29.99} & 0.8\textcolor{black}{5} & 26.1\textcolor{black}{4} & 0.6\textcolor{black}{6} &  $0.1\textcolor{black}{26}\pm0.010$   & $-2.\textcolor{black}{29}\pm0.\textcolor{black}{29}$ & 0.12 & ~~$0.0\textcolor{black}{49}\pm0.01\textcolor{black}{7}$    &  $0.\textcolor{black}{17}\pm0.4\textcolor{black}{3}$ & 0.16 \\
    \hline
    \end{tabular}}
    \label{tab:aox}
\end{table*}

\subsection{$\alpha_\mathrm{ox}$ related to black hole mass and accretion rate}
\label{sec:qsosed}
\begin{figure}
	\includegraphics[width=\columnwidth]{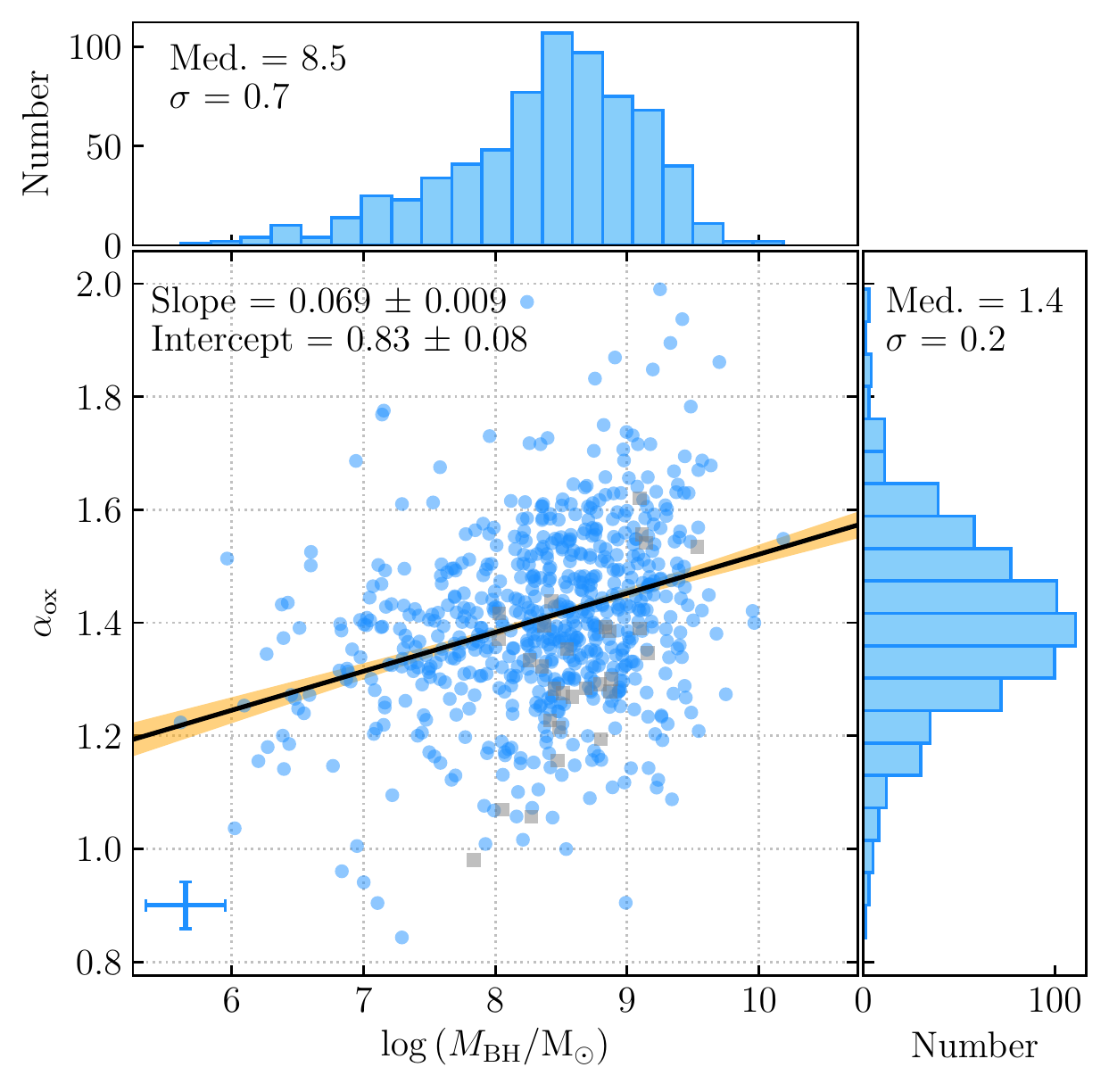}
    \caption{
    The relationship between $\alpha_\mathrm{ox}$ and the logarithm of black hole mass.
    We show the best fit relation found (solid black line) and its one sigma confidence region (in orange).
    Histograms show the distributions of the two quantities.
    A representative error bar is shown in the bottom left.
    }
    \label{fig:aox_mass}
\end{figure}

We also test the correlation between $\alpha_\mathrm{ox}$ and black hole mass. There is substantial scatter but 
\textsc{linmix} determines a significant relation with mass:
\begin{equation}
\alpha_\mathrm{ox} =  (0.\textcolor{black}{83}\pm0.08) + (0.0\textcolor{black}{6}9\pm0.00\textcolor{black}{9})\times \log\left(M_\mathrm{BH}/\mathrm{M}_\odot\right)\textcolor{black}{,}   
\end{equation}
\textcolor{black}{which} is shown in Fig.\ref{fig:aox_mass}.
\textcolor{black}{Whilst the correlation between $\alpha_\mathrm{ox}$ and $M_\mathrm{BH}$ is not particularly strong (the Pearson correlation coefficient is $\approx0.3$), we can exclude the the absence of a correlation with high confidence: the gradient $0.069\pm0.009$ is $>7\sigma$ inconsistent with zero and the Pearson correlation has a \textit{p}-value $\ll0.05$ for a null hypothesis of the distributions being uncorrelated.}

\begin{figure*}
	\includegraphics[width=\textwidth]{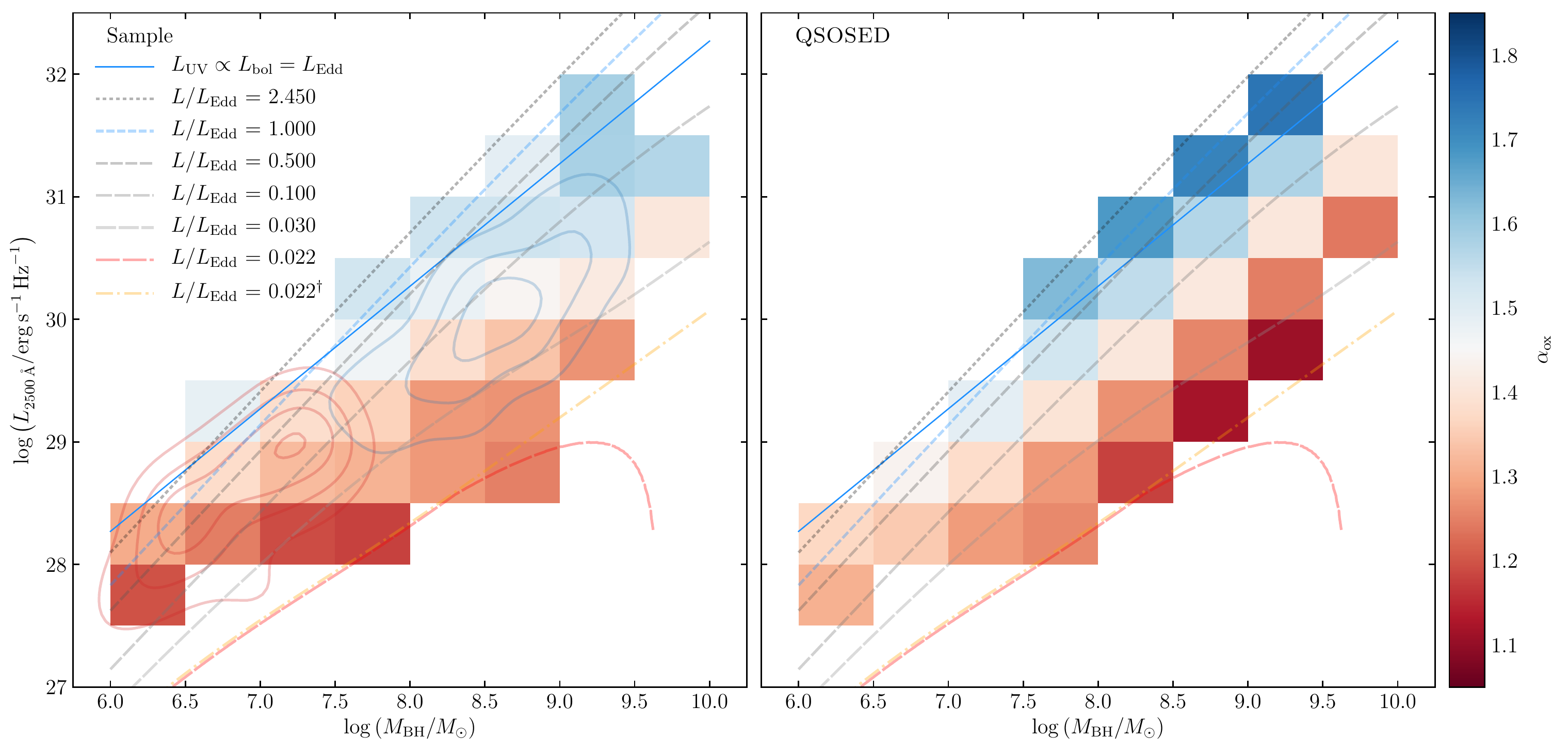}
    \caption{
    The relationship between $L_\mathrm{2500}$, black hole mass and $\alpha_{\mathrm{ox}}$.
    The left panel shows the average values of $\alpha_\mathrm{ox}$ for sources in our sample binned by mass and UV luminosity (only bins containing three or more AGN are drawn).
    The right panel shows the value of $\alpha_{\mathrm{ox}}$ taken from an SED generated by the \protect\cite{Kubota18} model \textsc{qsosed}, for mass and UV luminosity at the centre of each grid point.
    Dashed lines correspond to curves of constant $L/L_{\mathrm{Edd}}$ as determined by \textsc{qsosed}.
    $^\dagger$The dot-dashed \textcolor{black}{orange} line is computed for a model in which the outer disc radius $\log(R_\mathrm{out}/R_\mathrm{g})=5$, whereas in all other models the outer radius is the disc self-gravity radius \textcolor{black}{(cf.\ the red dashed line at the same $L/L_{\mathrm{Edd}}$)}.
    The solid \textcolor{black}{blue} line shows a curve of constant $L/L_\mathrm{Edd}=1$ assuming that $L_\mathrm{2500}\propto L_\mathrm{bol} \propto M_\mathrm{BH}$; clearly this relation is shallower than that predicted by \protect\textsc{qsosed} \textcolor{black}{(the dashed blue line)} in which $L_\mathrm{2500}\propto M_\mathrm{BH}^{4/3}$.
    Blue and red contours in the left-hand panel indicate the parameter space spanned by AGN drawn from the SDSS quasar catalogue and the supplementary sources added from the \protect\citetalias{Rakshit17} NLS1 catalogue, respectively.}
    \label{fig:aox_mass_qsosed}
\end{figure*}

We now investigate how much of the scatter in $\alpha_\mathrm{ox}$-$L_{2500}$
can be explained by a two-parameter dependence 
on both mass and mass accretion rate (or equivalently, mass and $L/L_\mathrm{Edd}$). 
The left panel of Fig.~\ref{fig:aox_mass_qsosed} shows the range of 
$\log L_{2500}$ (a tracer of mass accretion rate) and black hole mass spanned by our sample, binned onto a grid with two squares per decade. Within each grid square, we calculate the average
$\alpha_\mathrm{ox}$, where this should be a more reliable estimate as we are using only AGN with approximately the same mass and mass accretion rate. We remove squares containing fewer than 3 AGN, so 
the dispersion is dominated by the intrinsic scatter. 
This shows a subtle but systematic trend in the data, where 
$\alpha_\mathrm{ox}$ depends on both mass and $L_{2500}$, rather than $L_{2500}$ alone. 

For comparison, the right panel of Fig.~\ref{fig:aox_mass_qsosed} shows the predicted 
 $\alpha_\mathrm{ox}$ over this parameter space from the quasar SED model \textsc{qsosed} (\citealt{Kubota18}).
This model was built from extrapolating trends seen in 
a much smaller sample of AGN than used here (\citealt{Jin1,Jin3}). 
These earlier studies showed that optical-X-ray AGN SEDs are typically fit by three components: an outer standard disc, emitting thermal blackbody down to some radius, where it transitions to incomplete thermalisation (warm Comptonisation). 
This optically thick structure truncates below the radius $R_\mathrm{hot}$ 
so that the gravitational power within some $R_\mathrm{hot}>R_\mathrm{isco}$ (where $R_\mathrm{isco}$ is the radius of the innermost stable circular orbit) is dissipated instead in heating optically thin plasma which produces the power law tail by Comptonisation. 
The \textsc{qsosed} model assumes that the total X-ray luminosity is 
fixed at $L_\mathrm{X}=0.02L_\mathrm{Edd}$, i.e.\ the maximal ADAF limit. This 
captures the main trends seen in the data: at low mass accretion rates, only just above $L_\mathrm{bol}=0.02L_\mathrm{Edd}$, then 
almost all of the accretion power is required to power the X-rays, and $R_\mathrm{hot}$ is large. Conversely, at high mass accretion rates, $L_\mathrm{bol}\approx L_\mathrm{Edd}$
then almost all of the accretion power is dissipated in the disc,
and the X-rays at $L_\mathrm{X}=0.02L_\mathrm{Edd}$ are weak in comparison and 
$R_\mathrm{hot}\to R_\mathrm{isco}$. 
The changing size of $R_\mathrm{hot}$ and changing luminosity of the disc also sets the spectral slope of the power-law tail self consistently, so that the model predicts the entire SED (with some additional assumptions about the warm Compton region). 

We use this model, with spin fixed at 0 and the inclination at $30^\circ$ to generate a set of SEDs from black hole masses in the range $\log(M_\mathrm{BH}/\mathrm{M}_\odot)=5.75$--9.75 and adjust the mass-normalised accretion rate between $\dot{m}=0.022-2.5$ (the limits of the \textsc{qsosed} model) to match 
$\log L_\mathrm{2500}$ to the midpoint of each grid square. We extract $\alpha_\mathrm{ox}$ from these model SEDs, and colour code each grid point using the same colour scale as from the real data. 
We also overplot lines of constant $L/L_\mathrm{Edd}$ on both panels. 
At high Eddington ratio, the UV is always dominated by the standard disc over this mass range, so the monochromatic flux $L_{2500}\propto (M_\mathrm{BH}\dot{M})^{2/3}$, giving rise to the straight line\textcolor{black}{s for high $L/L_\mathrm{Edd}$} on the $\log(M_\mathrm{BH}/\mathrm{M}_\odot)-\log L_{2500}$ plot. Importantly this is not a linear relationship even for a standard disc. $L_{2500}\propto \dot{M}^{2/3}\propto L_\mathrm{bol}^{2/3}$ rather than being directly proportional to $L_\mathrm{bol}$. 
Thus using monochromatic flux (or flux in a narrow band) with a constant K-correction to convert to bolometric flux and hence derive Eddington ratio is subtly incorrect even for standard disc models.
The solid \textcolor{black}{blue} line in Fig.~\ref{fig:aox_mass_qsosed} shows the Eddington luminosity assuming a bolometric correction of $L_\mathrm{bol}=2.75\times L_{2500}$ (\citealt{Krawczyk13}). 

The \textsc{qsosed} model assumes the disc extends outwards only to the self-gravity radius, which may be only a few hundred $R_\mathrm{g}$ at the lowest accretion rates for the highest masses (see Mitchell et al., submitted). This causes the very sharp drop in $L_{2500}$ for these objects.
If instead we set the outer disc radius to $10^5$~$R_\mathrm{g}$ (as is usually done in the near-infrared spectroscopic modelling of AGN, e.g.\ \citealt{Landt11b}), the $0.022L_\mathrm{Edd}$ curve more closely resembles those of higher Eddington ratios (the \textcolor{black}{orange} dot-dashed line in Fig.~\ref{fig:aox_mass_qsosed}).

It is very clear that the data sit between the Eddington limits hardwired into the {\sc qsosed} code. The model assumes that below $L/L_\mathrm{Edd}\sim 0.02$, the flow is completely dominated by the optically thin, geometrically thick, hot plasma (ADAF/RIAF flow). This transition is seen directly in the `changing-look' AGN (\citealt{Noda18}, \citealt{Ruan19}).
Above $L/L_\mathrm{Edd}\sim 2.5$, the disc photosphere becomes brighter than its local Eddington flux (\citealt{Kubota19}), so its structure is probably modified by strong winds and/or radial advection of the excess radiation.
Both these probably lead to vertical structure so the radiation escaping preferentially only in a narrow funnel, making these objects rare (as well as probably short lived). 

\textcolor{black}{In the left-hand panel of Fig.~\ref{fig:aox_mass_qsosed} we see that the contours indicating the location of our data do not extend above the $L/L_\mathrm{Edd}=0.5$ line (taken from the \textsc{qsosed} model for the same mass and UV luminosity) at the high mass end; this implies} that high Eddington fraction flows are rare at the highest black hole masses in our sample. These objects are very rare in the local Universe, and are only likely to be present at the peak of the quasar epoch
at $z\sim 2-4$, beyond our sample redshift limit. 
Conversely, we only see the lowest mass black holes ($M\sim 10^6M_\odot$ ) at the highest Eddington ratios as these are otherwise diluted by their host galaxy (e.g.\ \citealt{Done12}). 

\begin{figure}
	\includegraphics[width=\columnwidth]{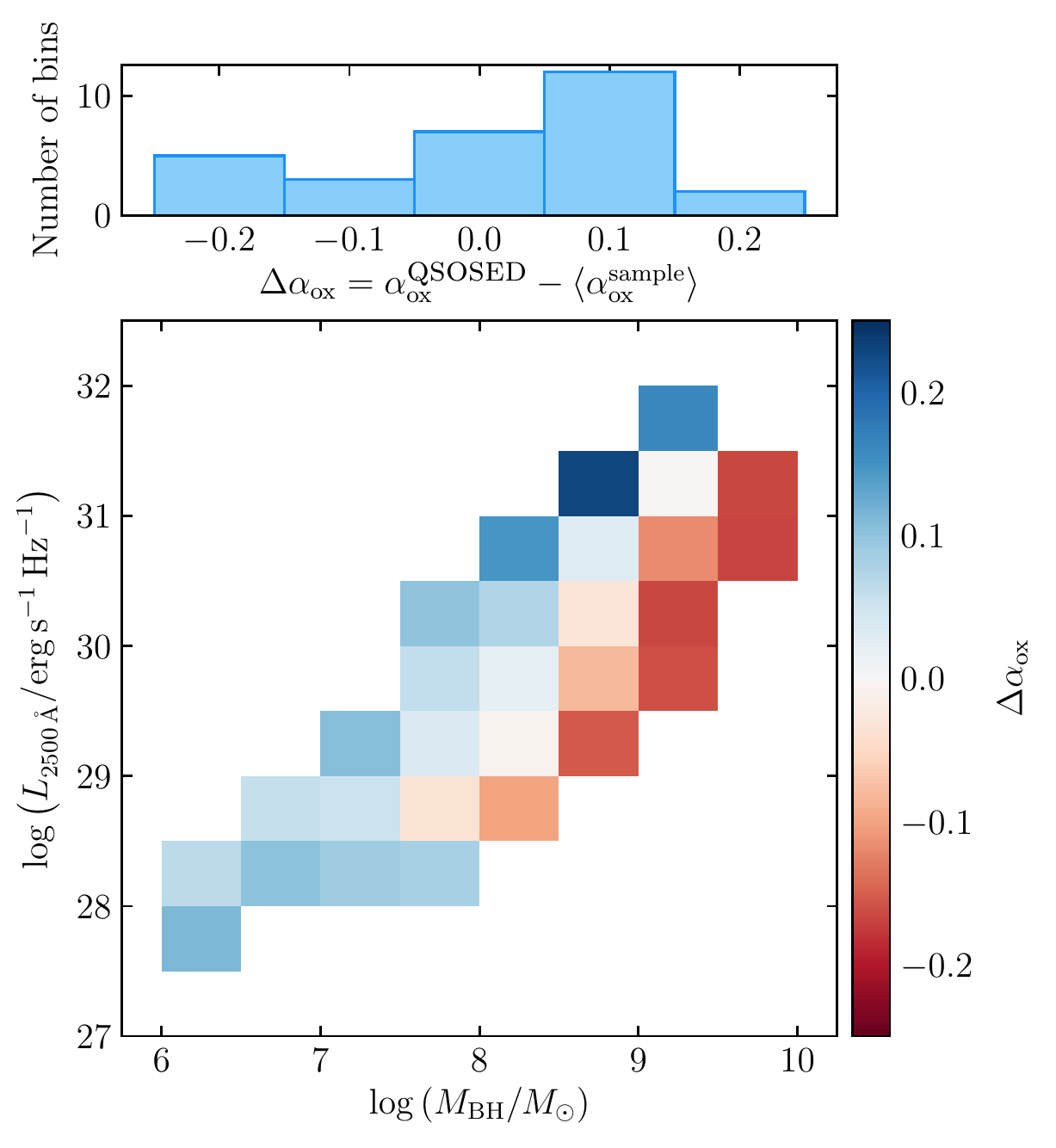}
    \caption{
    The difference between $\alpha_\mathrm{ox}$ as predicted by the \textsc{qsosed} model and as found from the data in our sample.
    The bottom panel shows the difference in each mass-luminosity bin, the top panel shows the distribution of $\Delta\alpha_\mathrm{ox}$ values.
    }
    \label{fig:delta_aox}
\end{figure}

In Fig.~\ref{fig:delta_aox} we show the difference between $\alpha_\mathrm{ox}$ as predicted by \textsc{qsosed} and as determined in our sample.
This shows that there is generally reasonable agreement between the model predictions and $\alpha_\mathrm{ox}$ estimated from our data,
In the majority of bins (97~per~cent) $|\Delta\alpha_\mathrm{ox}|<0.2$ and in \textcolor{black}{52}~per cent of the bins $|\Delta\alpha_\mathrm{ox}|<0.1$.
A detailed exploration of the differences between the data and the model predictions as a function of black hole mass, luminosity and accretion rate will be performed in future work (Mitchell et al., submitted). 

\section{Discussion}
\label{sec:disc}
The SOUX sample is a medium-sized selection of $\approx700$ AGN up to $z=2.5$ with quality multiwavelength data.
Although it is not a complete or representative AGN sample, it has been assembled with the intention of exploring a wide range of parameter space.
This first study has focussed on making key measurements from the data (the broad optical emission line properties, multiwavelength luminosities etc.) 
and on determining a consistent set of black hole masses from which we will be able to determine Eddington ratios.
We have taken a preliminary look at changes in the shape of the AGN SED (characterised by the parameter $\alpha_\mathrm{ox}$) as a function of black hole mass and AGN luminosity.
To first order, $\alpha_\mathrm{ox}$ is found to vary with both mass and luminosity as broadly expected from a theoretical model.
However, the systematic differences seen at second-order will require a deeper exploration.
Our main findings are discussed below.

\subsection{Broad emission line properties}
\label{sec:disc-blr}
\textcolor{black}{In Section~\ref{sec:lines} we found that the luminosities of the broad emission lines \ha\ and \hb\ were very well-correlated, with the mean luminosity ratio (the Balmer decrement) being $\langle \mathrm{H}\,\upalpha / \mathrm{H}\,\upbeta \rangle=3.9$ (Fig.~\ref{fig:balmer_decs}).
The analysed spectra have all been corrected for line-of-sight Galactic reddening, therefore if (as proposed by \citealt{Gaskell17}) the intrinsic broad line ratio is close to the Case B value ($\approx2.7$) then the typically higher values that we find imply that a large fraction of the AGN in our sample are reddened by dust in their host galaxies or galactic nucleus.
In Fig.~\ref{fig:hbd} we show the spectra of two sources with high Balmer decrements.  
One spectrum (6369-56217-0193) exhibits both high broad line and narrow line Balmer decrements and lacks a dominant blue AGN continuum.
(Note however that \textsc{PyQSOFit} performs a principal component analysis using galaxy and AGN eigenspectra to obtain a best estimate of the AGN continuum level.)
This AGN may plausibly be reddened by dust in its host galaxy affecting both the BLR and NLR; 
however the presence of reddening in many other spectra with high Balmer decrements is more ambiguous.
For example, the spectrum 5694-56213-0636 (also shown in Fig.~\ref{fig:hbd}) has a high broad line Balmer decrement but a narrow line decrement consistent with Case B and a very strong, blue continuum.
The modified spectrum, dereddened using the \cite{Gaskell04} AGN curve and an $A_\mathrm{v}$ estimated from the Balmer decrement, is shown in orange in Fig.~\ref{fig:hbd}.  
This spectrum is $\sim1.25$~dex brighter than the spectrum from which we have measured emission line and continuum properties, but it is still quasar-like.
Whilst we cannot definitively rule out the intrinsic reddening of such individual sources, Fig.~\ref{fig:balmer_decs} shows that across our sample Balmer decrements and the ratio of 5100 and 3000~\si\angstrom\ continuum luminosities are uncorrelated: we would expect there to be a correlation in the reddening scenario.  
Alternatively, it is possible that the intrinsic Balmer decrements in AGN BLRs are not always close to Case B.
This was a conclusion of \cite{Schnorr16}, who found that the intrinsic Balmer decrements of type 1 AGN spanned a wide range (2.5--6.6), derived from photoionization modelling of their data; the vast majority of the AGN in our sample fall within this range.
A range of intrinsic broad line Balmer decrements would partly explain why we do not find a correlation between Balmer decrements and continuum luminosity ratios in our sample.}

We showed in Section~\ref{sec:lines} that the luminosities of the three broad emission lines \ha, \hb\ and \mg\ are strongly correlated with the AGN continuum luminosities.  
Of all emission line-continuum relations we tested, the least scatter is found between $L_\mathrm{H\,\upbeta}$ and $L_{5100}$, although the relationship between the two continuum luminosities $L_{5100}$ and $L_{3000\,\si\angstrom}$ had least scatter overall. 
Whilst the luminosities of the two Balmer lines are approximately linearly proportional to $L_{5100}$, the $\log(L_\mathrm{Mg\,\textsc{ii}})$-$\log(L_{5100})$ relation has a gradient noticeably lower than 1 ($m=0.87\pm0.03$).
These results are similar to those reported by \cite{SL12}, who also found least scatter between $L_{5100}$ and $L_{3000\,\si\angstrom}$.
\cite{SL12} report a $\log(L_\mathrm{H\,\upalpha})$-$\log(L_{5100})$ relation with a gradient consistent with 1 and a $\log(L_\mathrm{Mg\,\textsc{ii}})$-$\log(L_{5100})$ proportionality consistent with the one we find here ($0.86\pm0.07$).
However, in their work they find a much steeper relation between $\log(L_\mathrm{H\,\upbeta})$-$\log(L_{5100})$ ($1.25\pm0.07$).
\cite{Paper2} also find a shallower gradient between $\log(L_\mathrm{Mg\,\textsc{ii}})$ and $\log(L_{5100})$ than between the Balmer lines and $\log(L_{5100})$.
In their work, the gradients of all three emission line luminosites with respect to $\log(L_{5100})$ were shallower than 1, although they were working with a much smaller sample (11 AGN) and much more limited range in luminosities ($\sim1$~dex).
It has been shown both in observations of reverberation-mapped and `changing-look' AGN (e.g.\ \citealt{Sun15}; \citealt{Zhu17}; \citealt{Kynoch19}) and photoionisation models (e.g.\ \citealt{Guo20}) that \mg\ responds more weakly to changes in the continuum than does \hb.
Here, we appear to see a similar trend in a statistical sense across several hundred snapshots of individual AGN.  
Overall, the strong correlations between the emission line and continuum luminosities demonstrate that the former are also reasonable proxies for the BLR radius, which we discuss below.

\subsection{Relations for virial black hole masses}
In Section~\ref{sec:mass} we computed the black hole masses of the AGN in our sample from the broad emission line widths and either a nearby continuum luminosity or the broad emission line luminosity.
We used several single-epoch virial black hole mass relations taken from recent literature (\citetalias{Mejia16}; \citealt{Woo18}; \citealt{Greene10}).
Mass estimates from the FWHM of the Balmer lines and $L_{5100}$ are generally in very good agreement, with a minor systematic difference $\vert\Delta{\log(M_\mathrm{BH})}\vert=0.2$ (i.e.\ a 5~per cent difference in $M_\mathrm{BH}$) and $\sigma=0.22$~dex.

We noticed that estimates made using the emission line luminosities were systematically lower than those made using the continuum luminosities.
The masses were $\approx60$~per cent lower in the case of \hb\ and a factor two lower in the case of \ha.
Discrepancies of a similar magnitude have been observed in previous studies calculating the black hole mass of single objects (e.g.\ \citealt{Landt17}; \citealt{Kynoch19b}) and also in large samples (e.g.\ \citealt{SL12}).
\cite{SL12} compared masses calculated with FWHM(\ha)-$L_\mathrm{H\,\upalpha}$ \citep{Shen11} and FWHM(\hb)-$L_\mathrm{H\,\upbeta}$ \citep{GH05} with fiducial masses calculated with FWHM(\hb)-$L_{5100}$ \citep{VP06} and again estimates using the line luminosities were lower, although to a lesser extent than we find here (\citealt{SL12} report $\langle\Delta\log[M_\mathrm{BH}]\rangle=-0.08$ and $-0.04$~dex for \ha\ and \hb, respectively).
The cause of these systematic discrepancies is not apparent.
Since, as we noted above, there is generally very good correspondence between the emission line and continuum luminosities, in principle it should be possible to determine a reliable mass estimate using either luminosity as a proxy for the BLR radius.
It is not simply the case that we have taken scaling relations from different papers in which the relations were derived for different samples of AGN: both relations involving \ha\ are taken from \citetalias{Mejia16}.
Such large systematic difference in mass are difficult to explain since by Eqn.~(\ref{eqn:virial}) $M_\mathrm{BH}\propto L^{1/2}$, a factor four difference in luminosity is required to reconcile a factor two difference in mass.
It is very unlikely that the 5100~\si\angstrom\ luminosity is contaminated by stellar light from the AGN host galaxies, resulting in masses determined using $L_{5100}$ to be overestimates.
\textsc{PyQSOFit} performs a subtraction of the host galaxy spectrum where necessary, so in principle the stellar contamination has been removed from our measure\textcolor{black}{ment}s.
Also, we see a systematic offset even for the highest-mass sources: highly luminous quasars with optical luminosities far in excess of their host galaxies.
It is possible that the procedure we used to fit the broad emission lines to determine their width and luminosity could differ from those used to determine the $M_\mathrm{BH}$ relations (\citetalias{Mejia16} used only two Gaussians for the Balmer lines, for example). 
As described in Section~\ref{sec:pyqsofit}, in this work the broad emission lines were fit with three Gaussians.
In many cases, the total line profile was decomposed into broad, very broad an\textcolor{black}{d} intermediate Gaussians. 
The shallow but very broad component (of \ha\ in particular) which was often present can add a substantial amount of flux to the line without greatly increasing the FWHM.
However, this would have the effect of increasing the emission line luminosities, whereas what we find is that the Balmer lines are apparently \textit{underluminous} compared to the continuum for a given FWHM, resulting in the lower mass estimates.
Differences in the assumed virial factor $f$ will also result in equivalently larger or smaller estimates of the mass.
However, the relations we test here use $f\approx1$ so the mass differences ought to be very minor.

To avoid problems introduced by systematic differences in black hole masses, we have calculated new mass scaling relations tailored to our sample.
As described in Section~\ref{sec:mass_measurements} these minimise the discrepancy with respect to estimates made from FWHM(\hb) and $L_{5100}$ (which we judge is the most reliable estimate).
The new scaling relations are presented in Table~\ref{tab:new_mass_rel}.
We note here that all of our relations have a slope $\alpha$ consistent with 0.5, the expected value for an inverse-square scaling of radius with luminosity.
The majority of the AGN in our sample (52~per cent) have their adopted masses derived using \textcolor{black}{the FWHM(\hb)-$L_{5100}$} relation from \citetalias{Mejia16}.
45~per cent of the sample have their preferred mass estimate taken from our FWHM(\mg) and $L_{3000\,\si\angstrom}$ relation; these were mostly the high-$z$ AGN in our sample, lacking optical spectral coverage of \hb. 
Our newly-derived FWHM(\mg)-$L_{3000\,\si\angstrom}$ mass relation has a very minor systematic offset with respect to the fiducial masses calculated via the FWHM(\hb)-$L_{5100}$ relation of \citetalias{Mejia16}: the median difference in mass is negligible, although there is substantial scatter ($\sigma=0.26$~dex).
Following these recalibrations, we do not expect that using different mass relations for different AGN will have an adverse impact on our results. 

\subsection{$\alpha_\mathrm{ox}$}\label{sec:disc-aox}
In Section~\ref{sec:aox} we explored the relationship between the UV-X-ray energy index $\alpha_\mathrm{ox}$ with the UV and X-ray luminosities and with black hole masses and accretion rates.
\textcolor{black}{As mentioned in Section~\ref{sec:uv_lum}, we have used a mixture of spectroscopic and photometric measurements of $L_{2500}$ (from which we derive $\alpha_\mathrm{ox}$) and there are subtle differences between the two luminosity estimates.  
In Appendix~\ref{sec:mix_lum}, we have performed a detailed investigation of the potential impact of mixing different luminosity measures on our results, in particular on the $\log(L_{2500})$-$\alpha_\mathrm{ox}$ correlation.
To summarise, the average difference between spectroscopic and photometric luminosities is small ($\langle\Delta\log(L_{2500})\rangle=-0.02$~dex) 
and we derive very similar distributions of $\alpha_\mathrm{ox}$ from both the spectroscopic and photometric estimates (the $\alpha_\mathrm{ox}$ distributions are statistically indistinguishable). 
We have recalculated $\log(L_{2500})$-$\alpha_\mathrm{ox}$ relations for the spectroscopically- and photometrically-derived quantities separately.
Each subset shows a statistically significant positive correlation between $\log(L_{2500})$ and $\alpha_\mathrm{ox}$, with similar dispersions ($\approx0.1$~dex) so it does not appear that photometric measurements of $L_{2500}$ are substantially less reliable than the spectroscopic ones.
The photometric $\log(L_{2500})$-$\alpha_\mathrm{ox}$ relation differs most from the spectroscopic one at the highest luminosities (Fig.~\ref{fig:mix_lum_aox}).
However, this difference will have little effect on our analysis in Section~\ref{sec:aox}.  
The high-luminosity sources are mostly at high redshift ($z>0.5$) and for these we have adopted $\log(L_{2500})$ from spectroscopy rather than photometry.  
Very few sources lie in the region of parameter space in which the photometric $\log(L_{2500})$-$\alpha_\mathrm{ox}$ relation is substantially different from the spectroscopic one \textit{and} for which $\log(L_{2500})$ and $\alpha_\mathrm{ox}$ were estimated from photometry.
We conclude that the mixing of spectroscopically- and photometrically-derived UV luminosities does not greatly affect our results and note that several of our relations are in good agreement with those of previous studies.}

\textcolor{black}{Our analysis of the final sample in Section~\ref{sec:aox} demonstrates} a highly significant correlation of $\alpha_\mathrm{ox}$ with the monochromatic UV luminosity $\log(L_{2500})$ for all AGN as well as the four subsets (narrow- and broad-line; radio loud and quiet) taken separately.
No highly significant $\log(L_\mathrm{2\,keV})$--$\alpha_\mathrm{ox}$ correlation was found for any individual subset, or for the full sample, although \textcolor{black}{a marginally significant correlation ($>3\sigma$) was found for the RQ subset only (see Table~\ref{tab:aox} and Fig.~\ref{fig:aox}).  I}n a previous study of quasars, \cite{Lusso10} found no significant relation between $\alpha_\mathrm{ox}$ and $\log(L_\mathrm{2\,keV})$\textcolor{black}{, consistent with our results from a sample containing both quasars and NLS1s}.

\textcolor{black}{Figs.~\ref{fig:aox} and \ref{fig:aox_subsamp_lumbin} illustrate differences in the $\alpha_\mathrm{ox}$ distributions, and the relationship between $\alpha_\mathrm{ox}$ and UV luminosity, between different subsets of our sample.
Whilst \textit{on average} narrow-line AGN have lower $\alpha_\mathrm{ox}$ than broad-line AGN they}
have slightly higher $\alpha_\mathrm{ox}$ \textcolor{black}{when} compared with broad-line AGN in the same $L_{2500}$ bin, likely as a result of the stronger soft X-ray excess (warm corona) component in the former.
\textcolor{black}{Figs.~\ref{fig:aox} and \ref{fig:aox_subsamp_lumbin} show a subtle difference in the dependence of $\alpha_\mathrm{ox}$ on the UV luminosity between the NL and BL subsets, with the best fitting $\alpha_\mathrm{ox}$-$L_{2500}$ relation being slightly steeper for NL AGN.  Starker differnces are apparent}
between our radio-loud and radio-quiet AGN: the radio-loud AGN are on average more X-ray luminous than the radio-quiet sources although both sets have similar UV luminosities (Fig.~\ref{fig:aox}). 
\textcolor{black}{We can see in Fig.~\ref{fig:aox} that} radio-loud AGN are systematically offset to lower $\alpha_\mathrm{ox}$ with respect to the radio-quiet sources, and very few radio-loud AGN are found to have higher $\alpha_\mathrm{ox}$ than the average for radio-quiet AGN at a given $\log{L_{2500}}$. 
Even when binning by $L_{2500}$, radio-loud AGN have systematically lower $\alpha_\mathrm{ox}$ (harder SEDs) than the radio-quiet AGN.
\textcolor{black}{This trend has been noted in previous studies.
\cite{Lawther17} found that radio-loud objects in their sample of $z\approx2$ AGN were offset from the best-fitting $L_{2500}$-$\alpha_\mathrm{ox}$ relation determined by \cite{Strateva05} for radio-quiet quasars.
More recently, \cite{Zhu20} found that their large sample of radio-loud quasars were offset from a comparison radio-quiet sample, also in the same sense that we find here.
In this work we show that the same trend extends to lower luminosities and redshifts than probed by \cite{Lawther17}, consistent with the findings of \cite{Zhu20}.}
Often this \textcolor{black}{systematic offset between RL and RQ AGN} is attributed to the radio jet contributing to the hard X-ray band, resulting in an increased X-ray flux and lower UV/X-ray ratio.
However, \cite{Zhu20} argue that for steep radio spectrum sources the excess X-ray luminosity of radio-loud AGN is likely \textit{not} due to contamination of the X-ray emission by the jet, but rather some evolution of the disc-corona configuration.
\cite{Zhu20} proposed that the best-fitting relation between $\alpha_\mathrm{ox}$ and $\log{L_{2500}}$ for radio-quiet quasars \textcolor{black}{(which we show in Fig.~\ref{fig:aox})} represented an approximate `jet line' for AGN, in analogy to the jet line of accreting stellar-mass black holes (e.g.\ \citealt{Fender04}): jets are quenched for AGN on the soft-SED side of the line and the most powerful jets are found in sources with the hardest SEDs for a given $L_{2500}$.
The best-fitting $\alpha_\mathrm{ox}$-$\log{L_{2500}}$ relation for radio-quiet quasars found by \citealt{Zhu20} is entirely consistent with our relation for radio-quiet AGN (Table~\ref{tab:aox} and Fig.~\ref{fig:aox}).

In Section~\ref{sec:qsosed} we calculated the mean $\alpha_\mathrm{ox}$ for 3\textcolor{black}{1} subsets of AGN, binned in both $\log(M_\mathrm{BH})$ and $\log(L_{2500})$, ranging over four orders of magnitude in each quantity (Fig.~\ref{fig:aox_mass_qsosed}).
Fig.~\ref{fig:aox_mass_qsosed} illustrates the diversity of our sample - blue and red contours show the location in parameter space of AGN drawn from the quasar and NLS1 catalogues, respectively.
The inclusion of AGN from the NLS1 catalogue extends our sample to, on average, lower black hole masses and higher accretion rates than probed by the quasars.
This wide coverage of parameter space allows us to track changes in the spectral shape (characterised by $\alpha_\mathrm{ox}$) with both mass and accretion rate.
We compared the mean $\alpha_\mathrm{ox}$ for each bin to the prediction made from the AGN SED model \textsc{qsosed}. 
The correspondence between the model predictions and our sample data is remarkably good: nearly two thirds of bins have $\vert\Delta\alpha_\mathrm{ox}\vert<0.1$ and in all but one bin $\vert\Delta\alpha_\mathrm{ox}\vert<0.2$.
Fig~\ref{fig:delta_aox} shows that the largest differences between the data and the model are at the extremes of the parameter space: for high-mass/high-luminosity AGN, the model predicts SEDs which are too red and for low-mass/low-luminosity AGN the predictions are too blue.
Of course, if there is a systematic error in the calculation of the black hole masses then the \textsc{qsosed} predictions are not being compared with the appropriate AGN (however, as described previously, we have made efforts to ensure our mass estimates are reliable).
Alternatively, whilst \textsc{qsosed} makes good predictions for moderate masses and luminosities, it is possible that the actual disc-corona coupling is more complex than assumed by the model.
We will explore these issues in more detail by modelling the full broadband SEDs of our sample (Mitchell et al., submitted). 

\section{Summary and conclusions}
\label{sec:conc}
We have assembled a sample of 696 AGN with SDSS optical spectra which also have X-ray spectra and simultaneous UV photometry recorded by \xmm.
Although this is not a complete or representative sample of AGN, the objects in our sample cover a wide range in black hole mass, accretion rate, spectral hardness and redshift.
The majority of sources in our sample are luminous AGN found in the SDSS-DR14 quasar catalog; 
these are supplemented \textcolor{black}{with a} selection of lower-luminosity, narrow-line type 1 AGN in the local Universe (taken from \citetalias{Rakshit17}).

We have performed new optical spectral fits so that we have comparable measurements of the AGN continuum and at least one broad emission line (\ha, \hb, or \mg) for the narrow-line AGN (and broad-line AGN that lacked a measurement of the \mg\ spectral region). 
We computed single-epoch virial black hole masses for the AGN in our sample using these broad emission lines with both the local continuum luminosity and the emission line luminosity.
However, we found some large systematic differences between mass estimated made using different relations (masses calculated with FWHM[\ha] and $L_\mathrm{H\,\upalpha}$ are on average a factor two lower than those made using $L_{5100}$).
We have calculated new virial black hole mass relations which minimise the systematic offsets with respect to fiducial black hole masses calculated using FWHM(\hb) and $L_{5100}$. 

From the X-ray catalogue fluxes, we have estimated rest-frame 2~keV luminosities and X-ray photon indices for the full sample.
Combining these with ou\textcolor{black}{r} UV measurements, we have calculated the energy index $\alpha_\mathrm{ox}$, an indicator of the broadband SED shape.
We found that narrow-line sources have steeper X-ray spectra (higher $\Gamma$) and, when controlling for $L_\mathrm{UV}$, have softer broad-band SEDs (higher $\alpha_\mathrm{ox}$) than typical broad-line AGN.
Differences were also found with respect to radio properties with radio-loud AGN being on average more X-ray luminous and having harder SEDs (lower $\alpha_\mathrm{ox}$) than the radio-quiet subset.
Fig.~\ref{fig:aox_mass_qsosed} illustrates that the majority of the AGN in our sample lie between $0.02<L/L_\mathrm{Edd}<2$, with narrow-line type 1 offset to lower masses and higher accretion rates compared with the broad-line quasars.
We have shown that\textcolor{black}{, to first order,} the model \textsc{qsosed} makes very good predictions of $\alpha_\mathrm{ox}$ for a given black hole mass and \textcolor{black}{UV luminosity (assumed to be a proxy for the} accretion rate\textcolor{black}{)}.
\textcolor{black}{Importantly h}owever, there are systematic differences between the data and model predictions at the extremes of the mass-accretion rate parameter space.
In a subsequent paper (Mitchell et al., submitted) we will model the broad-band SEDs, testing the physical assumptions made in the \textsc{qsosed} model and further explore the relationships between the UV/optical and X-ray spectra.

\section*{Acknowledgements}
DK acknowledges support from the Czech Science Foundation project No. 19-05599Y, funding from the Czech Academy of Sciences, and the receipt of a UK Science and Technology Facilities Council (STFC) studentship ST/N50404X/1.
JAJM acknowledges the support of the STFC studentship ST/S505365/1.  
CD and MJW acknowledge support from STFC grant ST/T000244/1. 
MJW acknowledges an Emeritus Fellowship award from the Leverhulme Trust.
HL acknowledges a Daphne Jackson Fellowship sponsored by the STFC.
 
We thank the anonymous referee for their careful reading of the original manuscript and their comments and suggestions that improved the quality of this paper. Collaboration on this project was partly facilitated by meetings of team 481 held at the International Space Science Institute (ISSI), Bern, Switzerland, and we thank ISSI for their hospitality.  Many thanks also go to Suv Rakshit, Vicky Fawcett and Hengxiao Guo for their help and advice on using \textsc{PyQSOFit}.

Funding for the Sloan Digital Sky Survey IV has been provided by the Alfred P. Sloan Foundation, the U.S. Department of Energy Office of Science, and the Participating Institutions. SDSS-IV acknowledges
support and resources from the Center for High-Performance Computing at
the University of Utah. The SDSS web site is \url{www.sdss.org}.

SDSS-IV is managed by the Astrophysical Research Consortium for the 
Participating Institutions of the SDSS Collaboration including the 
Brazilian Participation Group, the Carnegie Institution for Science, 
Carnegie Mellon University, the Chilean Participation Group, the French Participation Group, Harvard-Smithsonian Center for Astrophysics, 
Instituto de Astrof\'isica de Canarias, The Johns Hopkins University, 
Kavli Institute for the Physics and Mathematics of the Universe (IPMU) / 
University of Tokyo, the Korean Participation Group, Lawrence Berkeley National Laboratory, 
Leibniz Institut f\"ur Astrophysik Potsdam (AIP),  
Max-Planck-Institut f\"ur Astronomie (MPIA Heidelberg), 
Max-Planck-Institut f\"ur Astrophysik (MPA Garching), 
Max-Planck-Institut f\"ur Extraterrestrische Physik (MPE), 
National Astronomical Observatories of China, New Mexico State University, 
New York University, University of Notre Dame, 
Observat\'ario Nacional / MCTI, The Ohio State University, 
Pennsylvania State University, Shanghai Astronomical Observatory, 
United Kingdom Participation Group,
Universidad Nacional Aut\'onoma de M\'exico, University of Arizona, 
University of Colorado Boulder, University of Oxford, University of Portsmouth, 
University of Utah, University of Virginia, University of Washington, University of Wisconsin, 
Vanderbilt University, and Yale University.

This research has made use of data obtained from the 4XMM \xmm\ serendipitous source catalogue compiled by the 10 institutes of the \xmm\ Survey Science Centre selected by ESA.

This research made use of Astropy,\footnote{\url{http://www.astropy.org}} a community-developed core Python package for Astronomy \citep{astropy:2013, astropy:2018}.
Some of the results in this paper have been derived using the HEALPix\footnote{\url{https://healpix.sourceforge.io}.} (\citealt{Gorski05}) package.

\section*{data availability}
The data underlying this work are available from the catalogues and archives described in Section~\ref{sec:selection}.
The details of the AGN in the SOUX sample, relevant measures and derived quantities are presented in the table available online \textcolor{black}{which is described in Appendix~\ref{sec:table}}.




\bibliographystyle{mnras}
\bibliography{refs} 


\appendix
\section{UV and optical luminosities}
\label{sec:mix_lum}
In Section~\ref{sec:uv_lum} we estimated the UV luminosities at rest-frame 2500~\si\angstrom\ from either the optical spectrum of a source, or its \xmm\ OM or SDSS photometry and in Section~\ref{sec:aox} we used these to calculate $\alpha_\mathrm{ox}$.
Similarly, in Section~\ref{sec:uv_lum} we estimated the optical luminosities at rest-frame 4400~\si\angstrom\ from either the optical spectrum of a source or its SDSS photometry and used these to calculate the radio loudness parameter $R$ in Section~\ref{sec:radio}.
This mixture of luminosity measures from different sources is not ideal and raises the following issues:
\begin{itemize}
    \item Variability: only the OM photometry was recorded simultaneously with the X-ray data but the SDSS spectra and photometry were not. 
    \item Aperture effects: the \textit{XMM} OM photometric aperture (12--35$^{\prime\prime}$ diameter) is larger than the SDSS spectral fiber diameter (3$^{\prime\prime}$ for the SDSS spectrograph and 2$^{\prime\prime}$ for the BOSS spectrograph).
    \item Flux sampling: the luminosities are arguably best determined from the spectra, from which the flux is measured in a narrow (100~\si\angstrom\ wide) window centred on the relevant rest-frame wavelength.
    In contrast the OM and SDSS photometric bands sample flux from a larger wavelength window: the effective widths of the bands are several hundred \AA\ and therefore they possibly contain some broad line emission.  Whilst we took photometric fluxes from a band containing either 2500~\si\angstrom\ or 4400~\si\angstrom, the bands will not generally be centred on these wavelengths.
    The Galactic dust dereddening we have applied to the fluxes will be more accurate for the spectroscopic measures than the photometric ones (although Galactic extinction is very low in the SDSS survey area, and the effect of reddening at 4400~\si\angstrom\ is relatively minor anyway).
\end{itemize}
Here we assess the possible scale of the discrepancies that may be introduced, and their effect on $\alpha_\mathrm{ox}$ and $R$ across the sample.

\begin{figure*}
    \centering
    \begin{tabular}{c}
         \includegraphics[width=2\columnwidth]{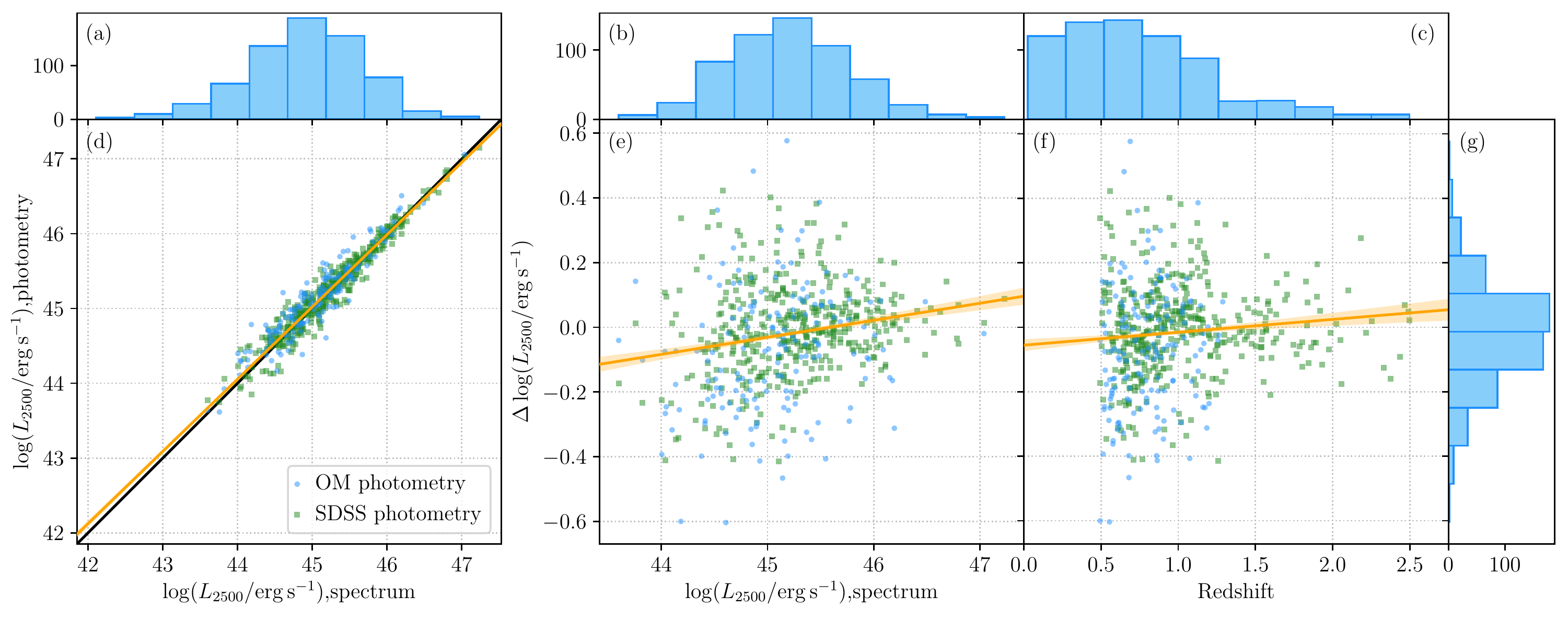}  \\
         \includegraphics[width=2\columnwidth]{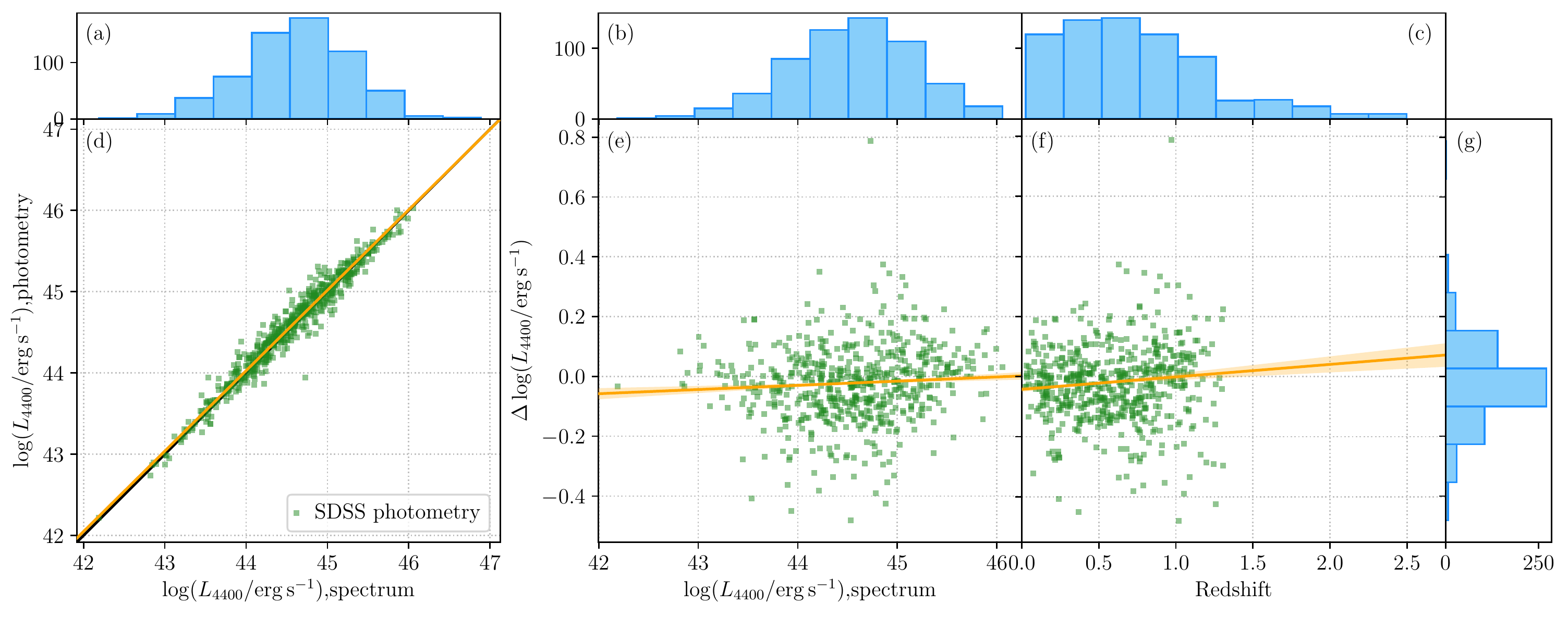}  \\
    \end{tabular}
    \caption{A comparison of spectroscopic and photometric measures of the luminosities at 2500~\si\angstrom\ (top) and 4400~\si\angstrom\ (bottom).  Panels (d) show a scatter plot of the spectroscopic and photometric luminosities; the black line is the one-to-one relation and the orange line in panels (d), (e) and (f) shows the best-fitting linear regression, with the 1$\sigma$ confidence regions shaded orange.  Panels (e) and (f) are scatter plots of the luminosity difference $\Delta\log(L) = \log(L^\mathrm{spec.})-\log(L^\mathrm{phot.})$ as a function of $\log(L^\mathrm{spec.})$ and redshift, respectively.
    Histograms of $\Delta\log(L)$ are shown in panels (g).
    Panels (a) show the distributions of $\log(L)$ for all possible sources.  Panels (b) show the distributions of the spectroscopic measures of $\log(L)$ for sources plotted in panels (e). Panels (c) show the redshift distribution of all 696 AGN.
    Note that we used both \textit{XMM} OM and SDSS photometry to determine $L_{2500}$ but only SDSS photometry to determine $L_{4400}$.}
    \label{fig:mix_lum}
\end{figure*}

For $L_{2500}$ 162 sources have a measurement derived from both the SDSS spectrum and \textit{XMM} OM photometry; 414 sources have a measurement derived from both the spectrum and SDSS photometry.  
For $L_{4400}$ 588 sources have a measurement derived from both the SDSS spectrum and SDSS photometry (note that we did not use any \textit{XMM} OM photometry to determine $L_{4400}$).
In Fig~\ref{fig:mix_lum} we compare the spectroscopic and photometric measures of the luminosities at 2500 and 4400~\si\angstrom; we have used \textsc{linmix} to determine the best-fitting linear regressions and present the values in Table~\ref{tab:uv_lum_check}.
Panels (d) in Fig~\ref{fig:mix_lum} show that the relationship between the spectroscopic and photometric luminosities is approximately one-to-one across several orders of magnitude in luminosity, however there is substantial scatter ($\sigma=0.14$~dex for $L_{2500}$ and 0.09~dex for $L_{4400}$).
We have also calculated the luminosity differences $\Delta\log(L) = \log(L^\mathrm{spec.})-\log(L^\mathrm{phot.})$ and do not find, on average, strong evidence for a systematic offset between the spectroscopic and photometric luminosities with $\langle\Delta\log(L_{2500})\rangle=-0.02\pm0.16$ and $\langle\Delta\log(L_{4400})\rangle=-0.03\pm0.13$.
This is clear from panels (g) of Fig.~\ref{fig:mix_lum} which show the distributions of $\Delta\log(L)$.
Panels (e) show $\Delta\log(L)$ as a function of $\log(L^\mathrm{spec.})$; in these panels we see a weak tendency for photometric measurements to be higher than spectroscopic ones at low spectroscopic luminosities and vice versa.
Both $\Delta\log(L_{2500})$ and $\Delta\log(L_{4400})$ are weakly but significantly correlated with their corresponding spectroscoptic luminosities\footnote{The Pearson correlation coefficient of $\Delta\log(L_{2500})$ with $\log\left(L_{2500}^\mathrm{spec.}\right)$ is 0.21 with a $p$-value $\ll0.05$; for the correlation of $\Delta\log(L_{4400})$ with $\log\left(L_{4400}^\mathrm{spec.}\right)$ we determine a coefficient 0.13 and $p$-value 0.002.}.

As panels (f) in Fig.~\ref{fig:mix_lum} show, we can directly compare spectroscopic and photometric measures of $L_{4400}$, but not $L_{2500}$, for AGN at $z\lesssim0.5$.
We would expect to see greater discrepancies at 4400~\si\angstrom\ for these low redshifts because (i) aperture effects will be greatest for AGN in resolved host galaxies and (ii) galaxies will be relatively brighter in the optical than the UV.
Panel (f) in the $L_{4400}$ plot shows that there is no significant correlation between $\Delta\log(L_{4400})$ and $z$: the Pearson correlation coefficient is 0.05 with $p$-value 0.20.
This provides some reassurance that the photometric UV fluxes will not be strongly enhanced either.
The extremes of the discrepancies along the best-fitting linear regressions are $\Delta\log(L_{2500})\lesssim\pm0.1$~dex and $\Delta\log(L_{4400})\lesssim\pm0.05$~dex which are small when compared with the scatter between the luminosities.
We therefore judge that any systematic differences between spectroscopic and photometric luminosities have a negligible effect on our results. 
\begin{table}
    \caption{Linear regression parameters for the comparisons of $L_{2500}$ and $L_{4400}$ measurements from spectra and photometry}
    \centering
    \resizebox{\columnwidth}{!}{
    \begin{tabular}{llccc}
    \hline
    Indep.\ var.\ & Dep.\ var.\ & Slope & Intercept & $\sigma$ \\
    \hline
    $\log\left(L_{2500}^\mathrm{spec.}\right)$ & $\log\left(L_{2500}^\mathrm{phot.}\right)$ & $0.966\pm0.012$ & ~$1.55\pm0.54$   & 0.14 \\
    $\log\left(L_{2500}^\mathrm{spec.}\right)$ & $\Delta\log(L_{2500})$                     & $0.053\pm0.012$ & $-2.40\pm0.53$   & 0.14 \\
    Redshift                                   & $\Delta\log(L_{2500})$                     & $0.040\pm0.018$ & $-1.06\pm0.02$   & 0.14 \\
    
    $\log\left(L_{4400}^\mathrm{spec.}\right)$ & $\log\left(L_{4400}^\mathrm{phot.}\right)$ & $0.990\pm0.007$ & ~$0.47\pm0.33$   & 0.09 \\
    $\log\left(L_{4400}^\mathrm{spec.}\right)$ & $\Delta\log(L_{4400})$                     & $0.014\pm0.007$ & $-0.65\pm0.33$   & 0.09 \\
    Redshift                                   & $\Delta\log(L_{4400})$                     & $0.042\pm0.017$ & $-0.042\pm0.009$ & 0.09 \\
    \hline
    \end{tabular}}
    \parbox[]{8cm}{We define $\Delta\log(L_{2500})=\log\left(L_{2500}^\mathrm{spec.}\right)-\log\left(L_{4400}^\mathrm{phot.}\right)$.}
    \label{tab:uv_lum_check}
\end{table}

We examined the impact on the radio loudness categorisation by determining the number of sources which would change category if their photometric $L_{4400}$ were `wrong' on the scale of the scatter between spectroscopic and photometric measurements.
Only 35 AGN with radio information (6~per cent) have $L_{4400}$ derived from photometry, compared with 593 which have $L_{4400}$ derived from their optical spectrum.
We have checked these 35 sources and only three have $\vert [\log(L_\mathrm{5GHz}) - 1] - \log(L_{4400}) \vert > 0.09$~dex (where 0.09~dex is the scatter between measurements determined above) meaning their radio loudness classification would change if the optical luminosities change by $\pm0.09$~dex.
All three were undetected in FIRST, so the radio loudness classifications are only estimates based on the FIRST flux limit (as described in Section~\ref{sec:radio}).
One source, which we have classified as radio quiet, would have undetermined radio loudness if it were 0.09~dex optically fainter.
The other two have undetermined radio loudness but would be classified as radio quiet if they were 0.09~dex optically brighter.
A change in the radio loudness category of three AGN out of 611 will not have a noticeable effect on our results.

\begin{figure}
    \centering
    \includegraphics[width=\columnwidth]{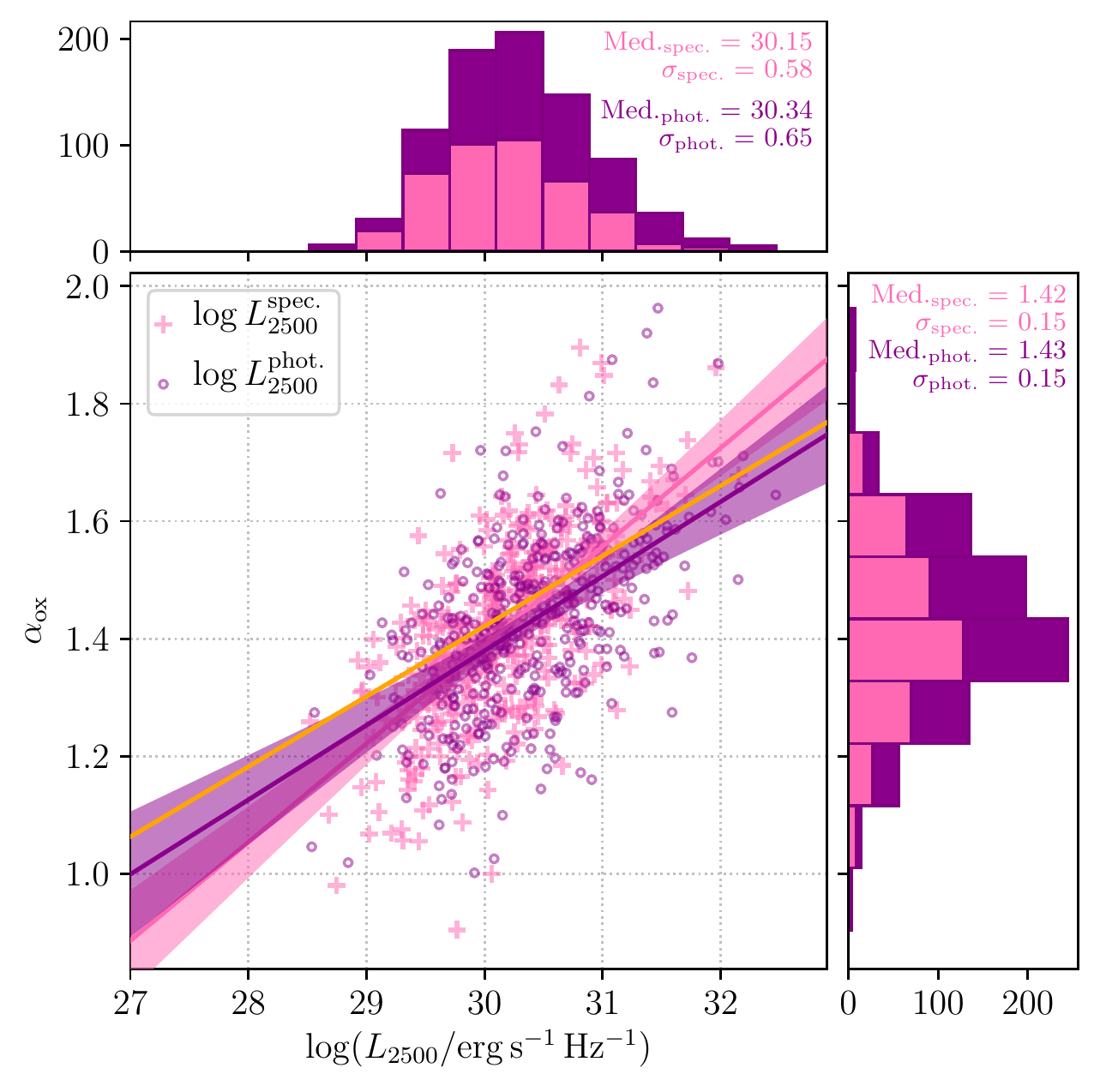}
    \caption{Relationships of $\alpha_\mathrm{ox}$ and $\log(L_{2500})$ for $\log(L_{2500})$ derived from the optical spectra (pink pluses) and from photometry (purple circles).
    All 416 sources shown have a measure of $\log(L_{2500})$ from \textit{both} the spectrum and photometry, so they are cover the same range in redshift.
    The pink and purple solid lines show the best-fitting linear regressions for $\log\left(L_{2500}^\mathrm{spec.}\right)$-$\alpha_\mathrm{ox}$ and $\alpha_\mathrm{ox}$-$\log\left(L_{2500}^\mathrm{phot.}\right)$, respectively with 3$\sigma$ uncertainty regions shaded in the same colours.
    The orange line shows the best-fitting linear regression for the full sample (as in Fig.~\ref{fig:aox}).
    The top and right-hand panels show the distributions of the two quantities for each subset, and we quote the median value and standard deviation.}
    \label{fig:mix_lum_aox}
\end{figure}

There was a greater mixture of $L_{2500}$ measurements: as stated in Section~\ref{sec:uv_lum}, for 421 AGN we have used $L_{2500}$ from the optical spectrum and for 264 AGN (39~per cent) we have used photometry.
We therefore test for possible discrepancies in $\alpha_\mathrm{ox}$ that may result from this mixture.
Optical sectroscopic coverage of rest-frame 2500~\si\angstrom\ is only possible for sources at $z\gtrsim0.5$, whereas the wide wavelength range spanned by the UV-optical-infrared photometric filters mean that a $L_{2500}$ determination is possible across the full range of redshifts covered by our sample.  
To avoid biases introduced by selecting different subsets of AGN in our parameter space, we test only the set of 416 AGN which have \textit{both} a spectroscopic and photometric value of $L_{2500}$.
For each of these AGN we have two values of $L_{2500}$ and two corresponding values of $\alpha_\mathrm{ox}$; these are shown in Fig.~\ref{fig:mix_lum_aox}.
The plot shows that the distributions of $L_{2500}$ and $\alpha_\mathrm{ox}$ are slightly different for the spectroscopic and photometric subsets.
The median $\log(L_{2500}/\mathrm{erg\,s^{-1}\,Hz^{-1}})=30.15$ for the spectroscopic measurements and is slightly higher for the photometric measurements with a median of 30.34.
KS and Anderson-Darling tests both indicate that the $\log(L_{2500})$ measurements are drawn from different distributions (the $p$-values of the tests are both $<0.001$).
However, whilst the $\alpha_\mathrm{ox}$ distributions are not identical, the same statistical tests cannot rule out that the spectroscopically- and photometrically-derived quantities are drawn from the same distribution (the $p$-values are 0.44 and 0.25, respectively).
So we see consistency in the derived $\alpha_\mathrm{ox}$ values despite a slight difference in $\log(L_{2500})$.

\begin{figure}
    \centering
    \includegraphics[width=\columnwidth]{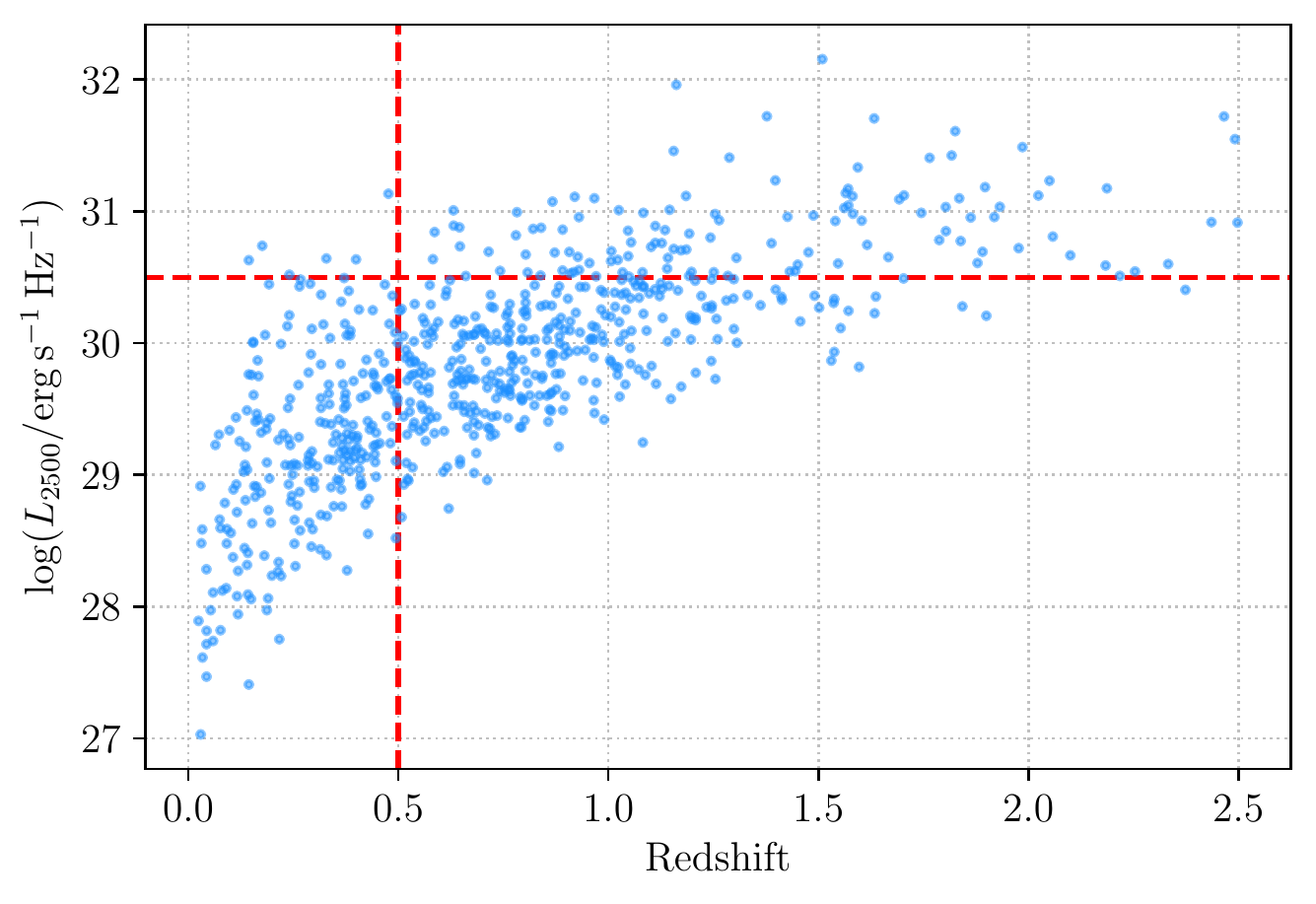}
    \caption{The UV luminosity at 2500~\si\angstrom\ versus source redshift.  The vertical red dashed line indicates the approximate redshift above which rest-frame 2500~\si\angstrom\ is contained within an SDSS spectrum.
    The horizontal red dashed line indicates the luminosity above which we see a substantial discrepancy in the $\log(L_{2500})$-$\alpha_\mathrm{ox}$ relations for spectroscopically- and photometrically-derived quantities (see Fig.~\ref{fig:mix_lum_aox}).}
    \label{fig:luv_z}
\end{figure}

We have recomputed $\log(L_{2500})$-$\alpha_\mathrm{ox}$ relations for the spectroscopic and photometric quantities separately.
The best-fitting linear regressions for the two subsets both show a statistically significant positive correlation between $\alpha_\mathrm{ox}$ and $\log(L_{2500})$, as was found for the full sample in Section~\ref{sec:aox} (see Table~\ref{tab:aox_spec_phot}).
In Fig.~\ref{fig:mix_lum_aox} we show these best-fitting relations and their $3\sigma$ uncertainty regions.
The relation for the spectroscopic subset is slightly steeper (slope $0.168\pm0.011$) than that of the photometric subset (slope $0.127\pm0.009$).
In the plot we see that the two relations agree to within $3\sigma$ up to $\log(L_{2500}/\mathrm{erg\,s^{-1}\,Hz^{-1}})\approx30.5$, above which we see a greater discrepancy.
Again, we judge that this will have only a minor effect on our results.  
As stated in Section~\ref{sec:uv_lum}, we have taken $L_{2500}$ from the optical spectrum where it is available (which it is for sources at $z\gtrsim0.5$) and used this value to determine $\alpha_\mathrm{ox}$.
Only where $L_{2500}$ is not available in an optical spectrum will we have estimated $L_{2500}$ and $\alpha_\mathrm{ox}$ from photometry.
Fig.~\ref{fig:luv_z} demonstrates that there are very few sources in the region of parameter space in which $L_{2500}$ can only be determined from photometry and the spectroscopic and photometric $\log(L_{2500})$-$\alpha_\mathrm{ox}$ relations are substantially different.
Since there is very good agreement between the spectroscopic and photometric $\log(L_{2500})$-$\alpha_\mathrm{ox}$ relations in the region of parameter space that contains the majority of our sources, we do not believe that mixing the two types of measurement will have a substantial impact on our results and conclusions.
This is corroborated by the fact that we see good agreement between our $\log(L_{2500})$-$\alpha_\mathrm{ox}$ relations and those of previous studies (e.g.\ \citealt{Strateva05}; \citealt{Lusso10}; \citealt{Zhu20}).

Based on the findings of previous work, we do not expect the scale of aperture effects to be very substantial in the SOUX sample.
In a sample of 51 $z<0.4$ AGN, \cite{Jin1} noticed a discrepancy between flux levels of the SDSS spectra and \textit{XMM} photometry for 17 sources, all with $z\lesssim0.24$. 
The photometric points were above the extrapolation from the optical spectra, so it was proposed that the discrepancies were due to aperture effects, as described above.
To obtain better agreement, \cite{Jin1} reanalysed the optical-UV images for affected sources and remeasured the photometry using a smaller aperture, excluding more of the extended host galaxy flux.
The optical-UV photometric fluxes were generally reduced by $\sim50$~per cent when using a smaller aperture (see Fig.1 of \citealt{Jin1}).
A bespoke reanalysis of the OM photometry for AGN in the SOUX sample is not feasible since it contains hundreds of sources and our investigation here is based on the analysis of standard catalogue and pipeline data, rather than tailored data reductions.
Unlike the low-redshift AGN sample of \cite{Jin1}, the SOUX sample consists of AGN drawn from NLS1 and quasar catalogues, so our AGN are generally UV-bright and not strongly contaminated by their host galaxy flux in the UV.

\begin{table}
    \caption{Linear regression parameters for $\log(L_{2500})$-$\alpha_\mathrm{ox}$ relations when $L_{2500}$ is derived from either spectra or photometry}
    \centering
    \begin{tabular}{llccc}
    \hline
    Subset                                     & $N$ & Slope           & Intercept        & $\sigma$ \\
    \hline
    $\log\left(L_{2500}^\mathrm{spec.}\right)$ & 416 & $0.168\pm0.011$ & $-3.64\pm0.32$   & 0.11 \\
    $\log\left(L_{2500}^\mathrm{phot.}\right)$ & 416 & $0.127\pm0.009$ & $-2.43\pm0.28$   & 0.12 \\
    All with $\alpha_\mathrm{ox}$              & 685 & $0.120\pm0.006$ & $-2.16\pm0.19$   & 0.13 \\
    \hline
    \end{tabular}
    \parbox[]{7.5cm}{For the $\log\left(L_{2500}^\mathrm{spec.}\right)$ subset $\log(L_{2500})$ has been derived from the optical spectrum and this value is used to determine $\alpha_\mathrm{ox}$.
    The $\log\left(L_{2500}^\mathrm{phot.}\right)$ subset contains the same 416 sources, but $\alpha_\mathrm{ox}$ has been determined from the photometrically-derived $\log(L_{2500})$.
    For comparison we also give values for the full set of 685 AGN with $\alpha_\mathrm{ox}$, which contains a mixture of spectroscopically (421) and photometrically (264) derived quantities.}
    \label{tab:aox_spec_phot}
\end{table}

\section{AGN with multiple optical spectra}
\label{sec:multispec}
\begin{figure}
    \centering
    \includegraphics[width=.8\columnwidth]{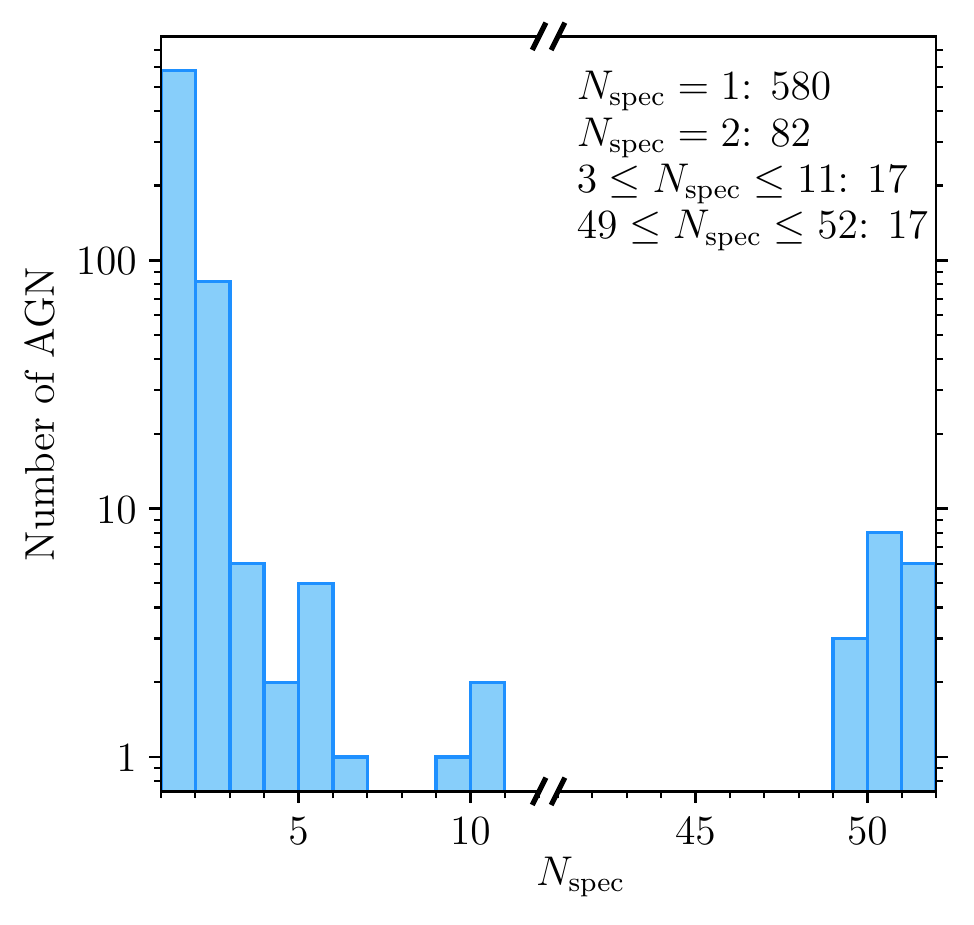}
    \caption{The number of SDSS optical spectra ($N_\mathrm{spec}$) available for the 696 AGN in our sample, up to and including DR14.  The majority of our sample (83~per cent) have only one spectrum, but a small number have $\approx50$ spectra.}
    \label{fig:multispec}
\end{figure}

Whilst in this study we consider only one optical spectrum per AGN, many sources have been observed multiple times.
Here we show quantitatively the number of spectra for the AGN in the SOUX sample.
We use \textsc{astroquery} (\citealt{astroquery}) to search for all optical spectra of our 696 AGN in SDSS DR14.
The query performs a 2$^{\prime\prime}$-radius search around the input coordinates (taken from the \citealt{SDSS-DR14Q} and \citetalias{Rakshit17} catalogues) for sources with spectroscopy.
We checked that all spectra returned by the query relate to the same object (i.e.\ they have the same \texttt{ObjID}) and that the specific spectrum analysed in this work is one of those.
No results were returned by \textsc{astroquery} for seven AGN; for these we manually checked for all spectra in the SDSS Science Archive Server\footnote{\url{https://dr14.sdss.org/optical/spectrum/search}}.
We show the distribution of the number of spectra in Fig.~\ref{fig:multispec}.
Whilst 83~per cent of the AGN in our sample have only one optical spectrum, 17 have many ($>49$).
The wealth of optical data for these AGN may be explored further in future studies.

\section{Quality flags}
\label{sec:qual}
We assess the quality of measurements determined from the spectral fits by assigning a quality flag.
If no issues are raised the quality flag value is 0 and the measurement may be considered to be reliable.
The quality flags are integer numbers which are the sum $2^\mathrm{Bit~0} + 2^\mathrm{Bit~1} ... + 2^{\mathrm{Bit}~n}$ for set bits.
The bit numbers and the conditions for setting them are listed in Tables~\ref{tab:cont_q_flags} and \ref{tab:line_q_flags}.
For example, if the fractional uncertainty of the FWHM of a measured broad \hb\ line  is greater than 2 and the uncertainty on the velocity offset of the line is greater than 1000~km\,s$^{-1}$ then bits 5 and 8 are set and the \hb\ quality flag has the value $2^5 + 2^8 = 32 + 256 = 288$.

\begin{table}
    \centering
    \caption{Quality flags for the fitted continuum}
    \resizebox{\columnwidth}{!}{\begin{tabular}{ll}
        \hline
         Bit  & Condition  \\
         \hline
         -    & no quality flags raised \\
         0    & the luminosity or its uncertainty are zero, NaN or infinite \\
         1    & fractional uncertainty of the luminosity > 1.5 \\
         2    & the continuum slope or its uncertainty are zero, NaN or infinite     \\
         3    & the continuum slope has reached the lower or upper limit in the fit     \\
         4    & the uncertainty on the continuum slope > 0.3      \\
         5    & the continuum fit $\chi^2_\nu>50$     \\
         \hline
    \end{tabular}}
    \label{tab:cont_q_flags}
\end{table}

\begin{table}
    \centering
    \caption{Quality flags for the fitted emission lines}
    \resizebox{\columnwidth}{!}{\begin{tabular}{ll}
        \hline
         Bit  & Condition  \\
         \hline
         -    & no quality flags raised \\
         0    & S/N in the continuum around the line < 3 \\
         1    & the luminosity or its uncertainty are zero, NaN or infinite \\
         2    & fractional uncertainty of the luminosity > 1.5     \\
         3    & the FWHM or its uncertainty are zero, NaN or infinite     \\
         4    & the $\mathrm{FWHM}\leqslant910$~km\,s$^{-1}$      \\
         5    & the fractional uncertainty of the FWHM > 2     \\
         6    & the velocity offset or its uncertainty are zero, NaN or infinite     \\
         7    & the velocity offset has reached the lower or upper limit in the fit     \\
         8    & the uncertainty on the velocity offset > 1000~km\,s$^{-1}$     \\
         \hline
    \end{tabular}}
    \label{tab:line_q_flags}
\end{table}

\section{Table information}
\label{sec:table}
We have made available online a table containing the details (RA, Dec., $z$, etc.), measurements (emission line widths, optical and X-ray continuum luminosities etc.) and derived quantities (black hole masses, $\alpha_\mathrm{ox}$ etc.) of the 696 AGN in the SOUX sample.
Table~\ref{tab:soux_table} gives an overview of the columns of that table. 

\begin{table*}
    \centering
    \caption{The SOUX AGN Sample - Column information for the table containing the details, measurements and derived quantities for the 696 AGN}
    \resizebox{\textwidth}{!}{\begin{tabular}{cllp{0.66\textwidth}}
         \hline
         \# & Column name            & Units         & Description  \\
         \hline
         \multicolumn{4}{l}{Source and SDSS optical spectrum details:} \\
         1  & SDSS\_ID               &               & Unique SDSS spectrum identifier (PLATE-MJD-FIBER) \\
         2  & RA                     & Degrees       & Right Ascension (J2000) \\
         3  & DEC                    & Degrees       & Declination (J2000) \\
         4  & Z                      &               & Spectroscopic redshift \\
         5  & DR14Q                  &               & 0: source is not in the SDSS DR14 quasar catalog; 1: source is in the SDSS DR14 quasar catalog \\
         6  & R17                    &               & 0: source is not in the \citetalias{Rakshit17} NLS1 catalog; 1: source is in the \citetalias{Rakshit17} NLS1 catalog \\
         7  & SN\_RATIO\_CONT        &               & Median S/N sampled in emission-line free windows of the spectrum  \\
         8  & MIN\_WAVE              & \AA\          & Shortest rest-frame wavelength covered in the optical spectrum \\
         9  & MAX\_WAVE              & \AA\          & Longest rest-frame wavelength covered in the optical spectrum \\
         \multicolumn{4}{l}{UV-optical continuum measurements:} \\
         10 & LOG\_L1350             & erg\,s$^{-1}$ & Logarithm of the continuum luminosity at 1350~\AA\ (rest-frame) from the SDSS spectrum \\
         11 & LOG\_L1350\_ERR        & erg\,s$^{-1}$ & Uncertainty on LOG\_L1350\_ERR \\
         12 & QUALITY\_L1350         &               & Quality flag on LOG\_L1350 (see Table~\ref{tab:cont_q_flags}) \\
         
         13 & LOG\_L2500\_SPEC             & erg\,s$^{-1}$ & Logarithm of the continuum luminosity at 2500~\AA\ (rest-frame) from the SDSS spectrum \\
         14 & LOG\_L2500\_SPEC\_ERR        & erg\,s$^{-1}$ & Uncertainty on LOG\_L2500\_SPEC \\
         15 & LOG\_L2500\_PHOT\_OM         & erg\,s$^{-1}$ & Logarithm of the continuum luminosity at 2500~\AA\ (rest-frame) from \xmm\ OM photometry \\
         16 & LOG\_L2500\_PHOT\_OM\_ERR    & erg\,s$^{-1}$ & Uncertainty on LOG\_L2500\_PHOT\_OM \\
         17 & LOG\_L2500\_PHOT\_SDSS       & erg\,s$^{-1}$ & Logarithm of the continuum luminosity at 2500~\AA\ (rest-frame) from SDSS photometry \\
         18 & LOG\_L2500\_PHOT\_SDSS\_ERR  & erg\,s$^{-1}$ & Uncertainty on LOG\_L2500\_PHOT\_SDSS \\
         19 & LOG\_L2500\_ALPHA\_OX        &               & Which $L_{2500}$ measurement was used to calculate ALPHA\_OX (\#{86}) \\
         
         20 & LOG\_L3000             & erg\,s$^{-1}$ & Logarithm of the continuum luminosity at 3000~\AA\ (rest-frame) from the SDSS spectrum \\
         21 & LOG\_L3000\_ERR        & erg\,s$^{-1}$ & Uncertainty on LOG\_L3000\_ERR \\
         22 & QUALITY\_L3000         &               & Quality flag on LOG\_L3000 (see Table~\ref{tab:cont_q_flags}) \\

         23 & LOG\_L4400\_SPEC             & erg\,s$^{-1}$ & Logarithm of the continuum luminosity at 4400~\AA\ (rest-frame) from the SDSS spectrum \\
         24 & LOG\_L4400\_SPEC\_ERR        & erg\,s$^{-1}$ & Uncertainty on LOG\_L4400\_SPEC \\
         25 & LOG\_L4400\_PHOT\_OM         & erg\,s$^{-1}$ & Logarithm of the continuum luminosity at 4400~\AA\ (rest-frame) from \xmm\ OM photometry \\
         26 & LOG\_L4400\_PHOT\_OM\_ERR    & erg\,s$^{-1}$ & Uncertainty on LOG\_L4400\_PHOT\_OM \\
         27 & LOG\_L4400\_PHOT\_SDSS       & erg\,s$^{-1}$ & Logarithm of the continuum luminosity at 4400~\AA\ (rest-frame) from SDSS photometry \\
         28 & LOG\_L4400\_PHOT\_SDSS\_ERR  & erg\,s$^{-1}$ & Uncertainty on LOG\_L4400\_PHOT\_SDSS \\         
         29 & LOG\_L4400\_R\_LOUD          &               & Which $L_{4400}$ measurement was used to calculate R\_LOUD (\#{90}) \\
         
         30 & LOG\_L5100             & erg\,s$^{-1}$ & Logarithm of the continuum luminosity at 5100~\AA\ (rest-frame) from the SDSS spectrum\\
         31 & LOG\_L5100\_ERR        & erg\,s$^{-1}$ & Uncertainty on LOG\_L5100\_ERR \\
         32 & QUALITY\_L5100         &               & Quality flag on LOG\_L5100 (see Table~\ref{tab:cont_q_flags}) \\
         \multicolumn{4}{l}{Broad emission line measurements:} \\
         33 & FWHM\_HA\_BR           & km\,s$^{-1}$  & Full width at half maximum of the broad \ha\ emission line \\
         34 & FWHM\_HA\_BR\_ERR      & km\,s$^{-1}$  & Uncertainty on FWHM\_HA\_BR \\
         35 & PEAK\_HA\_BR           & \AA\          & Peak wavelength of the broad \ha\ emission line \\
         36 & PEAK\_HA\_BR\_ERR      & \AA\          & Uncertainty on PEAK\_HA\_BR \\
         37 & EW\_HA\_BR             & \AA\          & Equivalent width of the broad \ha\ emission line \\
         38 & EW\_HA\_BR\_ERR        & \AA\          & Uncertainty on EW\_HA\_BR \\
         39 & LOGL\_HA\_BR           & erg\,s$^{-1}$ & Logarithm of the luminosity of the broad \ha\ emission line \\
         40 & LOGL\_HA\_BR\_ERR      & erg\,s$^{-1}$ & Uncertainty on LOGL\_HA\_BR \\
         41 & QUALITY\_HA            &               & Quality flag on the broad \ha\ emission line (see Table~\ref{tab:line_q_flags}) \\
         42 & FWHM\_HB\_BR           & km\,s$^{-1}$  & Full width at half maximum of the broad \hb\ emission line \\
         43 & FWHM\_HB\_BR\_ERR      & km\,s$^{-1}$  & Uncertainty on FWHM\_HB\_BR \\
         44 & PEAK\_HB\_BR           & \AA\          & Peak wavelength of the broad \hb\ emission line \\
         45 & PEAK\_HB\_BR\_ERR      & \AA\          & Uncertainty on PEAK\_HB\_BR \\
         46 & EW\_HB\_BR             & \AA\          & Equivalent width of the broad \hb\ emission line \\
         47 & EW\_HB\_BR\_ERR        & \AA\          & Uncertainty on EW\_HB\_BR \\
         48 & LOGL\_HB\_BR           & erg\,s$^{-1}$ & Logarithm of the luminosity of the broad \hb\ emission line \\
         49 & LOGL\_HB\_BR\_ERR      & erg\,s$^{-1}$ & Uncertainty on LOGL\_HB\_BR \\
         50 & QUALITY\_HB            &               & Quality flag on the broad \hb\ emission line (see Table~\ref{tab:line_q_flags}) \\
         
         51 & BALMER\_SOURCE         &               & K23: Balmer lines measurements from this work; R20: Balmer lines measured by \citetalias{Rakshit20} \\

         \hline 
    \end{tabular}}
    \label{tab:soux_table}
\end{table*}

\begin{table*}
\ContinuedFloat
    \centering
    \caption{Table~\ref{tab:soux_table} continued}
    \resizebox{\textwidth}{!}{\begin{tabular}{cllp{0.66\textwidth}}
    
         \hline
         \# & Column name             & Units         & Description  \\
         \hline
         52 & FWHM\_MGII\_BR         & km\,s$^{-1}$  & Full width at half maximum of the broad \mg\ emission line \\
         53 & FWHM\_MGII\_BR\_ERR    & km\,s$^{-1}$  & Uncertainty on FWHM\_MGII\_BR \\
         54 & PEAK\_MGII\_BR         & \AA\          & Peak wavelength of the broad \mg\ emission line \\
         55 & PEAK\_MGII\_BR\_ERR    & \AA\          & Uncertainty on PEAK\_MGII\_BR \\
         56 & EW\_MGII\_BR           & \AA\          & Equivalent width of the broad \mg\ emission line \\
         57 & EW\_MGII\_BR\_ERR      & \AA\          & Uncertainty on EW\_MGII\_BR \\
         58 & LOGL\_MGII\_BR         & erg\,s$^{-1}$ & Logarithm of the luminosity of the broad \mg\ emission line \\
         59 & LOGL\_MGII\_BR\_ERR    & erg\,s$^{-1}$ & Uncertainty on LOGL\_MGII\_BR \\
         60 & QUALITY\_MGII          &               & Quality flag on the broad \mg\ emission line (see Table~\ref{tab:line_q_flags}) \\
         61 & MGII\_SOURCE           &               & K23: \mg\ measurements from this work; R20: \mg\ measured by \citetalias{Rakshit20} \\
         62 & MGII\_SEP              &               & 0: \mg\ region modelled in global fit to full spectrum; 1: \mg\ region modelled separately to Balmer line regions (see Section~\ref{sec:pyqsofit}) \\         
         63 & LINEWIDTH              &               & N: narrow-line type 1 AGN; B: broad-line type 1 AGN (see Section~\ref{sec:bl_nl}) \\
         \multicolumn{4}{l}{Black hole mass estimates:} \\
         64 & MASS\_HA\_5100\_MR16   & M$_\odot$     & BH mass estimate from FWHM\_HA\_BR and LOG\_L5100 using the relation of \citetalias{Mejia16} \\
         65 & MASS\_HA\_HA\_MR16     & M$_\odot$     & BH mass estimate from FWHM\_HA\_BR and LOGL\_HA\_BR using the relation of \citetalias{Mejia16} \\
         66 & MASS\_HB\_5100\_MR16   & M$_\odot$     & BH mass estimate from FWHM\_HB\_BR and LOG\_L5100 using the relation of \citetalias{Mejia16} \\ 
         67 & MASS\_HB\_HB\_G10      & M$_\odot$     & BH mass estimate from FWHM\_HB\_BR and LOGL\_HB\_BR using the relation of \cite{Greene10} \\
         68 & MASS\_MGII\_3000\_MR16 & M$_\odot$     & BH mass estimate from FWHM\_MGII\_BR and LOG\_L3000 using the relation of \citetalias{Mejia16} \\ 
         69 & MASS\_MGII\_MGII\_W18  & M$_\odot$     & BH mass estimate from FWHM\_MGII\_BR and LOGL\_MGII\_BR using the relation of \cite{Woo18} \\

         70 & MASS\_HA\_5100\_K23    & M$_\odot$     & BH mass estimate from FWHM\_HA\_BR and LOG\_L5100 using the relation found in this work \\
         71 & MASS\_HA\_HA\_K23      & M$_\odot$     & BH mass estimate from FWHM\_HA\_BR and LOGL\_HA\_BR using the relation found in this work \\
         72 & MASS\_HB\_HB\_K23      & M$_\odot$     & BH mass estimate from FWHM\_HB\_BR and LOGL\_HB\_BR using the relation found in this work \\
         73 & MASS\_MGII\_3000\_K23  & M$_\odot$     & BH mass estimate from FWHM\_MGII\_BR and LOG\_L3000 using the relation found in this work \\ 
         74 & MASS\_MGII\_MGII\_K23  & M$_\odot$     & BH mass estimate from FWHM\_MGII\_BR and LOGL\_MGII\_BR using the relation found in this work \\
         
         75 & Q\_MASS\_HA\_5100       &               & Quality flag on \ha-5100\,\si\angstrom\ mass: $=1$ unless QUALITY\_HA and QUALITY\_L5100 are both 0 \\ 
         76 & Q\_MASS\_HB\_5100       &               & Quality flag on \hb-5100\,\si\angstrom\ mass: $=1$ unless QUALITY\_HB and QUALITY\_L5100 are both 0 \\ 
         77 & Q\_MASS\_MGII\_3000     &               & Quality flag on \mg-3000\,\si\angstrom\ mass: $=1$ unless QUALITY\_MGII and QUALITY\_L3000 are both 0 \\ 
         78 & LOG\_MASS\_PREF\_VALUE  & M$_\odot$     & Logarithm of our preferred BH mass estimate \\
         79 & LOG\_MASS\_PREF\_SOURCE &               & Which BH mass estimate is the preferred one \\
        
        \multicolumn{4}{l}{Multiwavelength measurements:} \\ 
         80 & XMM\_OBSID              &               & The \xmm\ Observation ID for the X-ray and OM photometric data \\
         81 & XMM\_SRCID              &               & The unique X-ray source number taken from the \xmm\ catalog \\
         82 & LOG\_L2KEV              & erg\,s$^{-1}$\,Hz$^{-1}$ & Logarithm of $L_\nu(2~\mathrm{keV~\mbox{rest-frame}})$ estimated from the \xmm\ band fluxes (see Section~\ref{sec:xray_lum}) \\
         83 & LOG\_L2KEV\_ERR         & erg\,s$^{-1}$\,Hz$^{-1}$ & Uncertainty on LOG\_L2KEV \\
         84 & GAMMA\_X                &               & Estimated X-ray photon index (see Section~\ref{sec:xray_lum}) \\
         85 & GAMMA\_X\_ERR           &               & Uncertainty on GAMMA\_X \\
         86 & ALPHA\_OX               &               & Optical-X-ray energy index $\alpha_\mathrm{ox}$ calculated from LOG\_L2500\_ALPHA\_OX and LOG\_L2KEV \\
         87 & ALPHA\_OX\_ERR          &               & Uncertainty on ALPHA\_OX \\
         88 & LOG\_L5GHZ              & W\,Hz$^{-1}$  & Logarithm of $L_\nu(5~\mathrm{GHz~\mbox{rest-frame}})$ estimated from the FIRST radio flux (see Section~\ref{sec:radio}) \\
         89 & LOG\_L5GHZ\_ERR         & W\,Hz$^{-1}$  & Uncertainty on L5GHZ \\
         90 & R\_LOUD                 &               & Radio loudness calculated from LOG\_L4400\_R\_LOUD and LOG\_L5GHZ \\
         91 & R\_LOUD\_FLAG           &               & RL: radio-loud; RQ: radio-detected and radio-quiet; RQu: radio-undetected and radio-quiet; RU: radio-undetected and undetermined radio-loudness; $-$999: outside FIRST footprint or no $L_{4400}$ \\

         \hline 
    \end{tabular}}
\end{table*}


\bsp	
\label{lastpage}
\end{document}